\documentclass{article}
\usepackage{bussproofs}
\usepackage{epsfig}
\usepackage[all, knot]{xy}
\usepackage{hyperref}
\usepackage{latexsym,url}
\usepackage{amssymb}
\usepackage{txfonts}
\usepackage{undertilde}
\usepackage{pb-diagram}
\usepackage{pstricks}
\usepackage{pstricks-add}
\usepackage{pst-grad}
\usepackage{pst-plot}
\usepackage[tiling]{pst-fill}
\usepackage{pstcol}
\psset{linewidth=0.3pt,dimen=middle}
\psset{xunit=.70cm,yunit=0.70cm}

\newtheorem{thm}{Theorem}
\newtheorem{definition}[thm]{Definition}

\newcommand{\NN}{\mathbb{N}}
\newcommand{\CC}{\mathbb{C}}

\newcommand{\Cob}{\mathrm{Cob}}
\newcommand{\Hilb}{\mathrm{Hilb}}
\newcommand{\Set}{\mathrm{Set}}

\newcommand{\Tang}{{\rm Tang}}

\newcommand{\Vect}{{\rm Vect}}

\newcommand{\maps}{\colon}
\newcommand{\iso}{\cong}
\newcommand{\isoto}{\xrightarrow{\sim}}
\newcommand{\To}{\Rightarrow}
\newcommand{\xrightarrow}[1]{\stackrel{#1}{\rightarrow}}

\newcommand{\lHom}{\vdash}
\newcommand{\lhom}{\multimap}

\renewcommand{\hom}{{\rm hom}}

\newcommand{\tensor}{\otimes}

\newcommand{\id}{{\rm i}}
\newcommand{\Id}{{\rm id}}
\newcommand{\ev}{{\rm ev}}

\newcommand{\eval}{{\rm eval}}

\newcommand{\assoc}{{\rm assoc}}
\newcommand{\unassoc}{{\rm unassoc}}
\newcommand{\braid}{{\rm braid}}
\newcommand{\Left}{{\rm left}}
\newcommand{\Right}{{\rm right}}
\newcommand{\unright}{{\rm unright}}
\newcommand{\unleft}{{\rm unleft}}

\newcommand{\cp}{{\rm cp}}
\newcommand{\vp}{{\rm vp}}
\newcommand{\cut}{{\circ}}
\newcommand{\op}{{\rm op}}   
\newcommand{\name}[1]{\ulcorner \! #1 \! \urcorner}

\newcommand{\Times}{{\rm times}}

\newcommand{\Day}{{\rm day}}
\newcommand{\Tuesday}{{\rm Tuesday}}
\newcommand{\integer}{{\rm integer}}
\newcommand{\duplicate}{{\rm duplicate}}

\renewcommand{\H}{H}
\renewcommand{\O}{O}

\newcommand{\cent}[1]{\begin{center} #1 \end{center}}

\newcommand{\di}[1]{\[\begin{diagram}#1\end{diagram}\]}

\newcommand{\text}{\mbox}

\newcommand{\multc}{
      \pscustom[fillstyle=gradient,
    gradbegin=white, gradend=gray,gradmidpoint=0,gradangle=70]{
        \psbezier(1.5,2.5)(1.5,1.1)(.4,1.6)(.5,0)
        \psbezier(.5,0)(.4,-.25)(-.4,-.25)(-.5,0)
        \psbezier(-0.5,0)(-.4,1.6)(-1.5,1.1)(-1.5,2.5)
        \psline(-.5,2.5)
        \psbezier(-.5,2.5)(-.6,1.5)(0.6,1.5)(.5,2.5)
        \psline(1.5,2.5)
    }
    \psellipse[fillcolor=lightgray,fillstyle=gradient,
        gradbegin=lightgray, gradend=gray,gradmidpoint=1,gradangle=110](-1,2.5)(.5,.2)
    \psellipse[fillcolor=lightgray,fillstyle=gradient,
        gradbegin=lightgray, gradend=gray,gradmidpoint=1,gradangle=110](1,2.5)(.5,.2)
     \begin{psclip}{
 \pspolygon[linestyle=none](.5,0)(.5,.3)(-.5,.3)(-.5,0)(.5,0)
 }
 \psellipse[linestyle=dotted](0,0)(.5,0.2)
 \end{psclip}
 }

\newcommand{\comultc}{
  \pscustom[fillstyle=gradient,
    gradbegin=white, gradend=gray,gradmidpoint=0,gradangle=110]{
        \psbezier(1.5,0)(1.5,1.4)(.4,.9)(.5,2.5)
        \psline(-0.5,2.5)
        \psbezier(-0.5,2.5)(-.4,.9)(-1.5,1.4)(-1.5,0)
        \psbezier(-1.5,0)(-1.4,-.25)(-.6,-.25)(-.5,0)
        \psbezier(-.5,0)(-.6,1)(0.6,1)(.5,0)
        \psbezier(.5,0)(.6,-.25)(1.4,-.25)(1.5,0)
    }
  \psellipse[fillcolor=lightgray,fillstyle=gradient,
        gradbegin=lightgray, gradend=gray,gradmidpoint=1,gradangle=110](0,2.5)(.5,.2)
\begin{psclip}{
 \pspolygon[linestyle=none](1.5,0)(1.5,.3)(-1.5,.3)(-1.5,0)(1.5,0)
 }
 \psellipse[linestyle=dotted](1,0)(.5,0.2)
 \psellipse[linestyle=dotted](-1,0)(.5,0.2)
 \end{psclip}
 }

\newcommand{\identc}{
 \pscustom[fillcolor=lightgray,fillstyle=gradient,
        gradbegin=white, gradend=gray,gradmidpoint=0,gradangle=88]{
 \psline(.5,0)(.5,2.5)
 \psline(-.5,2.5)
 \psline(-.5,0)
 \psbezier(-.5,0)(-.4,-.25)(.4,-.25)(.5,0)
 }
\psellipse[fillcolor=lightgray,fillstyle=gradient,
        gradbegin=lightgray, gradend=gray,gradmidpoint=1,gradangle=110](0,2.5)(.5,.2)
 \begin{psclip}{
 \pspolygon[linestyle=none](.5,0)(.5,.3)(-.5,.3)(-.5,0)(.5,0)
 }
 \psellipse[linestyle=dotted](0,0)(.5,0.2)
 \end{psclip}
 }

  \newcommand{\medidentc}{
     \pscustom[fillcolor=lightgray,fillstyle=gradient,
        gradbegin=white, gradend=gray,gradmidpoint=0,gradangle=88]{
        \psline(-.5,2)(-.5,0)
        \psbezier(-.5,0)(-.4,-.25)(.4,-.25)(.5,0)
        \psline(.5,2)
        \psline(-.5,2)
    }
\psellipse[fillcolor=lightgray,fillstyle=gradient,
        gradbegin=lightgray, gradend=gray,gradmidpoint=1,gradangle=110](0,2)(.5,.2)
 \begin{psclip}{
 \pspolygon[linestyle=none](.5,0)(.5,.3)(-.5,.3)(-.5,0)(.5,0)
 }
 \psellipse[linestyle=dotted](0,0)(.5,0.2)
 \end{psclip}
}

 \newcommand{\smallidentc}{
     \pscustom[fillcolor=lightgray,fillstyle=gradient,
        gradbegin=white, gradend=gray,gradmidpoint=0,gradangle=88]{
        \psline(-.5,1)(-.5,0)
        \psbezier(-.5,0)(-.4,-.25)(.4,-.25)(.5,0)
        \psline(.5,1)
        \psline(-.5,1)
    }
\psellipse[fillcolor=lightgray,fillstyle=gradient,
        gradbegin=lightgray, gradend=gray,gradmidpoint=1,gradangle=110](0,1)(.5,.2)
 \begin{psclip}{
 \pspolygon[linestyle=none](.5,0)(.5,.3)(-.5,.3)(-.5,0)(.5,0)
 }
 \psellipse[linestyle=dotted](0,0)(.5,0.2)
 \end{psclip}
}

\newcommand{\birthc}{
 \pscustom[fillstyle=gradient,
    gradbegin=white, gradend=gray,gradmidpoint=0,gradangle=110]{
        \psbezier(-.5,0)(-.5,.9)(0.5,.9)(.5,0)
        \psbezier(.5,0)(.4,-.25)(-.4,-.25)(-.5,0)
    }
 \begin{psclip}{
 \pspolygon[linestyle=none](.5,0)(.5,.3)(-.5,.3)(-.5,0)(.5,0)
 }
 \psellipse[linestyle=dotted](0,0)(.5,0.2)
 \end{psclip}
 }

\newcommand{\ucrossc}{
\pscustom[fillcolor=lightgray,fillstyle=gradient,
        gradbegin=white, gradend=gray,gradmidpoint=0,gradangle=125]{
 \psline(.5,0)(-1.5,2.5)
 \psline(-.5,2.5)
 \psline(1.5,0)
 \psbezier(1.5,0)(1.4,-.25)(.6,-.25)(.5,0)
 }
  \pscustom[fillcolor=lightgray,fillstyle=gradient,
        gradbegin=white, gradend=gray,gradmidpoint=0,gradangle=70]{
 \psline(-.5,0)(1.5,2.5)
 \psline(.5,2.5)
 \psline(-1.5,0)
 \psbezier(-1.5,0)(-1.4,-.25)(-.6,-.25)(-.5,0)
 }
\psellipse[fillcolor=lightgray,fillstyle=gradient,
        gradbegin=lightgray, gradend=gray,gradmidpoint=1,gradangle=110](-1,2.5)(.5,.2)
\psellipse[fillcolor=lightgray,fillstyle=gradient,
        gradbegin=lightgray, gradend=gray,gradmidpoint=1,gradangle=110](1,2.5)(.5,.2)
 \psline[linestyle=dotted](-1.5,2.5)(.5,0)
 \psline[linestyle=dotted](-.5,2.5)(1.5,0)
 \begin{psclip}{
 \pspolygon[linestyle=none](1.5,0)(1.5,.3)(-1.5,.3)(-1.5,0)(1.5,0)
 }
 \psellipse[linestyle=dotted](1,0)(.5,0.2)
 \psellipse[linestyle=dotted](-1,0)(.5,0.2)
 \end{psclip}
}

\newcommand{\zagc}{
   \pscustom[fillstyle=gradient,
    gradbegin=white, gradend=gray,gradmidpoint=0,gradangle=110]{
        \psbezier(1.5,0)(1.6,2)(-1.6,2)(-1.5,0)
        \psbezier(-1.5,0)(-1.4,-.25)(-.6,-.25)(-.5,0)
        \psbezier(-.5,0)(-.6,.8)(0.6,.8)(.5,0)
        \psbezier(.5,0)(.6,-.25)(1.4,-.25)(1.5,0)
    }
  \begin{psclip}{
 \pspolygon[linestyle=none](1.5,0)(1.5,.3)(-1.5,.3)(-1.5,0)(1.5,0)
 }
 \psellipse[linestyle=dotted](1,0)(.5,0.2)
 \psellipse[linestyle=dotted](-1,0)(.5,0.2)
 \end{psclip}
 }

\newcommand{\zigc}{
       \pscustom[fillstyle=gradient,
    gradbegin=white, gradend=gray,gradmidpoint=0,gradangle=70]{
        \psbezier(1.5,2)(1.6,0)(-1.6,0)(-1.5,2)
        \psline(-.5,2)
        \psbezier(-.5,2)(-.6,1.2)(0.6,1.2)(.5,2)
        \psline(1.5,2)
    }
 \psellipse[fillcolor=lightgray,fillstyle=gradient,
        gradbegin=lightgray, gradend=gray,gradmidpoint=1,gradangle=110](1,2)(.5,.2)
        \psellipse[fillcolor=lightgray,fillstyle=gradient,
        gradbegin=lightgray, gradend=gray,gradmidpoint=1,gradangle=110](-1,2)(.5,.2)
}

\title{Physics, Topology, Logic and Computation: \\
     A Rosetta Stone}
{\small
     \author{{\small John C.\ Baez} \\
     {\small Department of Mathematics,  University of California} \\
     {\small Riverside, California 92521, USA} \\
     \\  
    {\small Mike Stay} \\
    {\small Computer Science Department, University of Auckland} \\
    {\small \it and} \\
    {\small Google, 1600 Amphitheatre Pkwy} \\
    {\small Mountain View, California 94043, USA}
    \\
    \\ {\small email: baez@math.ucr.edu, stay@google.com}
    \\
    }
}

\date{\small {March 2, 2009}}
\begin{document}

\maketitle

\begin{abstract}
In physics, Feynman diagrams are used to reason about quantum
processes. In the 1980s, it became clear that underlying these
diagrams is a powerful analogy between quantum physics and topology.
Namely, a linear operator behaves very much like a `cobordism': a
manifold representing spacetime, going between two manifolds
representing space.  This led to a burst of work on topological
quantum field theory and `quantum topology'.  But this was just
the beginning: similar diagrams can be used to reason about
logic, where they represent proofs, and computation, where they
represent programs.  With the rise of interest in quantum cryptography
and quantum computation, it became clear that there is extensive
network of analogies between physics, topology, logic and
computation. In this expository paper, we make some of these analogies
precise using the concept of `closed symmetric monoidal
category'. We assume no prior knowledge of category theory, proof
theory or computer science.
\end{abstract}

\section{Introduction} 

Category theory is a very general formalism, but there is a certain
special way that physicists use categories which turns out to have
close analogues in topology, logic and computation.  A category has
\emph{objects} and \emph{morphisms}, which represent \emph{things} and
\emph{ways to go between things}.  In physics, the objects are often 
\emph{physical systems}, and the morphisms are \emph{processes} turning 
a state of one physical system into a state of another system ---
perhaps the same one.  In quantum physics we often formalize this 
by taking \emph{Hilbert spaces} as objects, and \emph{linear operators}
as morphisms.

Sometime around 1949, Feynman \cite{Kaiser} realized that in quantum
field theory it is useful to draw linear operators as diagrams: 
\vskip 1em
\centerline{\epsfysize=1.2in\epsfbox{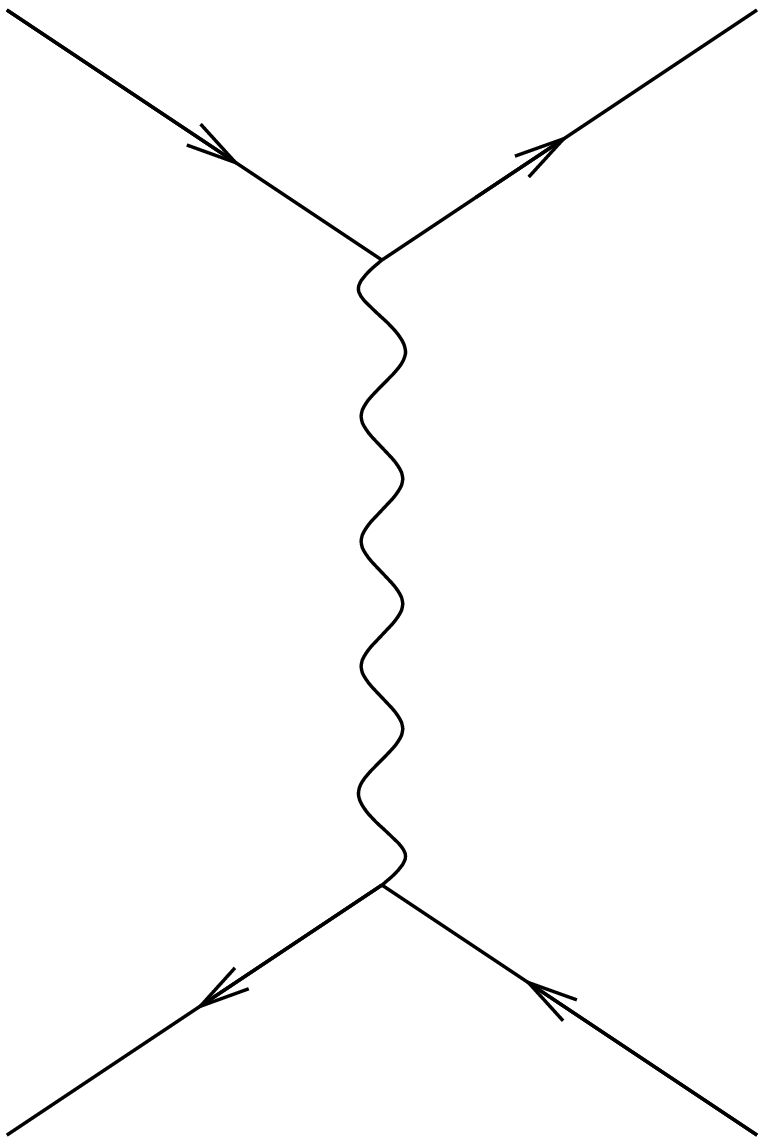}}
\vskip 1em
\noindent
This lets us reason with them pictorially.  We can warp a picture
without changing the operator it stands for: all that matters is the
topology, not the geometry.  In the 1970s, Penrose realized that
generalizations of Feynman diagrams arise throughout quantum theory,
and might even lead to revisions in our understanding of spacetime 
\cite{Penrose}.  In the 1980s, it became clear that underlying these 
diagrams is a powerful analogy between quantum physics and topology!  
Namely, a linear operator behaves very much like a `cobordism' ---
that is, an $n$-dimensional manifold going between manifolds of one
dimension less:

\[
\begin{pspicture}[0.5](2,5)
  \rput(1,0.5){\comultc}
  \rput(1,2.5){\multc}
\end{pspicture}
\]
\noindent
String theory exploits this analogy by replacing the Feynman diagrams
of ordinary quantum field theory with 2-dimensional cobordisms,
which represent  the worldsheets traced out by strings with the 
passage of time.  The analogy between operators and cobordisms 
is also important in loop quantum gravity and --- most of all ---
the more purely mathematical discipline of `topological quantum field 
theory'.  

Meanwhile, quite separately, logicians had begun using categories
where the objects represent \emph{propositions} and the morphisms
represent \emph{proofs}.  The idea is that a proof is a process going
from one proposition (the hypothesis) to another (the conclusion).  
Later, computer scientists started using categories where the 
objects represent \emph{data types} and the morphisms represent 
\emph{programs}.   They also started using `flow charts' to describe
programs.  Abstractly, these are very much like Feynman diagrams!

The logicians and computer scientists were never very far from each
other.  Indeed, the `Curry--Howard correspondence' relating proofs to
programs has been well-known at least since the early 1970s, with
roots stretching back earlier \cite{Curry,Howard}.  But, it is only
in the 1990s that the logicians and computer scientists bumped into
the physicists and topologists.  One reason is the rise of interest in 
quantum cryptography and quantum computation \cite{ChuangNielsen}. 
With this, people began to think of quantum processes as forms
of information processing, and apply ideas from computer science.  
It was then realized that the loose analogy between flow charts 
and Feynman diagrams could be made more precise and powerful with the 
aid of category theory \cite{AC}.

By now there is an extensive network of interlocking analogies
between physics, topology, logic and computer science.   They suggest
that research in the area of common overlap is actually trying to
build a new science: {\it a general science of systems and processes}.
Building this science will be very difficult.  
There are good reasons for this, but also bad ones.  One bad 
reason is that different fields use different terminology and notation.

The original Rosetta Stone, created in 196 BC, contains versions of
the same text in three languages: demotic Egyptian, hieroglyphic
script and classical Greek.  Its rediscovery by Napoleon's soldiers
let modern Egyptologists decipher the hieroglyphs.  Eventually this
led to a vast increase in our understanding of Egyptian culture.

At present, the deductive systems in mathematical logic look like hieroglyphs
to most physicists.  Similarly, quantum field theory is Greek to most
computer scientists, and so on.  So, there is a need for a new Rosetta
Stone to aid researchers attempting to translate between fields. 
Table \ref{analogy} shows our guess as to what this Rosetta
Stone might look like.  

\vskip 1em 
\begin{table}[h]
\begin{center}
\begin{tabular}{|c|c|c|c|c|}
\hline
Category Theory  &  Physics &   Topology  &  Logic       &  Computation \\
\hline
object           &  system  &  manifold   &  proposition &  data type \\ 
\hline
morphism         &  process &  cobordism  &  proof       &  program  \\
\hline
\end{tabular}
\\
\caption{The Rosetta Stone (pocket version)}
\label{analogy}
\end{center}
\end{table}
The rest of this paper expands on this table by comparing how
categories are used in physics, topology, logic, and computation.
Unfortunately, these different fields focus on slightly
different kinds of categories.  Though most physicists don't know it, 
quantum physics has long made use of `compact symmetric 
monoidal categories'.  Knot theory uses 
`compact braided monoidal categories', which are slightly more general.  
However, it became clear in the 1990's
that these more general gadgets are useful in physics too.  
Logic and computer science used to focus on `cartesian closed categories' --- 
where `cartesian' can be seen, roughly, as an antonym of `quantum'.  
However, thanks to work on linear logic and quantum computation, some 
logicians and computer scientists have dropped their insistence on 
cartesianness: now they study more general sorts of `closed symmetric 
monoidal categories'.

In Section \ref{physics_topology} we explain these concepts,
how they illuminate the analogy between physics and topology, and 
how to work with them using string diagrams.  We assume no prior 
knowledge of category theory, only a willingness to learn some.  
In Section \ref{logic} we explain how closed symmetric monoidal
categories correspond to a small fragment of ordinary 
propositional logic, which also happens to be a fragment of Girard's 
`linear logic' \cite{Girard1}.   In Section \ref{computation} 
we explain how closed symmetric monoidal categories correspond to a 
simple model of computation.   Each of these sections starts with
some background material.  In Section \ref{conclusions}, we conclude
by presenting a larger version of the Rosetta Stone.

Our treatment of all four subjects --- physics, topology, logic
and computation --- is bound to seem sketchy, narrowly focused
and idiosyncratic to practitioners of these subjects.  Our excuse 
is that we wish to emphasize certain analogies while saying no 
more than absolutely necessary.  To make up for this, we include 
many references for those who wish to dig deeper.

\section{The Analogy Between Physics and Topology}
\label{physics_topology}

\subsection{Background}

Currently our best theories of physics are general relativity 
and the Standard Model of particle physics.  The first describes
gravity without taking quantum theory into account; the second 
describes all the other forces taking quantum theory into account,
but ignores gravity.  So, our world-view is deeply schizophrenic.
The field where physicists struggle to solve this problem is 
called {\em quantum gravity}, since it is widely believed that the
solution requires treating gravity in a way that takes quantum
theory into account.

\begin{table}[h]
\begin{center}
\begin{tabular}{|c|c|}                    \hline
Physics                           &  Topology \\  \hline
Hilbert space                     &  $(n-1)$-dimensional manifold       \\  
(system)                          & (space)             \\  \hline
operator between                  &  cobordism between  \\
Hilbert spaces                    & $(n-1)$-dimensional manifolds \\
(process)                         &  (spacetime)           \\  \hline
composition of operators          &  composition of cobordisms \\  \hline
identity operator                 &  identity cobordism \\  \hline
\end{tabular} 
\\
\caption{Analogy between physics and topology}
\label{analogy2}
\end{center}
\end{table}

Nobody is sure how to do this, but there is a striking similarity
between two of the main approaches: string theory and loop quantum
gravity.  Both rely on the analogy between physics and
topology shown in Table \ref{analogy2}.
On the left we have a basic ingredient of quantum theory: 
the category $\Hilb$ whose objects are Hilbert spaces, used to describe
physical {\em systems}, and whose morphisms are linear operators,
used to describe physical {\em processes}.  On the right we have
a basic structure in differential topology: the category $n\Cob$.
Here the objects are $(n-1)$-dimensional manifolds, used to describe
{\em space}, and whose morphisms are $n$-dimensional cobordisms,
used to describe {\em spacetime}.  

As we shall see, $\Hilb$ and $n\Cob$ share many structural features.
Moreover, both are very different from the more familiar category $\Set$, 
whose objects are sets and whose morphisms are functions. Elsewhere 
we have argued at great length that this is important for better 
understanding quantum gravity \cite{B3} and even the foundations of
quantum theory \cite{B4}.  The idea is that if $\Hilb$ is more like 
$n\Cob$ than $\Set$, maybe we should stop thinking of a quantum process 
as a function from one set of states to another.  Instead, maybe we 
should think of it as resembling a `spacetime' going between spaces 
of dimension one less.   

This idea sounds strange, but the simplest example is something very 
practical, used by physicists every day: a Feynman diagram.  This is 
a 1-dimensional graph going between 0-dimensional collections 
of points, with edges and vertices labelled in certain ways.  Feynman
diagrams are topological entities, but they describe linear operators.
String theory uses 2-dimensional cobordisms equipped with extra
structure --- string worldsheets --- to do a similar job.  Loop quantum 
gravity uses 2d generalizations of Feynman diagrams called `spin 
foams' \cite{B2}.  Topological quantum field theory uses higher-dimensional
cobordisms \cite{BD}.  In each case, processes are described by
morphisms in a special sort of category: a `compact symmetric monoidal
category'.

In what follows, we shall not dwell on puzzles from quantum theory or
quantum gravity.  Instead we take a different tack, simply explaining 
some basic concepts from category theory and showing how $\Set$, 
$\Hilb$, $n\Cob$ and categories of tangles give examples.  A recurring 
theme, however, is that $\Set$ is very different from the other examples.

To help the reader safely navigate the sea of jargon, here is 
a chart of the concepts we shall explain in this section:
\[
\xymatrix{
*+[F]\txt{categories}\ar@{-}[d] 
\\
*+[F]\txt{monoidal categories}\ar@{-}[d] \ar@{-}[dr] 
\\
*+[F]\txt{braided \\ monoidal categories}\ar@{-}[d] \ar@{-}[dr] 
&*+[F]\txt{closed \\ monoidal categories}\ar@{-}[d] \ar@{-}[dr]
\\
*+[F]\txt{symmetric \\ monoidal categories} \ar@{-}[d] \ar@{-}[dr] 
&*+[F]\txt{closed braided \\ monoidal categories}\ar@{-}[d] \ar@{-}[dr] 
&*+[F]\txt{compact \\ monoidal categories}\ar@{-}[d] 
\\
*+[F]\txt{cartesian categories} \ar@{-}[d] 
&*+[F]\txt{closed symmetric \\ monoidal categories}\ar@{-}[dl] \ar@{-}[dr] 
&*+[F]\txt{compact braided \\ monoidal categories}\ar@{-}[d] 
\\
*+[F]\txt{cartesian \\ closed categories} 
& & *+[F]\txt{compact symmetric \\ monoidal categories} 
} 
\]

\vskip 1em
\noindent
The category $\Set$ is cartesian closed, while $\Hilb$ and $n\Cob$ are
compact symmetric monoidal.

\subsection{Categories}
\label{categories}

Category theory was born around 1945, with Eilenberg and Mac Lane
\cite{EM} defining `categories', `functors' between
categories, and `natural transformations' between functors.  By now
there are many introductions to the subject
\cite{Crole,MacLane2,McLarty}, including some available for free
online \cite{JLBell,Goldblatt}.  Nonetheless, we begin at the beginning:

\begin{definition} A {\bf category} $C$ consists of:
\begin{itemize}
    \item a collection of {\bf objects}, where if $X$ is an object of $C$
    we write $X \in C$, and
    \item for every pair of objects $(X,Y),$ a set $\hom(X,Y)$ of 
    {\bf morphisms} from $X$ to $Y$.  We call this set $\hom(X,Y)$ a
    {\bf homset}.  If $f \in \hom(X,Y),$ then we write $f\maps X\to Y.$
\end{itemize}
such that:
\begin{itemize}
    \item for every object $X$ there is an {\bf identity morphism} $1_X\maps 
     X\to X;$
    \item morphisms are composable: given $f\maps X\to Y$ and $g\maps
    Y\to Z,$ there is a {\bf composite morphism} $gf \maps X \to Z;$
    sometimes also written $g \circ f$.
    \item an identity morphism is both a {\bf left and a right unit} for 
   composition: if $f \maps X\to Y,$ then $f 1_X = f = 1_Y f;$ and
    \item composition is {\bf associative}: $(h g) f = 
   h (g f)$ whenever either side is well-defined.
\end{itemize}
\end{definition}

\begin{definition}
  We say a morphism $f\maps X \to Y$ is an \textbf{isomorphism} if it
  has an inverse--- that is, there exists another morphism $g\maps Y
  \to X$ such that $gf = 1_X$ and $fg = 1_Y.$
\end{definition}

A category is the simplest framework where we can talk about 
systems (objects) and processes (morphisms).  To visualize these,
we can use `Feynman diagrams' of a very primitive sort.  In applications
to linear algebra, these diagrams are often called `spin networks', but 
category theorists call them `string diagrams', and that is the term
we will use.  The term `string' here has little to do with string theory: 
instead, the idea is that objects of our category label `strings' or `wires':
\[\begin{pspicture}(0,0)(1,2)
\psset{angleA=-90,angleB=90,ArrowInside=->,arrowscale=2}
\pnode(0,2){A}
\pnode(1,0){B}
\nccurve{A}{B} \nbput{$X$}
\end{pspicture}\]
and morphisms $f \maps X \to Y$ label `black boxes' with 
an input wire of type $X$ and an output wire of type $Y$:
\[\begin{pspicture}(0,0)(2,4)
\psset{angleA=-90,angleB=90,ArrowInside=->,arrowscale=2}
\pnode(1,4){A}
\rput(1,2){\ovalnode{B}{$f$}}
\pnode(1,0){C}
\nccurve{A}{B} \nbput{$X$}
\nccurve{B}{C} \nbput{$Y$}
\end{pspicture}\]
We compose two morphisms by connecting the output of one black box
to the input of the next.  So, the composite of $f\maps X \to Y$
and $g\maps Y \to Z$ looks like this:
\begin{center}
    \begin{pspicture}(0,0)(2,6)
    \psset{angleA=-90,angleB=90,ArrowInside=->,arrowscale=2}
    \pnode(1,6){A}
    \rput(1,4){\ovalnode{B2}{$f$}}
    \rput(1,2){\ovalnode{B1}{$g$}}
    \pnode(1,0){C}
    \nccurve{A}{B2} \nbput{$X$}
    \nccurve{B2}{B1} \nbput{$Y$}
    \nccurve{B1}{C} \nbput{$Z$}
    \end{pspicture}
\end{center}
Associativity of composition is then implicit:
\begin{center}
    \begin{pspicture}(0,0)(2,8)
    \psset{angleA=-90,angleB=90,ArrowInside=->,arrowscale=2}
    \pnode(1,8){A}
    \rput(1,6){\ovalnode{B3}{$f$}}
    \rput(1,4){\ovalnode{B2}{$g$}}
    \rput(1,2){\ovalnode{B1}{$h$}}
    \pnode(1,0){C}
    \nccurve{A}{B3} \nbput{$X$}
    \nccurve{B3}{B2} \nbput{$Y$}
    \nccurve{B2}{B1} \nbput{$Z$}
    \nccurve{B1}{C} \nbput{$W$}
    \end{pspicture}
\end{center}
is our notation for both $h (g f)$ and 
$(h g) f$.  Similarly, if we draw the identity 
morphism $1_X \maps X \to X$ as a piece of wire of type $X$:
\[\begin{pspicture}(0,0)(1,2)
\psset{angleA=-90,angleB=90,ArrowInside=->,arrowscale=2}
\pnode(1,2){A}
\pnode(1,0){B}
\nccurve{A}{B} \nbput{$X$}
\end{pspicture}\]
then the left and right unit laws are also implicit.

There are countless examples of categories, but we will focus on 
four:
\begin{itemize}
\item $\Set$: the category where objects are sets.
\item $\Hilb$: the category where objects are finite-dimensional 
      Hilbert spaces.
\item $n\Cob$: the category where morphisms are $n$-dimensional 
      cobordisms.
\item $\Tang_k$: the category where morphisms are $k$-codimensional
      tangles.
\end{itemize}
As we shall see, all four are closed symmetric monoidal categories,
at least when $k$ is big enough.  However, the most familiar of the 
lot, namely $\Set$, is the odd man out: it is `cartesian'.

Traditionally, mathematics has been founded on the category $\Set$,
where the objects are {\em sets} and the morphisms are {\em
functions}.  So, when we study systems and processes in physics, it is
tempting to specify a system by giving its set of states, and a
process by giving a function from states of one system to states of
another.

However, in quantum physics we do something subtly different:
we use categories where objects are {\em Hilbert spaces} 
and morphisms are {\em bounded linear operators}.  We specify a
system by giving a Hilbert space, but this Hilbert space is 
not really the set of states of the system: a state is actually
a ray in Hilbert space.  Similarly, a bounded linear operator is
not precisely a function from states of one system to states of
another.  

In the day-to-day practice of quantum physics, what really matters is
not sets of states and functions between them, but Hilbert space and
operators.  One of the virtues of category theory is that it frees us
from the `$\Set$-centric' view of traditional mathematics and lets us
view quantum physics on its own terms.  As we shall see, this sheds
new light on the quandaries that have always plagued our understanding
of the quantum realm \cite{B4}.

To avoid technical issues that would take us far afield, we will
take $\Hilb$ to be the category where objects are {\em finite-dimensional
Hilbert spaces} and morphisms are {\em linear operators}
(automatically bounded in this case).  Finite-dimensional
Hilbert spaces suffice for some purposes; infinite-dimensional ones
are often important, but treating them correctly would 
require some significant extensions of the ideas we want to explain here.

In physics we also use categories where the objects represent
choices of {\em space}, and the morphisms represent choices of {\em
spacetime}.  The simplest is $n\Cob$, where the objects are {\em
$(n-1)$-dimensional manifolds}, and the morphisms are {\em
$n$-dimensional cobordisms}.  Glossing over some subtleties 
that a careful treatment would discuss \cite{Sawin}, a cobordism 
$f \maps X \to Y$ is an $n$-dimensional manifold whose boundary is 
the disjoint union of the $(n-1)$-dimensional manifolds $X$ and $Y$.  
Here are a couple of cobordisms in the case $n = 2$:
\[
\begin{pspicture}[0.5](4,4)
  \rput(3,1){\birthc}
  \rput(1,1){\medidentc}
\end{pspicture}
\qquad
 \xy
 {\ar_f (0,10)*+{X}; (0,-11)*+{Y}};
  \endxy
\qquad \qquad
 \begin{pspicture}[.5](4,2.5)
  \rput(2,0){\multc}
\end{pspicture}
\qquad
 \xy
 {\ar_g (0,10)*+{Y}; (0,-11)*+{Z}};
 \endxy
\]
We compose them by gluing the `output' of one to the `input' of
the other.   So, in the above example $gf \maps X \to Z$ looks like this:
\[
\begin{pspicture}[0.5](4,4)
  \rput(2,0){\multc}
  \rput(1,2.49){\smallidentc}
  \rput(3,2.49){\birthc}
\end{pspicture}
\qquad
 \xy
 {\ar_{gf} (0,15)*+{X}; (0,-16)*+{Z}};
 \endxy
\]

Another kind of category important in physics has objects representing
{\em collections of particles}, and morphisms representing their {\em
worldlines and interactions}.  Feynman diagrams are the classic
example, but in these diagrams the `edges' are not taken literally as
particle trajectories.  An example with closer ties to topology is
$\Tang_k$.  

Very roughly speaking, an object in $\Tang_k$ is a collection
of points in a $k$-dimensional cube, while a morphism is a `tangle':
a collection of arcs and circles smoothly embedded in a 
$(k+1)$-dimensional cube, such that the circles lie in the interior 
of the cube, while the arcs touch the boundary of the cube only at 
its top and bottom, and only at their endpoints.  A bit more precisely, 
tangles are `isotopy classes' of such embedded arcs and circles: this 
equivalence relation means that only the topology of the tangle matters, 
not its geometry.  We compose tangles by attaching one cube to another 
top to bottom.

More precise definitions can be found in many sources, at least for
$k = 2$, which gives tangles in a 3-dimensional cube \cite{FY,
Kassel,Sawin,Shum,Turaev,Yetter}.  But since a picture is worth a thousand 
words, here is a picture of a morphism in $\Tang_2$:
\[  
\epsfysize=1.5in\epsfbox{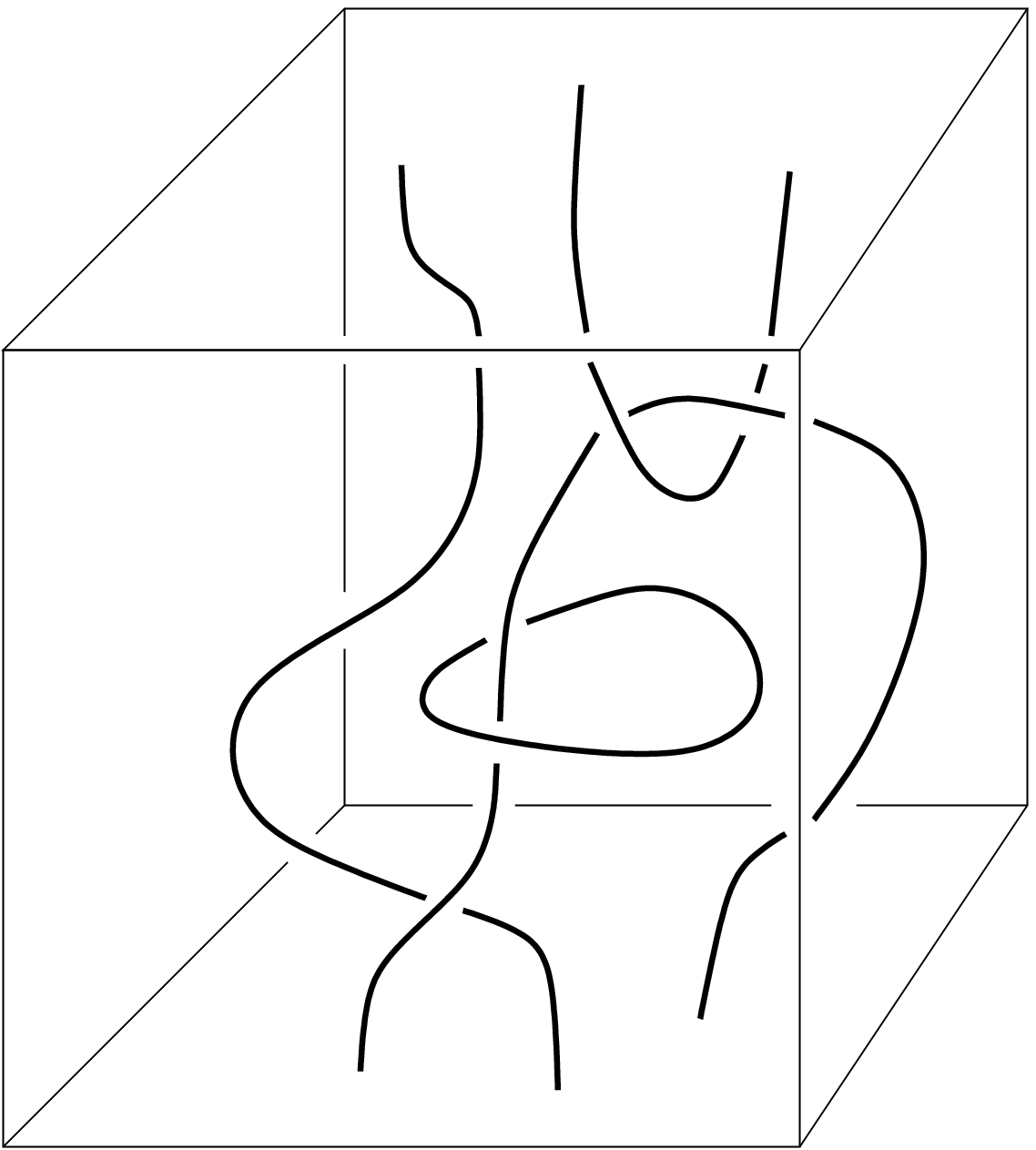}
\quad
 \xy
 {\ar_{f} (0,38)*+{X}; (0,0)*+{Y}};
 \endxy
\]
\noindent
Note that we can think of a morphism in $\Tang_k$ as a 1-dimensional
cobordism {\it embedded in a $k$-dimensional cube}.  This is why 
$\Tang_k$ and $n\Cob$ behave similarly in some respects.

Here are two composable morphisms in $\Tang_1$:
\[
\text{\epsfysize=1.2in\epsfbox{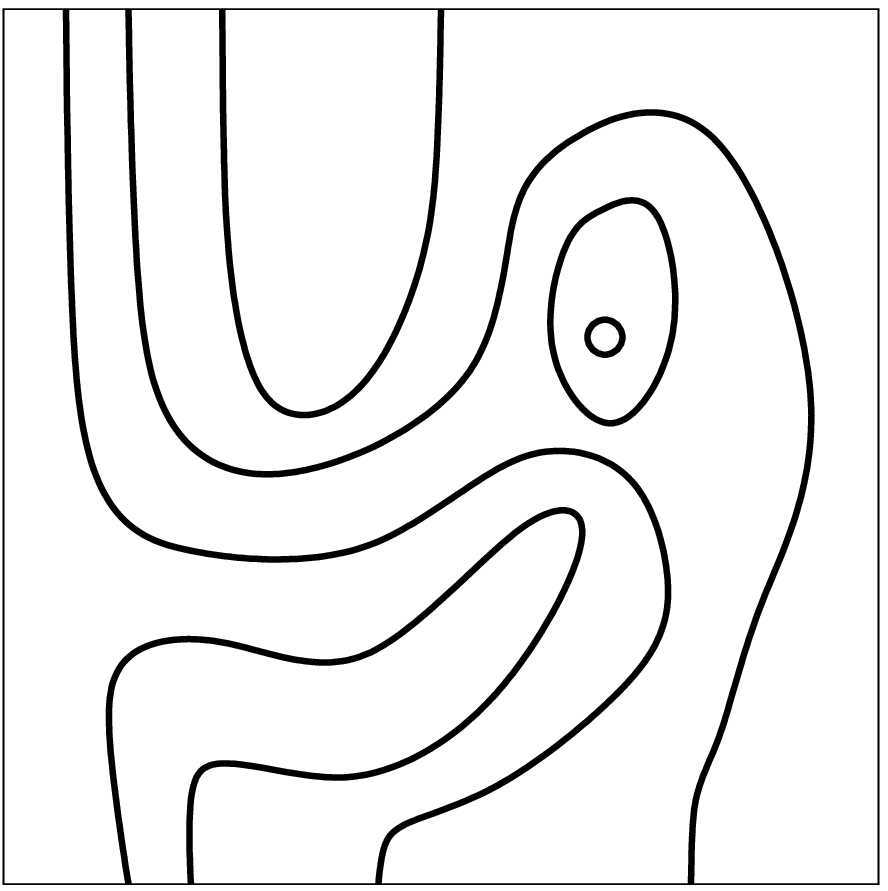}} 
\quad
 \xy
 {\ar_{f} (0,30)*+{X}; (0,0)*+{Y}};
 \endxy
\qquad \qquad
\text{\epsfysize=1.2in\epsfbox{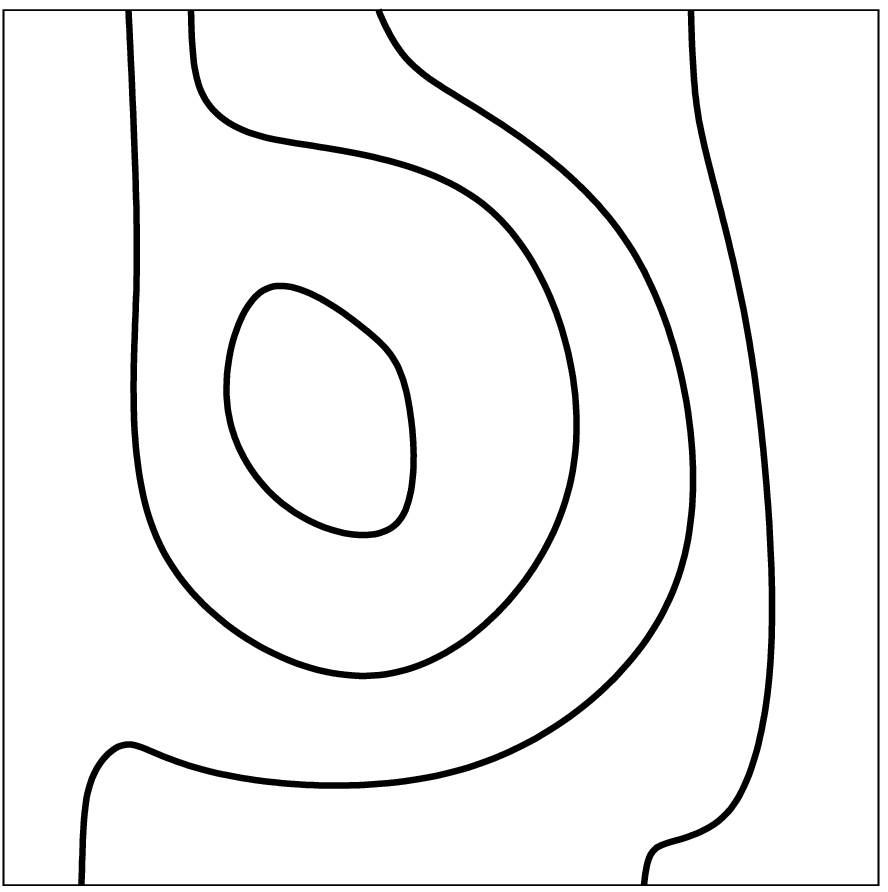}}
\quad
 \xy
 {\ar_{g} (0,30)*+{Y}; (0,0)*+{Z}};
 \endxy
\]
and here is their composite:
\[  
\epsfysize=2.2in\epsfbox{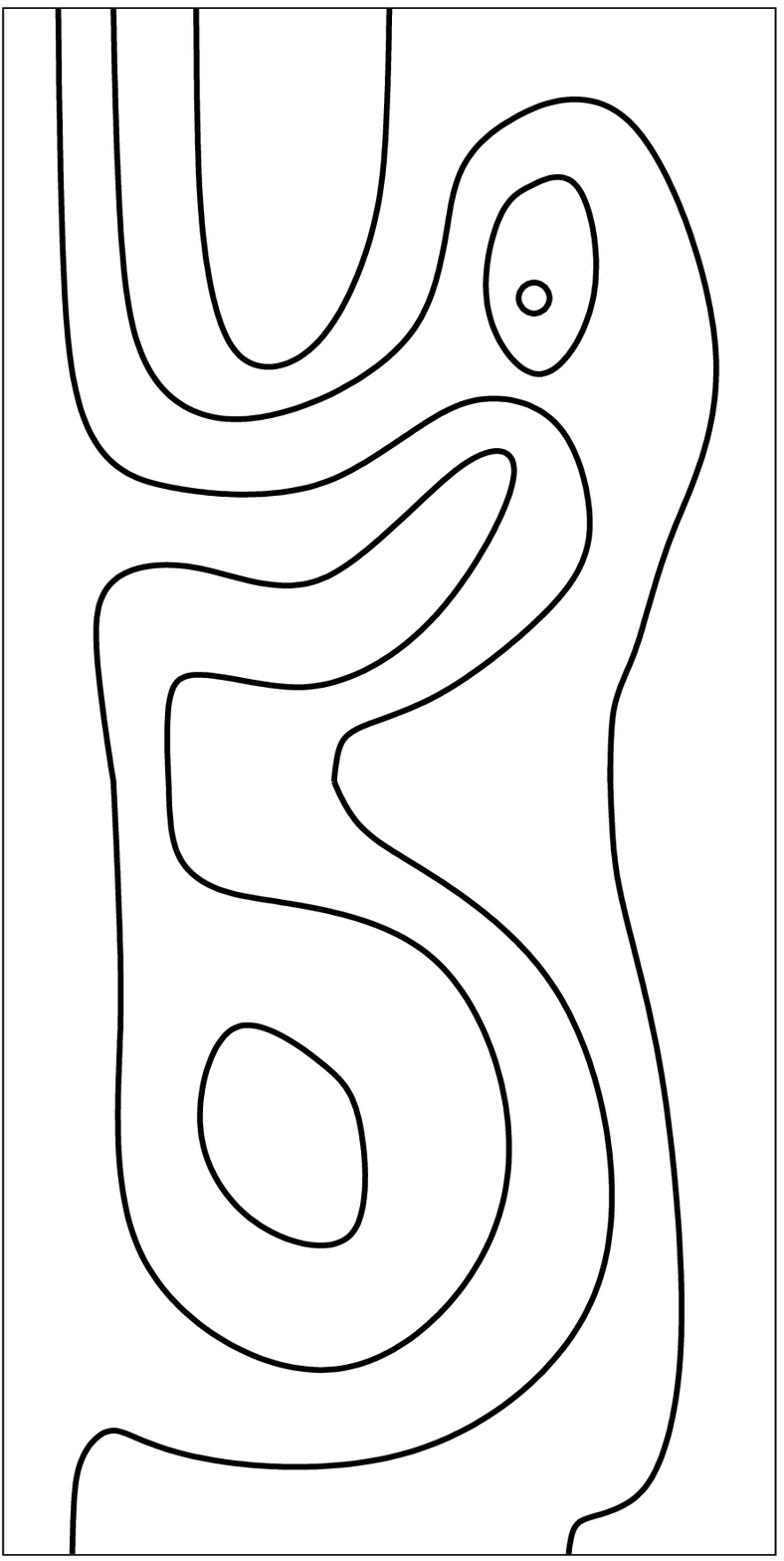}
\quad
 \xy
 {\ar_{gf} (0,55)*+{X}; (0,0)*+{Z}};
 \endxy
\]
Since only the tangle's topology matters, we are free to squash
this rectangle into a square if we want, but we do not need to.

It is often useful to consider tangles that are decorated in 
various ways.  For example, in an `oriented' tangle, each arc and 
circle is equipped with an orientation.  We can indicate this by
drawing a little arrow on each curve in the tangle.  In applications 
to physics, these curves represent worldlines of particles, and the arrows 
say whether each particle is going forwards or backwards in time, following 
Feynman's idea that antiparticles are particles going backwards in time.
We can also consider `framed' tangles.  Here each curve is replaced by 
a `ribbon'.  In applications to physics, this keeps track of how each 
particle twists.  This is especially important for fermions, where a 
$2\pi$ twist acts nontrivially.  Mathematically, the best-behaved tangles
are both framed and oriented \cite{BD,Shum}, and these are what we
should use to define $\Tang_k$.  The category $n\Cob$ also has a framed
oriented version.  However, these details will barely matter in what
is to come.

It is difficult to do much with categories without discussing the maps
between them.  A map between categories is called a `functor':

\begin{definition} 
A {\bf functor} $F\maps C\to D$ from a category $C$ to a category $D$
is a map sending:
\begin{itemize}
\item any object $X \in C$ to an object $F(X) \in D$,
\item any morphism $f\maps X\to Y$ in $C$ to a morphism $F(f)\maps F(X)\to F(Y)$ 
in $D$,
\end{itemize}
in such a way that:
\begin{itemize}
\item $F$ {\bf preserves identities}: for any object $X \in C$,
$F(1_X) = 1_{F(X)}$;
\item $F$ {\bf preserves composition}: for
any pair of morphisms $f \maps X \to Y$, $g \maps Y \to Z$ in $C$,
$F(gf) = F(g) F(f)$.
\end{itemize}
\end{definition}

In the sections to come, we will see that functors and natural
transformations are useful for putting extra structure on categories.
Here is a rather different use for functors: we can think of a
functor $F \maps C \to D$ as giving a picture, or `representation',
of $C$ in $D$.  The idea is that $F$ can map objects and morphisms of
some `abstract' category $C$ to objects and morphisms of a
more `concrete' category $D$.  

For example, consider an abstract group $G$.  This is the same as
a category with one object and with all morphisms invertible.
The object is uninteresting, so we can just call it $\bullet$, but the
morphisms are the elements of $G$, and we compose them by multiplying
them.  From this perspective, a {\bf representation} of $G$ on a 
finite-dimensional Hilbert space is the same as a functor 
$F \maps G \to \Hilb$.  Similarly, an {\bf action} of $G$ 
on a set is the same as a functor $F \maps G \to \Set$.  Both notions 
are ways of making an abstract group more concrete.

Ever since Lawvere's 1963 thesis on functorial semantics
\cite{Lawvere}, the idea of functors as representations has become
pervasive.  However, the terminology varies from field to field.
Following Lawvere, logicians often call the category $C$ a `theory', 
and call the functor $F \maps C \to D$ a `model' of this theory.
Other mathematicians might call $F$ an `algebra' of the theory.  
In this work, the default choice of $D$ is usually the category
$\Set$.  

In physics, it is the functor $F \maps C \to D$ that is called 
the `theory'.  Here the default choice of $D$ is either the category
we are calling $\Hilb$ or a similar category of \emph{infinite-dimensional}
Hilbert spaces.  For example, both `conformal field theories' \cite{Segal} 
and topological quantum field theories \cite{Atiyah} can be seen as
functors of this sort.

If we think of functors as models, natural transformations are 
maps between models:

\begin{definition} 
\label{naturality}
Given two functors $F,F'\maps C\to D,$ a {\bf natural 
transformation} $\alpha\maps F \Rightarrow F'$ assigns to every object
$X$ in $C$ a morphism $\alpha_X\maps F(X)\to F'(X)$ such that for any
morphism $f\maps X \to Y$ in $C,$ the equation $\alpha_Y \, F(f) = F'(f)
\, \alpha_X$ holds in $D.$   In other words, this square commutes:
\[\begin{diagram}
\node{F(X)}\arrow{e,t}{F(f)}\arrow{s,l}{\alpha_X}\node{F(Y)}
\arrow{s,r}{\alpha_Y}\\
\node{F'(X)}\arrow{e,b}{F'(f)}\node{F'(Y)}
\end{diagram}\]
(Going across and then down equals going down and then across.)
\end{definition}

\begin{definition} A {\bf natural isomorphism} between functors
$F,F' \maps C \to D$ is a natural transformation $\alpha \maps F
\Rightarrow F'$ such that $\alpha_X$ is an isomorphism for every $X \in
C$.
\end{definition}

For example, suppose $F, F' \maps G \to \Hilb$ are functors where
$G$ is a group, thought of as a category with one object, say $\bullet$.
Then, as already mentioned, $F$ and $F'$ are secretly just representations 
of $G$ on the Hilbert spaces $F(\bullet)$ and $F'(\bullet)$.  A natural 
transformation $\alpha \maps F \To F'$ is then the same as an {\bf 
intertwining operator} from one representation to another: that is, a 
linear operator 
\[  A \maps F(\bullet) \to F'(\bullet) \]
satisfying
\[  A F(g) = F'(g) A \]
for all group elements $g$. 

\subsection{Monoidal Categories}
\label{monoidal}

In physics, it is often useful to think of two systems sitting side by
side as forming a single system.  In topology, the disjoint union of
two manifolds is again a manifold in its own right.  In logic, the
conjunction of two statement is again a statement.  In programming we
can combine two data types into a single `product type'.
The concept of `monoidal category' unifies all these examples in a
single framework.

A monoidal category $C$ has a functor $\tensor \maps C \times C \to C$
that takes two objects $X$ and $Y$ and puts them together to give a new
object $X \tensor Y$.  To make this precise, we need the cartesian
product of categories:

\begin{definition} The {\bf cartesian product} $C \times C'$ of
categories $C$ and $C'$ is the category where:
\begin{itemize}
\item an object is 
a pair $(X,X')$ consisting of an object $X \in C$ and an object 
$X' \in C'$;
\item a morphism from $(X,X')$ to $(Y,Y')$ is a pair $(f,f')$ consisting
of a morphism $f \maps X \to Y$ and a morphism $f' \maps X' \to Y'$;
\item composition is done componentwise: $(g,g')(f,f') = (gf,g'f')$;
\item identity morphisms are defined componentwise:
$1_{(X,X')} = (1_X, 1_{X'})$.
\end{itemize}
\end{definition}

Mac Lane \cite{MacLane} defined monoidal categories in 1963.  The
subtlety of the definition lies in the fact that $(X \tensor Y)
\tensor Z$ and $X \tensor (Y \tensor Z)$ are not usually equal.
Instead, we should specify an isomorphism between them, called the
`associator'.  Similarly, while a monoidal category has a `unit
object' $I$, it is not usually true that $I \tensor X$ and $X \tensor I$ 
equal $X$.  Instead, we should specify isomorphisms $I \tensor X \cong X$
and $X \tensor I \cong X$.  To be manageable, all these isomorphisms
must then satisfy certain equations:

\begin{definition}
\label{mon.cat}
A {\bf monoidal category} consists of:
    \begin{itemize}
        \item a category $C,$
        \item a {\bf tensor product} functor $\tensor \maps
       C\times C\to C,$ 
        \item a {\bf unit object} $I \in C$,
        \item a natural isomorphism called the {\bf associator},
assigning to each triple of objects $X,Y,Z \in C$ an isomorphism
        \[a_{X,Y,Z}:(X \tensor Y)\tensor Z \isoto X \tensor (Y \tensor Z),\]
        \item natural isomorphisms called the {\bf left} and {\bf right 
unitors}, assigning to each object $X \in C$ isomorphisms 
        \[l_X:I\tensor X \isoto X\]
        \[r_X:X\tensor I \isoto X,\]
    \end{itemize}
    such that:
    \begin{itemize}
        \item for all $X,Y \in C$ the {\bf triangle equation} holds:
        \di{
            \node{(X\tensor I)\tensor Y}\arrow[2]{e,t}{a_{X,I,Y}}\arrow{se,b}{r_X\tensor 1_Y}\node[2]{X\tensor(I\tensor Y)}\arrow{sw,b}{1_X\tensor l_Y}\\
            \node[2]{X\tensor Y}
        }
        \item for all $W,X,Y,Z \in C$, the {\bf pentagon equation} holds:
        \di{
            \node[2]{((W\tensor X)\tensor Y)\tensor Z}\arrow{ssw,t}{a_{W\tensor X,Y,Z}}\arrow{se,t}{a_{W,X,Y}\tensor 1_Z}\\
            \node[3]{(W\tensor (X\tensor Y))\tensor Z}\arrow[2]{s,r}{a_{W,X\tensor Y,Z}}\\
            \node{(W\tensor X)\tensor(Y\tensor Z)}\arrow{sse,b}{a_{W,X,Y\tensor Z}}\\
            \node[3]{W\tensor ((X\tensor Y)\tensor Z)}\arrow{sw,r}{1_W\tensor a_{X,Y,Z}}\\
            \node[2]{W\tensor(X\tensor(Y\tensor Z))}
        }
    \end{itemize}
\end{definition}

When we have a tensor product of four objects, there are five ways to
parenthesize it, and at first glance the associator lets us build 
two isomorphisms from $W \tensor (X \tensor (Y \tensor Z))$ to
$((W \tensor X) \tensor Y) \tensor Z$.  But, the pentagon equation says
these isomorphisms are equal.  When we have tensor products of
even more objects there are even more ways to parenthesize them, and
even more isomorphisms between them built from the associator.
However, Mac Lane showed that the pentagon identity implies these
isomorphisms are all the same.  Similarly, if we also assume the
triangle equation, all isomorphisms with the same source and target
built from the associator, left and right unit laws are equal.

In a monoidal category we can do processes in `parallel' as well as
in `series'.  Doing processes in series is just composition of
morphisms, which works in any category.  But in a monoidal category
we can also tensor morphisms $f \maps X \to Y$ and $f' \maps 
X' \to Y'$ and obtain a `parallel process' $f \tensor f' \maps
X \tensor X' \to Y \tensor Y'$.  We can draw this in various ways:

\[\begin{pspicture}(0,0)(15,4)
\psset{angleA=-90,angleB=90,ArrowInside=->,arrowscale=2}
\pnode(1,4){A}
\rput(1,2){\ovalnode{B}{$f$}}
\pnode(1,0){C}
\nccurve{A}{B} \nbput{$X$}
\nccurve{B}{C} \nbput{$Y$}

\pnode(3,4){A}
\rput(3,2){\ovalnode{B}{$f'$}}
\pnode(3,0){C}
\nccurve{A}{B} \nbput{$X'$}
\nccurve{B}{C} \nbput{$Y'$}

\rput(5,2){=}

\pnode(7,4){A}
\pnode(9,4){A2}
\rput(8,2){\ovalnode{B}{$f \tensor f'$}}
\pnode(7,0){C}
\pnode(9,0){C2}
\nccurve[angleB=135]{A}{B} \nbput{$X$}
\nccurve[angleA=-135]{B}{C} \nbput{$Y$}
\nccurve[angleB=45]{A2}{B} \naput{$X'$}
\nccurve[angleA=-45]{B}{C2} \naput{$Y'$}

\rput(11,2){=}

\pnode(13,4){A}
\rput(13,2){\ovalnode{B}{$f \tensor f'$}}
\pnode(13,0){C}
\nccurve{A}{B} \nbput{$X \tensor X'$}
\nccurve{B}{C} \nbput{$Y \tensor Y'$}

\end{pspicture}\]

More generally, we can draw any morphism 
\[f \maps X_1 \tensor \cdots \tensor X_n \to 
          Y_1 \tensor \cdots \tensor Y_m \]
as a black box with $n$ input wires and $m$ output wires:
\[\begin{pspicture}(0,0)(4,4)
\psset{angleA=-90,angleB=90,ArrowInside=->,arrowscale=2}
    \pnode(0,4){X1}
    \pnode(2,4){X2}
    \pnode(4,4){X3}
    \rput(2,2){\ovalnode{F}{$f$}}
    \pnode(1,0){Y1}
    \pnode(3,0){Y2}
    \nccurve[angleB=135]{X1}{F} \nbput[npos=0]{$X_1$}
    \nccurve{X2}{F} \nbput[npos=0]{$X_2$}
    \nccurve[angleB=45]{X3}{F} \naput[npos=0]{$X_3$}
    \nccurve[angleA=-120]{F}{Y1} \nbput[npos=1]{$Y_1$}
    \nccurve[angleA=-60]{F}{Y2} \naput[npos=1]{$Y_2$}
\end{pspicture}\]
We draw the unit object $I$ as a blank space.  So, for example,
we draw a morphism $f \maps I \to X$ as follows:
\[\begin{pspicture}(0,0)(2,3)
\psset{angleA=-90,angleB=90,ArrowInside=->,arrowscale=2}
\pnode(1,3){A}
\rput(1,2){\ovalnode{B}{$f$}}
\pnode(1,0){C}
\nccurve{B}{C} \nbput{$X$}
\end{pspicture}\]
By composing and tensoring morphisms, we can build up
elaborate pictures resembling Feynman diagrams:
\[\begin{pspicture}(0,0)(8,6)
  \psset{angleA=-90,angleB=90,ArrowInside=->,arrowscale=2}
  \pnode(1,6){X1}
  \pnode(2,6){X2}
  \pnode(5,6){X3}
  \pnode(7,6){X4}
  \rput(1,2){\ovalnode{F}{$f$}}
  \rput(2,1){\ovalnode{G}{$g$}}
  \rput(5,2.5){\ovalnode{H}{$h$}}
  \rput(7,4){\ovalnode{J}{$j$}}
  \pnode(1,0){Y1}
  \pnode(2,0){Y2}
  \pnode(5,0){Y3}
  \pnode(7,0){Y4}
  \nccurve{X1}{F} \nbput[npos=0]{$X_1$}
  \nccurve[angleB=60]{X2}{G} \naput[npos=0]{$X_2$}
  \nccurve{X3}{H} \nbput[npos=0]{$X_3$}
  \nccurve{X4}{J} \nbput[npos=0]{$X_4$}
  \nccurve[angleA=-120]{F}{Y1} \nbput[npos=1]{$Y_1$}
  \nccurve{G}{Y2} \naput[npos=1]{$Y_2$}
  \nccurve{H}{Y3} \nbput[npos=1]{$Y_3$}
  \nccurve{J}{Y4} \nbput[npos=1]{$Y_4$}
  \nccurve[angleA=-60, angleB=120]{F}{G} \naput{$Z$}
\end{pspicture}\]
The laws governing a monoidal category allow us to neglect 
associators and unitors when drawing such pictures, without
getting in trouble.  The reason is that Mac Lane's Coherence
Theorem says any monoidal category is `equivalent', in a suitable
sense, to one where all associators and unitors are identity
morphisms \cite{MacLane}.

We can also deform the picture in a wide
variety of ways without changing the morphism it describes.  
For example, the above morphism equals this one:
\[\begin{pspicture}(0,0)(8,6)
    \psset{angleA=-90,angleB=90,ArrowInside=->,arrowscale=2}
    \pnode(1,6){X1}
    \pnode(2,6){X2}
    \pnode(5,6){X3}
    \pnode(7,6){X4}
    \rput(1,5){\ovalnode{F}{$f$}}
    \rput(2,3){\ovalnode{G}{$g$}}
    \rput(4,4){\ovalnode{H}{$h$}}
    \rput(6.5,2){\ovalnode{J}{$j$}}
    \pnode(1,0){Y1}
    \pnode(2,0){Y2}
    \pnode(5,0){Y3}
    \pnode(7,0){Y4}
    \nccurve[ArrowInsidePos=.75]{X1}{F} \nbput[npos=0]{$X_1$}
    \nccurve[angleB=60]{X2}{G} \naput[npos=0]{$X_2$}
    \nccurve{X3}{H} \nbput[npos=0]{$X_3$}
    \nccurve{X4}{J} \nbput[npos=0]{$X_4$}
    \nccurve[angleA=-120]{F}{Y1} \nbput[npos=1]{$Y_1$}
    \nccurve{G}{Y2} \naput[npos=1]{$Y_2$}
    \nccurve{H}{Y3} \nbput[npos=1]{$Y_3$}
    \nccurve{J}{Y4} \naput[npos=1]{$Y_4$}
    \nccurve[angleA=-60, angleB=120]{F}{G} \nbput{$Z$}
\end{pspicture}\]
Everyone who uses string diagrams for calculations in monoidal
categories starts by worrying about the rules of the game:
{\em precisely how} can we deform these pictures without changing 
the morphisms they describe?  Instead of stating the rules precisely
--- which gets a bit technical --- we urge you to explore for 
yourself what is allowed and what is not.  For example, show that
we can slide black boxes up and down like this:
\[\begin{pspicture}(0,0)(3,3)
    \psset{angleA=-90,angleB=90,ArrowInside=->,arrowscale=2}
    \pnode(1,3){X1}
    \pnode(2,3){X2}
    \rput(1,2){\ovalnode{F}{$f$}}
    \rput(2,1){\ovalnode{G}{$g$}}
    \pnode(1,0){Y1}
    \pnode(2,0){Y2}
    \nccurve{X1}{F} \nbput[npos=0]{$X_1$}
    \nccurve{F}{Y1} \nbput[npos=1]{$Y_1$}
    \nccurve{X2}{G} \naput[npos=0]{$X_2$}
    \nccurve{G}{Y2} \naput[npos=1]{$Y_2$}
\end{pspicture} \begin{pspicture}(0,0)(2,3)
    \rput(1,1.5){=}
\end{pspicture} \begin{pspicture}(0,0)(3,3)
    \psset{angleA=-90,angleB=90,ArrowInside=->,arrowscale=2}
    \pnode(1,3){X1}
    \pnode(2,3){X2}
    \rput(1,1.5){\ovalnode{F}{$f$}}
    \rput(2,1.5){\ovalnode{G}{$g$}}
    \pnode(1,0){Y1}
    \pnode(2,0){Y2}
    \nccurve{X1}{F} \nbput[npos=0]{$X_1$}
    \nccurve{F}{Y1} \nbput[npos=1]{$Y_1$}
    \nccurve{X2}{G} \naput[npos=0]{$X_2$}
    \nccurve{G}{Y2} \naput[npos=1]{$Y_2$}
\end{pspicture} \begin{pspicture}(0,0)(2,3)
    \rput(1,1.5){=}
\end{pspicture} \begin{pspicture}(0,0)(3,3)
    \psset{angleA=-90,angleB=90,ArrowInside=->,arrowscale=2}
    \pnode(1,3){X1}
    \pnode(2,3){X2}
    \rput(1,1){\ovalnode{F}{$f$}}
    \rput(2,2){\ovalnode{G}{$g$}}
    \pnode(1,0){Y1}
    \pnode(2,0){Y2}
    \nccurve{X1}{F} \nbput[npos=0]{$X_1$}
    \nccurve{F}{Y1} \nbput[npos=1]{$Y_1$}
    \nccurve{X2}{G} \naput[npos=0]{$X_2$}
    \nccurve{G}{Y2} \naput[npos=1]{$Y_2$}
\end{pspicture}\]
For a formal treatment of the rules governing string 
diagrams, try the original papers by Joyal and Street \cite{JS0}
and the book by Yetter \cite{Yetter}.

Now let us turn to examples.  Here it is crucial to realize that
the same category can often be equipped with different tensor products,
resulting in different monoidal categories:

\begin{itemize}

\item There is a way to make $\Set$ into a monoidal category where
$X \tensor Y$ is the cartesian product $X \times Y$ and the unit
object is any one-element set.  Note that this tensor product is not 
strictly associative, since $(x, (y, z)) \ne ((x, y), z),$ but there's 
a natural isomorphism $(X \times Y) \times Z \iso X \times (Y \times 
Z)$, and this is our associator.  Similar considerations give the left 
and right unitors.  In this monoidal category, the tensor 
product of $f \maps X \to Y$ and $f' \maps X' \to Y'$ is the function
\[
\begin{array}{rccl}
        f \times f' & \maps X\times X'& \to& Y\times Y' \\
                    & (x,x') &\mapsto & (f(x),f'(x')) .
\end{array}
\]

There is also a way to make $\Set$ into a monoidal category where
$X \tensor Y$ is the disjoint union of $X$ and $Y$, which we shall
denote by $X + Y$.  Here the unit object is the empty set.  
Again, as indeed with all these examples, the associative law and 
left/right unit laws hold only up to natural isomorphism.  In this 
monoidal category, the tensor product of $f \maps X \to Y$ and 
$f' \maps X' \to Y'$ is the function
\[
\begin{array}{rccl}
 f+f' \maps & X+X'  &\to&  Y+Y'  \\
           &      x &\mapsto& 
\left\{
\begin{array}{cl} 
f(x) & \text{if $x \in X$,} \\ 
f'(x) & \text{if $x \in X'$.}
\end{array} \right.
\end{array}
\]

{\em However, in what follows, when we speak of $\Set$ as a monoidal
category, we always use the cartesian product!}

\item There is a way to make $\Hilb$ into a monoidal category
with the usual tensor product of Hilbert spaces: $\CC^n \tensor \CC^m \cong 
\CC^{nm}.$  In this case the unit object $I$ can be taken to be a
1-dimensional Hilbert space, for example $\CC$.  

There is also a way to make $\Hilb$ into a monoidal category
where the tensor product is the direct sum: $\CC^n \oplus \CC^m \cong 
\CC^{n+m}.$  In this case the unit object is the zero-dimensional 
Hilbert space, $\{ 0\}.$

{\em However, in what follows, when we speak of $\Hilb$ as a monoidal
category, we always use the usual tensor product!}

\item The tensor product of objects and morphisms in $n\Cob$ is given 
by disjoint union.  For example, the tensor product of these two
morphisms:
\[
\begin{pspicture}[0.5](4,4)
  \rput(3,0.5){\birthc}
  \rput(1,0.5){\medidentc}
\end{pspicture}
\qquad
 \xy
 {\ar_f (0,10)*+{X}; (0,-11)*+{Y}};
 \endxy
\qquad \qquad
\begin{pspicture}[.5](4,2.5)
  \rput(2,0){\comultc}
\end{pspicture}
\qquad
 \xy
 {\ar_{f'} (0,10)*+{X'}; (0,-11)*+{Y'}};
 \endxy
\]
is this:
\[
\begin{pspicture}[.5](7,2.5)
  \rput(3,0){\birthc}
  \rput(1,0){\identc}
  \rput(6,0){\comultc}
\end{pspicture}
\qquad
 \xy
 {\ar_{f \tensor f'} (0,10)*+{X \tensor X'}; (0,-11)*+{Y \tensor Y'}};
 \endxy
\]

\item The category $\Tang_k$ is monoidal when $k \ge 1$, where the
the tensor product is given by disjoint union.  For example, given 
these two tangles:
\[
\text{\epsfysize=1.2in\epsfbox{2dtangle.eps}} 
\quad
 \xy
 {\ar_{f} (0,30)*+{X}; (0,0)*+{Y}};
 \endxy
\qquad \qquad
\text{\epsfysize=1.2in\epsfbox{another2dtangle.eps}}
\quad
 \xy
 {\ar_{f'} (0,30)*+{X'}; (0,0)*+{Y'}};
 \endxy
\]
their tensor product looks like this:
\[  
\epsfysize=1.2in\epsfbox{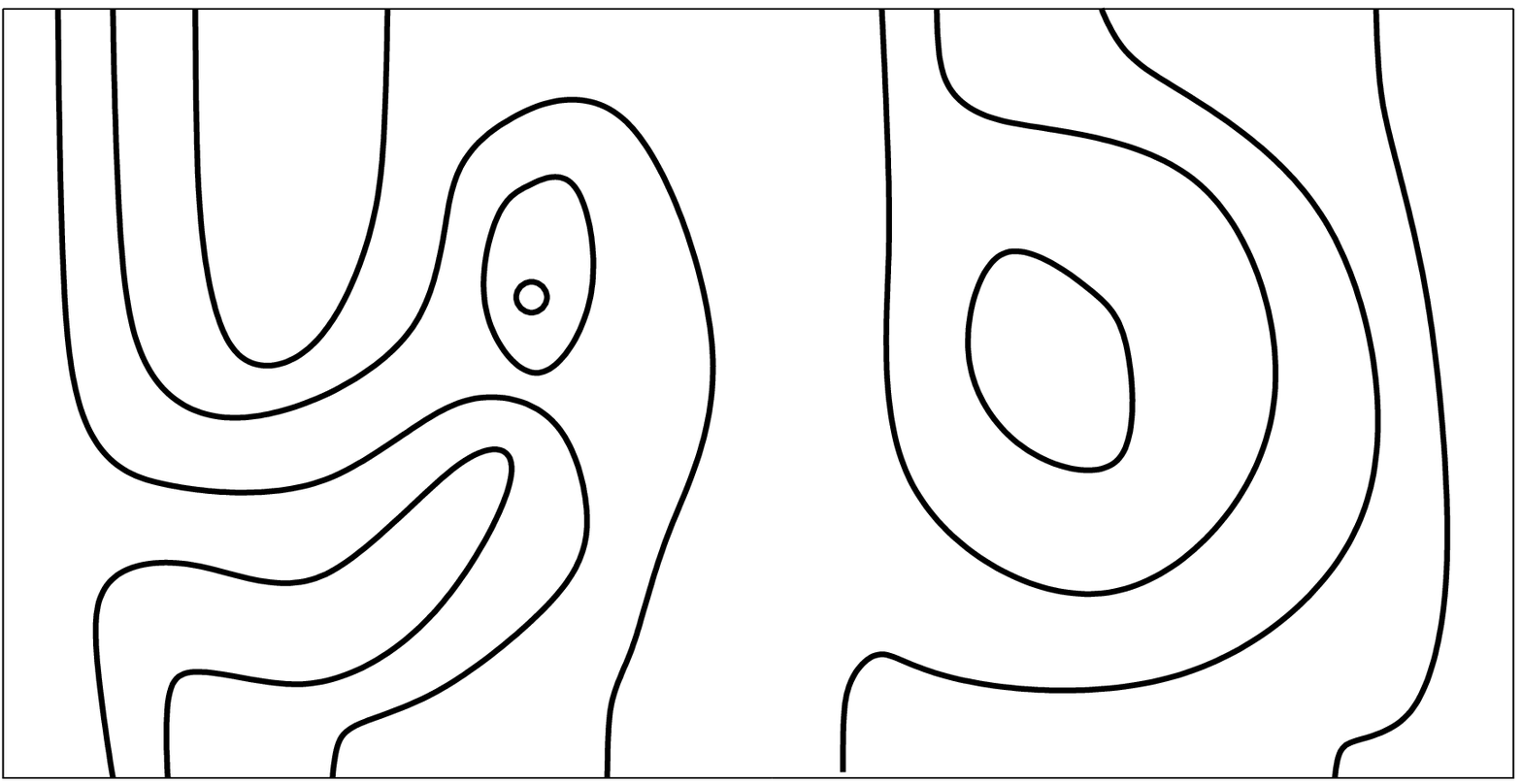}
\quad
 \xy
 {\ar_{f \tensor f'} (0,30)*+{X \tensor X'}; (0,0)*+{Y \tensor Y'}};
 \endxy
\]
\end{itemize}

The example of $\Set$ with its cartesian product is different from
our other three main examples, because the cartesian product of sets 
$X \times X'$ comes equipped with functions called `projections' to the 
sets $X$ and $X'$:
\[    
\begin{diagram}
\node{X} \node{X \times X'} \arrow{w,t}{p} \arrow{e,t}{p'} \node{X'}
\end{diagram}
\]
Our other main examples lack this feature --- though $\Hilb$ made
into a monoidal category using $\oplus$ has projections.
Also, every set has a unique function to the one-element set:
\[           !_X \maps X \to I. \]
Again, our other main examples lack this feature, though $\Hilb$ made
into a monoidal category using $\oplus$ has it.  A fascinating feature
of quantum mechanics is that we make $\Hilb$ into a monoidal category
using $\tensor$ instead of $\oplus$, even though the latter approach
would lead to a category more like $\Set$.

We can isolate the special features of the cartesian product of sets
and its projections, obtaining a definition that applies to any
category:

\begin{definition}
Given objects $X$ and $X'$ in some category,
we say an object $X \times X'$ equipped with morphisms 
\[    
\begin{diagram}
\node{X} \node{X \times X'} \arrow{w,t}{p} \arrow{e,t}{p'} \node{X'}
\end{diagram}
\]
is a {\bf cartesian product} (or simply {\bf product}) of $X$ and $X'$ 
if for any object $Q$ and morphisms 
\[          
\begin{diagram}
\node[2]{Q} \arrow{sw,t}{f} \arrow{se,t}{f'}  \\
\node{X} \node[2]{X'}
\end{diagram}
\]
there exists a unique morphism $g \maps Q \to X \times X'$ making
the following diagram commute:
\[          
\begin{diagram}
\node[2]{Q} \arrow{sw,t}{f} \arrow{se,t}{f'} 
\arrow{s,r}{g} \\
\node{X} \node{X \times X'} \arrow{w,b}{p} \arrow{e,b}{p'} \node{X'}
\end{diagram}
\]
(That is, $f = p g$ and $f' = p' g$.)  We say a category has
{\bf binary products} if every pair of objects has a product.
\end{definition}
\noindent
The product may not exist, and it may not be unique, but when it
exists it is unique up to a canonical isomorphism.  This justifies our
speaking of `the' product of objects $X$ and $Y$ when it exists, and
denoting it as $X \times Y$.

The definition of cartesian product, while absolutely fundamental, is
a bit scary at first sight.  To illustrate its power, let us do
something with it: combine two morphisms $f \maps X \to Y$ and $f'
\maps X' \to Y'$ into a single morphism
\[   f \times f' \maps X \times X' \to Y \times Y'. \]
The definition of cartesian product says how to build a morphism of
this sort out of a pair of morphisms: namely, morphisms from $X \times
X'$ to $Y$ and $Y'$.  If we take these to be $f p$ and $f' p'$, we
obtain $f \times f'$:
\[          
\begin{diagram}
\node[2]{X \times X'} \arrow{sw,t}{fp} \arrow{se,t}{f'p'} 
\arrow{s,r}{f \times f'} \\
\node{Y} \node{Y \times Y'} \arrow{w,t}{p} \arrow{e,t}{p'} \node{Y'}
\end{diagram}
\]

Next, let us isolate the special features of the one-element set:

\begin{definition}
An object $1$ in a category $C$ is {\bf terminal} if for any object
$Q \in C$ there exists a unique morphism from $Q$ to $1$, which we
denote as $!_Q \maps Q \to 1$.
\end{definition}
\noindent
Again, a terminal object may not exist and may not be unique, but it
is unique up to a canonical isomorphism.  This is why we can speak of `the'
terminal object of a category, and denote it by a specific symbol, $1$.

We have introduced the concept of binary products.  One can also talk
about $n$-ary products for other values of $n$, but a category with
binary products has $n$-ary products for all $n \ge 1$, since we can
construct these as iterated binary products.  The case $n = 1$ is
trivial, since the product of one object is just that object itself
(up to canonical isomorphism).  The remaining case is $n = 0$.  The
zero-ary product of objects, if it exists, is just the terminal object.
So, we make the following definition:

\begin{definition} 
\label{finite_products}
A category has {\bf finite products} if it has binary products
and a terminal object.  
\end{definition}
\noindent
A category with finite products can always be made into a monoidal
category by choosing a specific product $X \times Y$ to be the
tensor product $X \tensor Y$, and choosing a specific terminal object
to be the unit object.  It takes a bit of work to show this!  A
monoidal category of this form is called {\bf cartesian}.

In a cartesian category, we can `duplicate and delete information'.  
In general, the definition of cartesian products gives a way to take
two morphisms $f_1 \maps Q \to X$ and $f_2 \maps Q \to Y$ and combine them
into a single morphism from $Q$ to $X \times Y$.  If we take $Q = X = Y$ and
take $f_1$ and $f_2$ to be the identity, we obtain the {\bf diagonal}
or {\bf duplication} morphism:
\[         \Delta_X \maps X \to X \times X.  \]
In the category $\Set$ one can check that this maps any element $x \in X$ to 
the pair $(x,x)$.  In general, we can draw the diagonal as follows:
\[\begin{pspicture}(0,0)(2,4)
\psset{angleA=-90,angleB=90,ArrowInside=->,arrowscale=2}
\pnode(1,4){A}
\pnode(0,0){B1}
\pnode(2,0){B2}
\rput(1,2){\ovalnode{D}{$\Delta$}}
\nccurve{A}{D}\nbput{$X$}
\nccurve[angleA=-135]{D}{B1}\nbput{$X$}
\nccurve[angleA=-45]{D}{B2}\naput{$X$}
\end{pspicture}\]
Similarly, we call the unique map to the terminal object
\[      !_X \maps X \to 1  \]
the {\bf deletion} morphism, and draw it as follows:
\[\begin{pspicture}(0,0)(2,3)
\psset{angleA=-90,angleB=90,ArrowInside=->,arrowscale=2}
\pnode(1,3){A}
\rput(1,1){\ovalnode{D}{!}}
\nccurve{A}{D}\nbput{$X$}
\end{pspicture}\]
Note that we draw the unit object as an empty space.

A fundamental fact about cartesian categories is that duplicating 
something and then deleting either copy is the same as doing nothing 
at all!  In string diagrams, this says:
\[\begin{pspicture}(0,0)(8,7)
  \psset{angleA=-90,angleB=90,ArrowInside=->,arrowscale=2}
  \pnode(1,7){A}
  \rput(0,3){\ovalnode{B1}{!}}
  \pnode(2,3){B2}
  \rput(1,5){\ovalnode{D}{$\Delta$}}
  \nccurve{A}{D}\nbput{$X$}
  \nccurve[angleA=-135]{D}{B1}\nbput{$X$}
  \nccurve[angleA=-45,ArrowInside=-]{D}{B2}\naput{$X$}
  \pnode(2,1){B3}
  \nccurve{B2}{B3}    
  
  \rput(3,5){$=$}
  
  \pnode(5,7){A}
  \pnode(5,1){B}
  \nccurve{A}{B} \nbput{$X$}
  
  \rput(7,5){$=$}
  
  \pnode(9,7){A}
  \pnode(8,3){B1}
  \rput(10,3){\ovalnode{B2}{!}}
  \rput(9,5){\ovalnode{D}{$\Delta$}}
  \nccurve{A}{D}\nbput{$X$}
  \nccurve[angleA=-135]{D}{B1}\nbput{$X$}
  \nccurve[angleA=-45,ArrowInside=-]{D}{B2}\naput{$X$}
  \pnode(8,1){B3}
  \nccurve{B1}{B3}    
\end{pspicture}\]
We leave the proof as an exercise for the reader.

Many of the puzzling features of quantum theory come from the
noncartesianness of the usual tensor product in $\Hilb$.  
For example, in a cartesian category, every morphism 
\[\begin{pspicture}(0,0)(4,3)
\psset{angleA=-90,angleB=90,ArrowInside=->,arrowscale=2}
    \rput(2,2){\ovalnode{F}{$g$}}
    \pnode(1,0){Y1}
    \pnode(3,0){Y2}
    \nccurve[angleA=-120]{F}{Y1} \nbput[npos=1]{$X$}
    \nccurve[angleA=-60]{F}{Y2} \naput[npos=1]{$X'$}
\end{pspicture}\]
is actually of the form 
\[\begin{pspicture}(0,0)(4,3)
\psset{angleA=-90,angleB=90,ArrowInside=->,arrowscale=2}
\pnode(1,4){A}
\rput(1,2){\ovalnode{B}{$f$}}
\pnode(1,0){C}
\nccurve{B}{C} \nbput{$X$}
\rput(3,2){\ovalnode{B}{$f'$}}
\pnode(3,0){C}
\nccurve{B}{C} \naput{$X'$}
\end{pspicture}\]
In the case of $\Set$, this says that every point of the set $X \times X'$
comes from a point of $X$ and a point of $X'$.  In physics, this
would say that every state $g$ of the combined system $X \tensor X'$ 
is built by combining states of the systems $X$ and $X'$.  
Bell's theorem \cite{JSBell} says that is {\it not} true in quantum 
theory.   The reason is that quantum theory uses the noncartesian
monoidal category $\Hilb$!

Also, in quantum theory we {\it cannot} freely duplicate or delete 
information.  Wootters and Zurek \cite{WZ} proved a precise
theorem to this effect, focused on duplication: the `no-cloning 
theorem'.  One can also prove a `no-deletion theorem'.  Again, 
these results rely on the noncartesian tensor product in $\Hilb$.

\subsection{Braided Monoidal Categories}
\label{braided}

In physics, there is often a process that lets us `switch' two systems
by moving them around each other.  In topology, there is a tangle that 
describes the process of switching two points:
\begin{center}
\epsfysize=1.2in\epsfbox{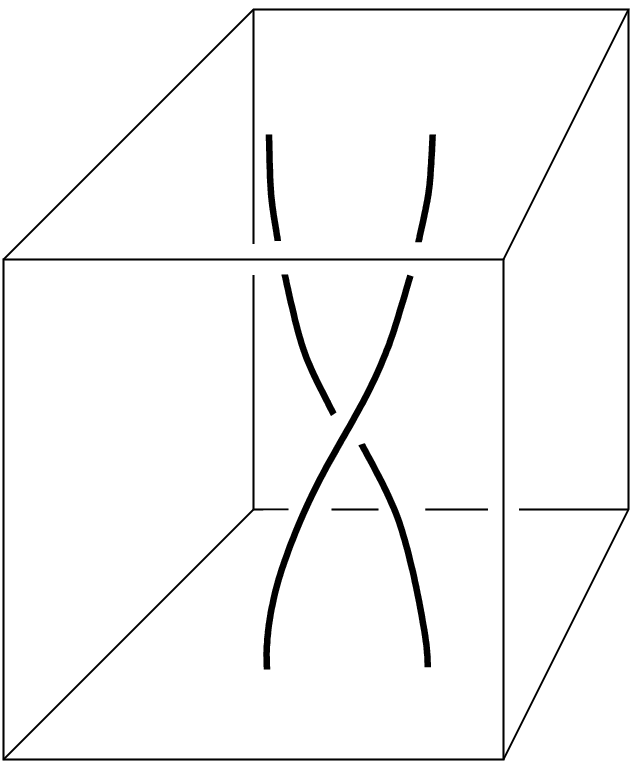} 
\end{center}
In logic, we can switch the order of two statements in a conjunction:
the statement `$X$ and $Y$' is isomorphic to `$Y$ and $X$'.  In 
computation, there is a simple program that switches the order of
two pieces of data.  A monoidal category in which we can do this sort
of thing is called `braided':

\begin{definition} A {\bf braided monoidal category} consists of:
\begin{itemize} 
\item a monoidal category $C$,
\item a natural isomorphism called the {\bf braiding} 
that assigns to every pair of objects $X, Y \in C$ an isomorphism
    \[b_{X,Y} \maps X\tensor Y \to Y \tensor X, \]
\end{itemize}
such that the {\bf hexagon equations} hold:
\[
    \begin{diagram}
 \node{X \tensor (Y \tensor Z)} \arrow{e,t}{a^{-1}_{X,Y,Z}} \arrow{s,l}{b_{X, Y\tensor Z}} 
 \node{(X \tensor Y) \tensor Z} \arrow{e,t}{b_{X,Y}\tensor 1_Z} 
 \node{(Y \tensor X) \tensor Z} \arrow{s,r}{a_{Y,X,Z}}\\
 \node{(Y \tensor Z) \tensor X} 
 \node{Y \tensor (Z \tensor X)} \arrow{w,b}{a^{-1}_{Y,Z,X}} 
 \node{Y\tensor (X\tensor Z)} \arrow{w,b}{1_Y \tensor b_{X, Z}}
\\
 \node{(X\tensor Y)\tensor Z} \arrow{e,t}{a_{X,Y,Z}} \arrow{s,l}{b_{X\tensor Y,Z}} 
 \node{X \tensor(Y\tensor Z)} \arrow{e,t}{1_X \tensor b_{Y,Z}} 
 \node{X \tensor(Z\tensor Y)} \arrow{s,r}{a^{-1}_{X,Z,Y}}\\
 \node{Z \tensor (X \tensor Y)} 
 \node{(Z \tensor X) \tensor Y} \arrow{w,b}{a_{Z,X,Y}} 
 \node{(X \tensor Z) \tensor Y} \arrow{w,b}{b_{X, Z}\tensor 1_Y} 
    \end{diagram}
\]
\end{definition}
\noindent
The first hexagon equation says that switching the object $X$ past 
$Y \tensor Z$ all at once is the same as switching it past $Y$ and then 
past $Z$ (with some associators thrown in to move the parentheses).
The second one is similar: it says switching $X \tensor Y$ past $Z$ 
all at once is the same as doing it in two steps.

In string diagrams, we
draw the braiding $b_{X,Y} \maps X \tensor Y \to Y \tensor X$ like this:
\[\begin{pspicture}(0,0)(2,3)
\psset{angleA=-90,angleB=90,ArrowInside=->,arrowscale=2}
\pnode(0,2){A1}
\pnode(0,0){B1}
\pnode(1,2){A2}
\pnode(1,0){B2}
\nccurve[ArrowInsidePos=.25]{A1}{B2} \nbput[npos=0]{$X$} \ncput{\cnode[linecolor=white,fillstyle=solid,fillcolor=white]{4pt}{C}}
\nccurve[ArrowInsidePos=.25]{A2}{B1} \naput[npos=0]{$Y$}
\end{pspicture}\]
We draw its inverse $b_{X,Y}^{-1}$ like this:
\[\begin{pspicture}(0,0)(2,3)
\psset{angleA=-90,angleB=90,ArrowInside=->,arrowscale=2}
\pnode(0,2){A1}
\pnode(0,0){B1}
\pnode(1,2){A2}
\pnode(1,0){B2}
\nccurve[ArrowInsidePos=.25]{A2}{B1} \naput[npos=0]{$Y$} \ncput{\cnode[linecolor=white,fillstyle=solid,fillcolor=white]{4pt}{C}}
\nccurve[ArrowInsidePos=.25]{A1}{B2} \nbput[npos=0]{$X$} 
\end{pspicture}\]
This is a nice notation, because it makes the equations
saying that $b_{X,Y}$ and $b_{X,Y}^{-1}$ are inverses 
`topologically true':
\[\begin{pspicture}(0,0)(2,5)
\psset{angleA=-90,angleB=90,ArrowInside=->,arrowscale=2}
\pnode(0,4){A1}
\pnode(0,2){A2}
\pnode(0,0){A3}
\pnode(1,4){B1}
\pnode(1,2){B2}
\pnode(1,0){B3}
\nccurve[ArrowInsidePos=.25]{A1}{B2} \nbput[npos=0]{$X$} \ncput{\cnode[linecolor=white,fillstyle=solid,fillcolor=white]{4pt}{C}}
\nccurve[ArrowInsidePos=.25]{B2}{A3} \nbput[npos=1]{$X$} \ncput{\cnode[linecolor=white,fillstyle=solid,fillcolor=white]{4pt}{C}}
\nccurve[ArrowInsidePos=.25]{B1}{A2} \naput[npos=0]{$Y$} 
\nccurve[ArrowInsidePos=.25]{A2}{B3} \naput[npos=1]{$Y$} 
\end{pspicture} \begin{pspicture}(0,0)(2,5)
\rput(0.5,2){=}
\end{pspicture} \begin{pspicture}(0,0)(2,5)
\psset{angleA=-90,angleB=90,ArrowInside=->,arrowscale=2}
\pnode(0,4){A1}
\pnode(0,0){A3}
\pnode(1,4){B1}
\pnode(1,0){B3}
\nccurve{A1}{A3} \nbput{$X$} 
\nccurve{B1}{B3} \naput{$Y$}
\end{pspicture} \begin{pspicture}(0,0)(2,5)
\rput(0.5,2){=}
\end{pspicture} \begin{pspicture}(0,0)(2,5)
\psset{angleA=-90,angleB=90,ArrowInside=->,arrowscale=2}
\pnode(0,4){A1}
\pnode(0,2){A2}
\pnode(0,0){A3}
\pnode(1,4){B1}
\pnode(1,2){B2}
\pnode(1,0){B3}
\nccurve[ArrowInsidePos=.25]{B1}{A2} \naput[npos=0]{$Y$} \ncput{\cnode[linecolor=white,fillstyle=solid,fillcolor=white]{4pt}{C}}
\nccurve[ArrowInsidePos=.25]{A2}{B3} \naput[npos=1]{$Y$} \ncput{\cnode[linecolor=white,fillstyle=solid,fillcolor=white]{4pt}{C}}
\nccurve[ArrowInsidePos=.25]{A1}{B2} \nbput[npos=0]{$X$} 
\nccurve[ArrowInsidePos=.25]{B2}{A3} \nbput[npos=1]{$X$} 
\end{pspicture}\]

Here are the hexagon equations as string diagrams:
\[\begin{pspicture}(0,0)(2,5)
  \psset{angleA=-90,angleB=90,ArrowInside=->,arrowscale=2}
  \pnode(0,4){A1}
  \pnode(2,4){B1}
  \pnode(0,0){A3}
  \pnode(2,0){B3}
  \nccurve[ArrowInsidePos=.4]{A1}{B3} \nbput[npos=0]{$X$} \naput[npos=1]{$X$} \ncput{\cnode[linecolor=white,fillstyle=solid,fillcolor=white]{4pt}{C}} 
  \nccurve[ArrowInsidePos=.4]{B1}{A3} \naput[npos=0]{$Y\tensor Z$} \nbput[npos=1]{$Y\tensor Z$}
  \end{pspicture} \begin{pspicture}(0,0)(4,5)
  \rput(2,2){=}
  \end{pspicture} \begin{pspicture}(0,0)(2,5)
  \psset{angleA=-90,angleB=90,ArrowInside=->,arrowscale=2}
  \pnode(0,4){A1}
  \pnode(1,4){B1}
  \pnode(2,4){C1}
  \pnode(0,2){A2}
  \pnode(1,2){B2}
  \pnode(2,2){C2}
  \pnode(0,0){A3}
  \pnode(1,0){B3}
  \pnode(2,0){C3}
  \nccurve[ArrowInsidePos=1]{A1}{B2} \nbput[npos=0]{$X$} \ncput{\cnode[linecolor=white,fillstyle=solid,fillcolor=white]{4pt}{C}}
  \nccurve[ArrowInsidePos=1]{B1}{A2} \naput[npos=0]{$Y$} 
  \nccurve[ArrowInsidePos=1]{C1}{C2} \naput[npos=0]{$Z$} 
  \nccurve[ArrowInside=]{A2}{A3} \nbput[npos=1]{$Y$} 
  \nccurve[ArrowInside=]{B2}{C3} \naput[npos=1]{$X$} \ncput{\cnode[linecolor=white,fillstyle=solid,fillcolor=white]{4pt}{C}}
  \nccurve[ArrowInside=]{C2}{B3} \nbput[npos=1]{$Z$}
\end{pspicture}\]

\[\begin{pspicture}(0,0)(2,5)
  \psset{angleA=-90,angleB=90,ArrowInside=->,arrowscale=2}
  \pnode(0,4){A1}
  \pnode(2,4){B1}
  \pnode(0,0){A3}
  \pnode(2,0){B3}
  \nccurve[ArrowInsidePos=.4]{A1}{B3} \nbput[npos=0]{$X\tensor Y$} \naput[npos=1]{$X\tensor Y$} \ncput{\cnode[linecolor=white,fillstyle=solid,fillcolor=white]{4pt}{C}}
  \nccurve[ArrowInsidePos=.4]{B1}{A3} \naput[npos=0]{$Z$} \nbput[npos=1]{$Z$} 
  \end{pspicture} \begin{pspicture}(0,0)(4,5)
  \rput(2,2){=}
  \end{pspicture} \begin{pspicture}(0,0)(2,5)
  \psset{angleA=-90,angleB=90,ArrowInside=->,arrowscale=2}
  \pnode(0,4){A1}
  \pnode(1,4){B1}
  \pnode(2,4){C1}
  \pnode(0,2){A2}
  \pnode(1,2){B2}
  \pnode(2,2){C2}
  \pnode(0,0){A3}
  \pnode(1,0){B3}
  \pnode(2,0){C3}
  \nccurve[ArrowInsidePos=1]{B1}{C2} \nbput[npos=0]{$Y$} \ncput{\cnode[linecolor=white,fillstyle=solid,fillcolor=white]{4pt}{C}}
  \nccurve[ArrowInsidePos=1]{C1}{B2} \naput[npos=0]{$Z$} 
  \nccurve[ArrowInsidePos=1]{A1}{A2} \nbput[npos=0]{$X$} 
  \nccurve[ArrowInside=]{C2}{C3} \naput[npos=1]{$Y$} 
  \nccurve[ArrowInside=]{A2}{B3} \naput[npos=1]{$X$} \ncput{\cnode[linecolor=white,fillstyle=solid,fillcolor=white]{4pt}{C}}
  \nccurve[ArrowInside=]{B2}{A3} \nbput[npos=1]{$Z$}
\end{pspicture}\]

For practice, we urge you to prove the following equations:
\[\begin{pspicture}(0,0)(2,5)
  \psset{angleA=-90,angleB=90,ArrowInside=->,arrowscale=2}
  \pnode(0,4){A1}
  \rput(0,3){\ovalnode{F}{$f$}}
  \pnode(0,2){A2}
  \pnode(0,0){A3}
  \pnode(1,4){B1}
  \rput(1,3){\ovalnode{G}{$g$}}
  \pnode(1,2){B2}
  \pnode(1,0){B3}
  \nccurve[ArrowInsidePos=.75]{A1}{F} \nbput[npos=0]{$X$}
  \nccurve[ArrowInside=]{F}{A2}
  \nccurve[ArrowInsidePos=.25]{A2}{B3} \naput[npos=1]{$X'$} \ncput{\cnode[linecolor=white,fillstyle=solid,fillcolor=white]{4pt}{C}}
  \nccurve[ArrowInsidePos=.75]{B1}{G} \naput[npos=0]{$Y$}
  \nccurve[ArrowInside=]{G}{B2}
  \nccurve[ArrowInsidePos=.25]{B2}{A3} \nbput[npos=1]{$Y'$}
  \end{pspicture} \begin{pspicture}(0,0)(2,5)
  \rput(0.5,2){=}
  \end{pspicture} \begin{pspicture}(0,0)(1,5)
  \psset{angleA=-90,angleB=90,ArrowInside=->,arrowscale=2}
  \pnode(0,4){A1}
  \pnode(0,2){A2}
  \rput(0,1){\ovalnode{G}{$g$}}
  \pnode(0,0){A3}
  \pnode(1,4){B1}
  \pnode(1,2){B2}
  \rput(1,1){\ovalnode{F}{$f$}}
  \pnode(1,0){B3}
  \nccurve[ArrowInsidePos=.75]{A1}{B2} \nbput[npos=0]{$X$} \ncput{\cnode[linecolor=white,fillstyle=solid,fillcolor=white]{4pt}{C}}
  \nccurve[ArrowInside=]{B2}{F}
  \nccurve[ArrowInsidePos=.75]{F}{B3} \naput[npos=1]{$X'$}
  \nccurve[ArrowInsidePos=.75]{B1}{A2} \naput[npos=0]{$Y$}
  \nccurve[ArrowInside=]{A2}{G}
  \nccurve[ArrowInsidePos=.75]{G}{A3} \nbput[npos=1]{$Y'$}
\end{pspicture}\]
\[\begin{pspicture}(0,0)(2,7)
\psset{angleA=-90,angleB=90,ArrowInside=->,arrowscale=2}
\pnode(0,6){A1}
\pnode(1,6){B1}
\pnode(2,6){C1}
\pnode(0,4){A2}
\pnode(1,4){B2}
\pnode(2,4){C2}
\pnode(0,2){A3}
\pnode(1,2){B3}
\pnode(2,2){C3}
\pnode(0,0){A4}
\pnode(1,0){B4}
\pnode(2,0){C4}
\nccurve[ArrowInsidePos=1]{C1}{C2} \naput[npos=0]{$Z$} 
\nccurve[ArrowInsidePos=1]{A1}{B2} \nbput[npos=0]{$X$} \ncput{\cnode[linecolor=white,fillstyle=solid,fillcolor=white]{4pt}{C}}
\nccurve[ArrowInsidePos=1]{B1}{A2} \nbput[npos=0]{$Y$}
\nccurve[ArrowInsidePos=1]{B2}{C3} \ncput{\cnode[linecolor=white,fillstyle=solid,fillcolor=white]{4pt}{C}}
\nccurve[ArrowInsidePos=1]{C2}{B3} 
\nccurve[ArrowInsidePos=1]{A2}{A3} 
\nccurve[ArrowInside=]{C3}{C4} \naput[npos=1]{$X$} 
\nccurve[ArrowInside=]{A3}{B4} \nbput[npos=1]{$Y$} \ncput{\cnode[linecolor=white,fillstyle=solid,fillcolor=white]{4pt}{C}}
\nccurve[ArrowInside=]{B3}{A4} \nbput[npos=1]{$Z$}
\end{pspicture} \begin{pspicture}(0,0)(4,5)
\rput(2,3){=}
\end{pspicture} \begin{pspicture}(0,0)(2,7)
\psset{angleA=-90,angleB=90,ArrowInside=->,arrowscale=2}
\pnode(0,6){A1}
\pnode(1,6){B1}
\pnode(2,6){C1}
\pnode(0,4){A2}
\pnode(1,4){B2}
\pnode(2,4){C2}
\pnode(0,2){A3}
\pnode(1,2){B3}
\pnode(2,2){C3}
\pnode(0,0){A4}
\pnode(1,0){B4}
\pnode(2,0){C4}
\nccurve[ArrowInsidePos=1]{B1}{C2} \nbput[npos=0]{$Y$} \ncput{\cnode[linecolor=white,fillstyle=solid,fillcolor=white]{4pt}{C}}
\nccurve[ArrowInsidePos=1]{C1}{B2} \naput[npos=0]{$Z$} 
\nccurve[ArrowInsidePos=1]{A1}{A2} \nbput[npos=0]{$X$}
\nccurve[ArrowInsidePos=1]{C2}{C3} 
\nccurve[ArrowInsidePos=1]{A2}{B3} \ncput{\cnode[linecolor=white,fillstyle=solid,fillcolor=white]{4pt}{C}}
\nccurve[ArrowInsidePos=1]{B2}{A3} 
\nccurve[ArrowInside=]{B3}{C4} \naput[npos=1]{$X$} \ncput{\cnode[linecolor=white,fillstyle=solid,fillcolor=white]{4pt}{C}}
\nccurve[ArrowInside=]{C3}{B4} \nbput[npos=1]{$Y$} 
\nccurve[ArrowInside=]{A3}{A4} \nbput[npos=1]{$Z$}
\end{pspicture}\]
If you get stuck, here are some hints.  The first equation
follows from the naturality of the braiding.  The second
is called the {\bf Yang--Baxter
equation} and follows from a combination of naturality and the
hexagon equations \cite{JS1}.

Next, here are some examples.  There can be many different ways to give a
monoidal category a braiding, or none.  However, most of our favorite
examples come with well-known `standard' braidings:

\begin{itemize}
\item Any cartesian category automatically becomes braided, and
in $\Set$ with its cartesian product, this standard braiding is given by:
\[
\begin{array}{rccl}
        b_{X,Y} & \maps X\times Y& \to& Y\times X \\
                    & (x,y) &\mapsto & (y,x) .
\end{array}
\]
\item In $\Hilb$ with its usual tensor product, the standard braiding
is given by:
\[
\begin{array}{rccl}
        b_{X,Y} & \maps X\tensor Y& \to& Y\tensor X \\
                    & x \tensor y &\mapsto & y \tensor x .
\end{array}
\]
\item The monoidal category $n\Cob$ has a standard braiding where
$b_{X,Y}$ is diffeomorphic to the disjoint union of cylinders $X
\times [0,1]$ and $Y \times [0,1]$.  For $2\Cob$ this braiding looks 
as follows when $X$ and $Y$ are circles:
\[
\begin{pspicture}[0.5](4,2.5)
  \rput(2,0){\ucrossc}
\end{pspicture}
\qquad
 \xy
 {\ar_{b_{X,Y}} (0,13)*+{X \tensor Y}; (0,-13)*+{Y \tensor X}};
 \endxy
\]
\item The monoidal category $\Tang_k$ has a standard braiding 
when $k \ge 2$.  For $k = 2$ this looks as follows when $X$ and 
$Y$ are each a single point:
\begin{center}
\epsfysize=1.2in\epsfbox{braid.eps}
\qquad
$
 \xy
 {\ar_{b_{X,Y}} (0,32)*+{X \tensor Y}; (0,-2)*+{Y \tensor X}};
 \endxy
$
\end{center}
\end{itemize}

The example of $\Tang_k$ illustrates an important pattern.  $\Tang_0$ is
just a category, because in 0-dimensional space we can only do
processes in `series': that is, compose morphisms.  $\Tang_1$ is a
monoidal category, because in 1-dimensional space we can also do
processes in `parallel': that is, tensor morphisms.  $\Tang_2$ is a
braided monoidal category, because in 2-dimensional space there is
room to move one object around another.  Next we shall see what happens when
space has 3 or more dimensions!

\subsection{Symmetric Monoidal Categories}
\label{symmetric}

Sometimes switching two objects and switching them again is the same
as doing nothing at all.  Indeed, this situation is very familiar.
So, the first braided monoidal categories to be discovered were
`symmetric' ones \cite{MacLane}:

\begin{definition}
    A {\bf symmetric monoidal category} is a braided monoidal
category where the braiding satisfies $b_{X,Y} = b_{Y,X}^{-1}$.
\end{definition}

So, in a symmetric monoidal category, 
\[\begin{pspicture}(0,0)(5,5)
\psset{angleA=-90,angleB=90,ArrowInside=->,arrowscale=2}
\pnode(0,4){A1}
\pnode(1,4){A2}
\pnode(0,2){B1}
\pnode(1,2){B2}
\nccurve[ArrowInsidePos=1]{A1}{B2} \nbput[npos=0]{$X$} \ncput{\cnode[linecolor=white,fillstyle=solid,fillcolor=white]{4pt}{C}}
\nccurve[ArrowInsidePos=1]{A2}{B1} \naput[npos=0]{$Y$}
\pnode(0,2){A1}
\pnode(1,2){A2}
\pnode(0,0){B1}
\pnode(1,0){B2}
\nccurve[ArrowInside=-]{A1}{B2} \naput[npos=1]{$Y$} \ncput{\cnode[linecolor=white,fillstyle=solid,fillcolor=white]{4pt}{C}}
\nccurve[ArrowInside=-]{A2}{B1} \nbput[npos=1]{$X$}

\rput(2.5,2){$=$}

\pnode(4,4){A1}
\pnode(5,4){A2}
\pnode(4,0){B1}
\pnode(5,0){B2}
\nccurve{A1}{B1} \nbput{$X$}
\nccurve{A2}{B2} \naput{$Y$}
\end{pspicture}\]
or equivalently:
\[\begin{pspicture}(0,0)(1,3)
\psset{angleA=-90,angleB=90,ArrowInside=->,arrowscale=2}
\pnode(0,2){A1}
\pnode(1,2){A2}
\pnode(0,0){B1}
\pnode(1,0){B2}
\nccurve[ArrowInsidePos=.25]{A1}{B2} \nbput[npos=0]{$X$} \ncput{\cnode[linecolor=white,fillstyle=solid,fillcolor=white]{4pt}{C}}
\nccurve[ArrowInsidePos=.25]{A2}{B1} \naput[npos=0]{$Y$}
\end{pspicture}
\begin{pspicture}(0,0)(2,2)
\rput(1,1){$=$}
\end{pspicture}
\begin{pspicture}(0,0)(1,2)
\psset{angleA=-90,angleB=90,ArrowInside=->,arrowscale=2}\pnode(0,2){A1}
\pnode(1,2){A2}
\pnode(0,0){B1}
\pnode(1,0){B2}
\nccurve[ArrowInsidePos=.25]{A2}{B1} \naput[npos=0]{$Y$} \ncput{\cnode[linecolor=white,fillstyle=solid,fillcolor=white]{4pt}{C}}
\nccurve[ArrowInsidePos=.25]{A1}{B2} \nbput[npos=0]{$X$} 
\end{pspicture}\]

Any cartesian category automatically becomes a symmetric monoidal
category, so $\Set$ is symmetric.  It is also easy to check that
$\Hilb$, $n\Cob$ are symmetric monoidal categories.  So is $\Tang_k$ 
for $k \ge 3$.   

Interestingly, $\Tang_k$ `stabilizes' at $k = 3$: increasing the value
of $k$ beyond this value merely gives a category equivalent to
$\Tang_3$.  The reason is that we can already untie all knots in
4-dimensional space; adding extra dimensions has no real effect.  
In fact, $\Tang_k$ for $k \ge 3$ is equivalent to $1\Cob$.
This is part of a conjectured larger pattern called the `Periodic 
Table' of $n$-categories \cite{BD}.  A piece of this is shown in 
Table \ref{periodic_table}.

An $n$-category has not only morphisms going between objects, but 
2-morphisms going between morphisms, 3-morphisms going between 
2-morphisms and so on up to $n$-morphisms.  In topology we can 
use $n$-categories to describe tangled higher-dimensional surfaces
\cite{HDA4}, and in physics we can use them to describe not just 
particles but also strings and higher-dimensional membranes \cite{BD,BL}.   
The Rosetta Stone we are describing concerns only the $n = 1$ column 
of the Periodic Table.   So, it is probably just a fragment 
of a larger, still buried $n$-categorical Rosetta Stone.  

\begin{table}[h]
\begin{center}
\begin{tabular}{|c|c|c|c|}  \hline
        & $\mathbf{\mathit n = 0}$ & $\mathbf{\mathit n = 1}$ &
$\mathbf{\mathit n = 2}$\\ \hline $\mathbf{\mathit k = 0}$ & sets
& categories & 2-categories     \\     \hline
$\mathbf{\mathit k = 1}$  & monoids   & monoidal   & monoidal         \\
        &           & categories & 2-categories     \\     \hline
$\mathbf{\mathit k = 2}$  &commutative& braided    & braided          \\
        & monoids   & monoidal   & monoidal         \\
        &           & categories & 2-categories     \\     \hline
$\mathbf{\mathit k = 3}$  &`'         & symmetric  & sylleptic \\
        &           & monoidal   & monoidal         \\
        &           & categories & 2-categories     \\     \hline
$\mathbf{\mathit k = 4}$  &`'         & `'         & symmetric \\
        &           &            & monoidal         \\
        &           &            & 2-categories     \\     \hline
$\mathbf{\mathit k = 5}$  &`'         &`'          & `'               \\
        &           &            &                  \\
        &           &            &                  \\     \hline
$\mathbf{\mathit k = 6}$  &`'         &`'          & `'               \\
        &           &            &                  \\
        &           &            &                  \\     \hline
\end{tabular}
\\
\caption{The Periodic Table: conjectured descriptions of $(n+k)$-categories
with only one $j$-morphism for $j < k$.}
\label{periodic_table}
\end{center}
\end{table}

\subsection{Closed Categories}
\label{closed}

In quantum mechanics, one can encode a linear operator $f \maps X \to
Y$ into a quantum state using a technique called `gate teleportation'
\cite{GC}.  In topology, there is a way to take any tangle $f \maps X
\to Y$ and bend the input back around to make it part of the output.
In logic, we can take a proof that goes from some assumption $X$ to
some conclusion $Y$ and turn it into a proof that goes from no
assumptions to the conclusion `$X$ implies $Y$'.  In computer science,
we can take any program that takes input of type $X$ and produces
output of type $Y$, and think of it as a piece of data of a new type:
a `function type'.  The underlying concept that unifies all these
examples is the concept of a `closed category'.

Given objects $X$ and $Y$ in any category $C$, there is a {\em set} of
morphisms from $X$ to $Y$, denoted $\hom(X,Y)$.  In a closed category
there is also an {\em object} of morphisms from $X$ to $Y$, which we
denote by $X \lhom Y$.  (Many other notations are also used.)  In this
situation we speak of an `internal hom', since the object $X \lhom Y$
lives inside $C$, instead of `outside', in the category of sets.

Closed categories were introduced in 1966, by Eilenberg and Kelly \cite{EK}.
While these authors were able to define a closed structure for any
category, it turns out that the internal hom is most easily understood
for monoidal categories.  The reason is that when our category has
a tensor product, it is closed precisely when morphisms from $X \tensor Y$
to $Z$ are in natural one-to-one correspondence with morphisms from $Y$
to $X \lhom Z$.  In other words, it is closed when we have a natural
isomorphism
\[  
\begin{array}{rcl}       
 \hom(X \tensor Y , Z) &\iso & \hom(Y, X \lhom Z)  \\
                    f &\mapsto & \tilde{f}  
\end{array}
\]
For example, in the category $\Set$, if we take
$X \tensor Y$ to be the cartesian product $X \times Y$, then 
$X \lhom Z$ is just the set of functions from $X$ to $Z$, and we 
have a one-to-one correspondence between
\begin{itemize}
\item functions $f$ that eat elements of
$X \times Y$ and spit out elements of $Z$ 
\end{itemize}
and 
\begin{itemize}
\item
functions $\tilde{f}$ that eat elements of $Y$ and spit out 
functions from $X$ to $Z$.
\end{itemize}
This correspondence goes as follows:
\[          \tilde{f}(x)(y) = f(x,y)  .\]

Before considering other examples, we should make the definition
of `closed monoidal category' completely precise.  For this we must
note that for any category $C$, there is a functor
\[            \hom \maps C^{\op} \times C \to \Set  .\]

\begin{definition} The {\bf opposite category} $C^{\op}$ of a category $C$ 
has the same objects as $C$, but a morphism $f \maps x \to y$ in $C^{\op}$ is 
a morphism $f \maps y \to x$ in $C$, and the composite $gf$ in
$C^{\op}$ is the composite $fg$ in $C$.  \end{definition}

\begin{definition} 
\label{hom.functor}
For any category $C$, the {\bf hom functor}
\[       \hom \maps C^\op \times C \to \Set  \]
sends any object $(X,Y) \in C^\op \times C$ to the set $\hom(X,Y)$,
and sends any morphism $(f,g) \in C^\op \times C$ to the function
\[ 
\begin{array}{rrcl}
\hom(f,g) \maps & \hom(X,Y) & \to     & \hom(X',Y')\\
                & h         & \mapsto & ghf
\end{array}
\]
when $f\maps X' \to X$ and $g\maps Y \to Y'$ are morphisms in $C$.
\end{definition}

\begin{definition}
\label{closed.definition}
A monoidal category $C$ is {\bf left closed} if there is an
{\bf internal hom} functor
\[     \lhom \maps C^\op \times C \to C \]
together with a natural isomorphism $c$ called
{\bf currying} that assigns to any objects $X, Y, Z \in C$ 
a bijection
\[ \begin{array}{rrcl}
 c_{X,Y,Z} \maps & \hom(X\tensor Y, Z) & \isoto & \hom(X,Y \lhom Z)  
\end{array}
\]
It is {\bf right closed} if there is an internal hom functor as above
and a natural isomorphism
\[ \begin{array}{rrcl}
 c_{X,Y,Z} \maps & \hom(X\tensor Y, Z) & \isoto & \hom(Y,X \lhom Z) . 
\end{array}
\]
\end{definition}
\noindent
The term `currying' is mainly used in computer science, after the 
work of Curry \cite{Curry}.  In the rest of this section we only 
consider {\it right} closed monoidal categories.   Luckily, there 
is no real difference between left and right closed for a braided 
monoidal category, as the braiding gives an isomorphism $X \tensor 
Y \cong Y \tensor X$.  

All our examples of monoidal categories are closed, but we shall see 
that, yet again, $\Set$ is different from the rest:

\begin{itemize}
\item The cartesian category $\Set$ is closed, where 
$X \lhom Y$ is just the set of functions from $X$ to $Y$.
In $\Set$ or any other cartesian closed category, 
the internal hom $X \lhom Y$ is usually denoted $Y^X$.
To minimize the number of different notations and emphasize
analogies between different contexts, we shall not do this:
we shall always use $X \lhom Y$.  To treat $\Set$ as {\it left}
closed, we define the curried version of $f \maps X \times Y \to Z$
as above:
\[        \tilde{f}(x)(y) = f(x,y)  .\]
To treat it as {\it right} closed, we instead define it by
\[        \tilde{f}(y)(x) = f(x,y)  .\]
This looks a bit awkward, but it will be nice for string diagrams.
\item The symmetric monoidal category $\Hilb$ with its usual tensor product 
is closed, where $X \lhom Y$ is the set of linear operators
from $X$ to $Y$, made into a Hilbert space in a standard way.  In 
this case we have an isomorphism
\[        X \lhom Y \iso X^* \tensor Y \]
where $X^*$ is the dual of the Hilbert space $X$, that is, the
set of linear operators $f \maps X \to \mathbb{C}$, made into
a Hilbert space in the usual way.  
\item The monoidal category $\Tang_k$ ($k \ge 1$) is closed.
As with $\Hilb$, we have
\[        X \lhom Y \iso X^* \tensor Y \]
where $X^*$ is the orientation-reversed version of $X$.
\item The symmetric monoidal category $n\Cob$ is also closed;
again
\[        X \lhom Y \iso X^* \tensor Y \]
where $X^*$ is the $(n-1)$-manifold $X$ with its orientation
reversed.
\end{itemize}

Except for $\Set$, all these examples are actually `compact'.  This
basically means that $X \lhom Y$ is isomorphic to $X^* \tensor Y$, where
$X^*$ is some object called the `dual' of $X$.  But to make this
precise, we need to define the `dual' of an object in an arbitrary 
monoidal category.

To do this, let us generalize from the case of $\Hilb$.  As 
already mentioned, each object $X \in \Hilb$ has a dual $X^*$ 
consisting of all linear operators $f \maps X \to I$, where the
unit object $I$ is just $\mathbb{C}$.  There is thus a 
linear operator 
\[
\begin{array}{rccl}
          e_X  \maps & X \tensor X^* & \to    & I  \\
                     & x \tensor f  & \mapsto & f(x)  
\end{array}
\]
called the {\bf counit} of $X$.  Furthermore, the space of all linear 
operators from $X$ to $Y \in \Hilb$ can be identified with $X^* \tensor Y$.  
So, there is also a linear operator called the {\bf unit} of $X$:
\[ 
\begin{array}{rccc}
          i_X  \maps & I & \to     & X^* \tensor X \\
                     & c & \mapsto & c\, 1_X  
\end{array}
\]   
sending any complex number $c$ to the corresponding multiple of the
identity operator. 

The significance of the unit and counit become clearer
if we borrow some ideas from Feynman.  In physics, if 
$X$ is the Hilbert space of internal states of some particle, 
$X^\ast$ is the Hilbert space for the corresponding antiparticle.  
Feynman realized that it is enlightening to think of antiparticles 
as particles going backwards in time.  So, we draw a wire labelled by 
$X^\ast$ as a wire labelled by $X$, but with an arrow pointing 
`backwards in time': that is,
up instead of down:
\[\begin{pspicture}(0,0)(2.5,2)
 \psset{angleA=-90,angleB=90,ArrowInside=->,arrowscale=2}
 \pnode(0,2){A1}
 \pnode(0,0){B1}
 \pnode(2.5,2){A2}
 \pnode(2.5,0){B2}
 \nccurve{A1}{B1} \nbput{$X^*$}
 \rput(1,1){=}
 \nccurve[angleA=90,angleB=-90]{B2}{A2} \naput{$X$}
 \end{pspicture}\]
(Here we should admit that most physicists use the opposite convention,
where time marches up the page.  Since we read from top to bottom,
we prefer to let time run down the page.)  

If we draw $X^*$ as $X$ going backwards in time, we can 
draw the unit as a {\bf cap}:
\[\begin{pspicture}(0,0)(1,2)
  \psset{ArrowInside=->,arrowscale=2}
  \pnode(0,0){A1}
  \pnode(1,0){B1}
  \pnode(.5,1){C1}
  \nccurve[angleA=90,angleB=180]{A1}{C1} \naput[npos=0]{$X$}
  \nccurve[angleA=0,angleB=90]{C1}{B1} \naput[npos=1]{$X$}
\end{pspicture}\]
and the counit as a {\bf cup}:
\[\begin{pspicture}(0,0)(1,2)
  \psset{ArrowInside=->,arrowscale=2}
  \pnode(0,1){A1}
  \pnode(1,1){B1}
  \pnode(.5,0){C1}
  \nccurve[angleA=-90,angleB=180]{A1}{C1} \nbput[npos=0]{$X$}
  \nccurve[angleA=0,angleB=-90]{C1}{B1} \nbput[npos=1]{$X$}
\end{pspicture}\]
In Feynman diagrams, these describe the {\em creation} and
{\em annihilation} of virtual particle-antiparticle pairs!

It then turns out that the unit and counit satisfy two equations,
the {\bf zig-zag equations}:
\[\begin{pspicture}(0,0)(2,4)
  \psset{angleA=-90,angleB=90,ArrowInside=->,arrowscale=2}
  \pnode(0,3){A}
  \pnode(.5,1){B}
  \pnode(1,2){C}
  \pnode(1.5,0){D}
  \nccurve[angleB=180]{A}{B} \nbput[npos=0]{$X$}
  \nccurve[angleA=0,angleB=180]{B}{C}
  \nccurve[angleA=0]{C}{D} \nbput[npos=1]{$X$}
  \end{pspicture}
  \begin{pspicture}(0,0)(2,4)\rput(0.5,1.5){=}\end{pspicture}
  \begin{pspicture}(0,0)(1,4)
  \psset{angleA=-90,angleB=90,ArrowInside=->,arrowscale=2}
  \pnode(0,3){A}
  \pnode(0,0){B}
  \nccurve{A}{B} \nbput{$X$}
  \end{pspicture}
\]
\[
  \begin{pspicture}(0,0)(2,4)
  \psset{angleA=90,angleB=-90,ArrowInside=->,arrowscale=2}
  \pnode(1.5,3){A}
  \pnode(1,1){B}
  \pnode(.5,2){C}
  \pnode(0,0){D}
  \nccurve[angleB=180]{D}{C} \naput[npos=0]{$X$}
  \nccurve[angleA=0,angleB=180]{C}{B}
  \nccurve[angleA=0]{B}{A} \naput[npos=1]{$X$}
  \end{pspicture}
  \begin{pspicture}(0,0)(2,4)\rput(0.5,1.5){=}\end{pspicture}
  \begin{pspicture}(0,0)(1,4)
  \psset{angleA=90,angleB=-90,ArrowInside=->,arrowscale=2}
  \pnode(0,3){A}
  \pnode(0,0){B}
  \nccurve{B}{A} \naput{$X$}
  \end{pspicture}
\]
Verifying these is a fun exercise in linear algebra, which we
leave to the reader.  If we write these equations as commutative diagrams,
we need to include some associators and unitors, and they become
a bit intimidating:
\[
    \begin{diagram}
\node{X \tensor I} \arrow{e,t}{1_X \tensor i_X} \arrow{s,l}{r_X} 
\node{X \tensor (X^* \tensor X)} \arrow{e,t}{a^{-1}_{X,X^*,X}}
\node{(X \tensor X^*) \tensor X} \arrow{s,r}{e_X \tensor 1_X}\\
\node{X} \node[2]{I \tensor X} \arrow[2]{w,b}{l_X} 
\\
\node{I \tensor X^*} \arrow{e,t}{i_X \tensor 1_X} \arrow{s,l}{l_X} 
\node{(X^* \tensor X) \tensor X^*} \arrow{e,t}{a_{X^*,X,X^*}}
\node{X^* \tensor (X \tensor X^*)} \arrow{s,r}{1_{X^*} \tensor e_X}\\
\node{X^*} \node[2]{X^* \tensor I} \arrow[2]{w,b}{r_{X^*}} 
    \end{diagram}
\]
But, they really just say that zig-zags in string diagrams
can be straightened out.

This is particularly vivid in examples like $\Tang_k$ and $n\Cob$.  
For example, in $2\Cob$, taking $X$ to be the circle, the unit
looks like this:
\[
\begin{pspicture}[0.5](2,2)
  \rput(1,0){\zagc}
\end{pspicture}
\qquad
 \xy
 {\ar_{i_X} (0,9)*+{I}; (0,-13)*+{X^* \tensor X}};
 \endxy
\]
while the counit looks like this:
\[
\begin{pspicture}[0.5](2,2)
  \rput(1,0){\zigc}
\end{pspicture}
\qquad
 \xy
 {\ar_{e_X} (0,12)*+{X \tensor X^*}; (0,-10)*+{I}};
 \endxy
\]
In this case, the zig-zag identities say we can straighten a 
wiggly piece of pipe.

Now we are ready for some definitions:

\begin{definition} Given objects $X^*$ and $X$ in a monoidal
category, we call $X^*$ a {\bf right dual} of $X$, and $X$ 
a {\bf left dual} of $X^*$, if there are morphisms
\[           i_X \maps I \to X^* \tensor X  \]
and 
\[           e_X \maps X \tensor X^* \to I, \]
called the {\bf unit} and {\bf counit} respectively, satisfying
the zig-zag equations.
\end{definition}
One can show that the left or right dual of an object is
unique up to canonical isomorphism.   So, we usually speak 
of `the' right or left dual of an object, when it exists.

\begin{definition} A monoidal category $C$ is {\bf compact}
if every object $X \in C$ has both a left dual and a right dual.
\end{definition}
Often the term `autonomous' is used instead of 
`compact' here.  Many authors reserve the term `compact' for
the case where $C$ is symmetric or at least braided; then
left duals are the same as right duals, and things simplify \cite{FY}.
To add to the confusion, compact symmetric monoidal categories 
are often called simply `compact closed categories'.

A partial explanation for the last piece of terminology is that 
any compact monoidal category is automatically closed!   For this,
we define the internal hom on objects by
\[                X \lhom Y = X^* \tensor Y  .\]
We must then show that the $*$ operation extends naturally to a functor
$* \maps C \to C$, so that $\lhom$ is actually a functor.  Finally, 
we must check that there is a natural isomorphism
\[          \hom(X \tensor Y , Z) \cong \hom(Y, X^* \tensor Z) \]
In terms of string diagrams, this isomorphism takes any morphism
\[\begin{pspicture}(0,0)(2,4)
  \psset{angleA=-90,angleB=90,ArrowInside=->,arrowscale=2}
  \pnode(0,4){A}
  \pnode(2,4){B}
  \rput(1,2){\ovalnode{C}{$f$}}
  \pnode(1,0){D}
  \nccurve[angleB=135]{A}{C} \nbput{$X$}
  \nccurve[angleB=45]{B}{C} \naput{$Y$}
  \nccurve{C}{D} \nbput{$Z$}
\end{pspicture}\]
and bends back the input wire labelled $X$ to make it an output:
\[\begin{pspicture}(0,0)(2,4)
  \psset{angleA=-90,angleB=90,ArrowInside=->,arrowscale=2}
  \pnode(0,0){A}
  \pnode(1,4){B}
  \rput(1,2){\ovalnode{C}{$f$}}
  \pnode(2,0){D}
  \nccurve[angleA=90,angleB=135]{A}{C} \naput[npos=.25]{$X$}
  \nccurve[angleB=45]{B}{C} \nbput[npos=.25]{$Y$}
  \nccurve{C}{D} \naput[npos=.75]{$Z$}
\end{pspicture}\]

Now, in a compact monoidal category, we have:
\[\begin{pspicture}(0,0)(4,2)
  \psset{angleA=-90,angleB=90,ArrowInside=->,arrowscale=2}
  \pnode(0,2){A1}
  \pnode(0,0){B1}
  \pnode(1,2){A2}
  \pnode(1,0){B2}
  \nccurve[angleA=90,angleB=-90]{B1}{A1} \naput{$X$}
  \nccurve{A2}{B2} \nbput{$Z$}
  \rput(2,1){=}
  \pnode(3,2){A3}
  \pnode(3,0){B3}
  \nccurve{A3}{B3} \naput{$X\lhom Z$}
\end{pspicture}\]
But in general, closed monoidal categories don't allow arrows pointing
up!  So for these, drawing the internal hom is more of a challenge.  
We can use the same style of notation as long as we add a decoration --- a
{\bf clasp} --- that binds two strings together:
\[\begin{pspicture}(0,0)(4,2)
  \psset{angleA=-90,angleB=90,ArrowInside=->,arrowscale=2}
  \pnode(0,2){A1}
  \pnode(0,0){B1}
  \pnode(1,2){A2}
  \pnode(1,0){B2}
  \nccurve[angleA=90,angleB=-90]{B1}{A1} \naput{$X$} \ncput[npos=.8]{\pnode{A3}}
  \nccurve{A2}{B2} \nbput{$Z$} \ncput[npos=.2]{\cnode{4pt}{B3}}
  \nccurve[angleA=0,angleB=180,ArrowInside=]{A3}{B3}
  \rput(2,1){:=}
  \pnode(3,2){A3}
  \pnode(3,0){B3}
  \nccurve{A3}{B3} \naput{$X\lhom Z$}
\end{pspicture}\]
Only when our closed monoidal category happens to be compact can we
eliminate the clasp. 

Suppose we are working in a closed monoidal category.
Since we draw a morphism $f \maps X\tensor Y\to Z$ like this:
\[\begin{pspicture}(0,0)(2,4)
  \psset{angleA=-90,angleB=90,ArrowInside=->,arrowscale=2}
  \pnode(0,4){A}
  \pnode(2,4){B}
  \rput(1,2){\ovalnode{C}{$f$}}
  \pnode(1,0){D}
  \nccurve[angleB=135]{A}{C} \nbput{$X$}
  \nccurve[angleB=45]{B}{C} \naput{$Y$}
  \nccurve{C}{D} \nbput{$Z$}
\end{pspicture}\]
we can draw its curried version $\tilde f \maps Y\to X\lhom Z$ by
bending down the input wire labelled $X$ to make it part of the output:
\[\begin{pspicture}(0,0)(2,4)
  \psset{angleA=-90,angleB=90,ArrowInside=->,arrowscale=2}
  \pnode(0,0){A}
  \pnode(1,4){B}
  \rput(1,2){\ovalnode{C}{$f$}}
  \pnode(2,0){D}
  \nccurve[angleA=90,angleB=135]{A}{C} \naput[npos=.25]{$X$} \ncput[npos=.25]{\pnode{E}}
  \nccurve[angleB=45]{B}{C} \naput[npos=.25]{$Y$}
  \nccurve{C}{D} \naput[npos=.75]{$Z$} \ncput[npos=.75]{\cnode{4pt}{F}}
  \nccurve[linestyle=none,ArrowInside=-]{C}{C} \ncput{\pscircle{.8}}
  \nccurve[angleA=0,angleB=180,ArrowInside=-]{E}{F}
\end{pspicture}\]
Note that where we bent back the wire labelled $X$, a cap like this 
appeared:
\[\begin{pspicture}(0,0)(1,1)
  \psset{ArrowInside=->,arrowscale=2}
  \pnode(0,0){A1}
  \pnode(1,0){B1}
  \pnode(.5,1){C1}
  \nccurve[angleA=90,angleB=180]{A1}{C1} \naput[npos=0]{$X$}
  \nccurve[angleA=0,angleB=90]{C1}{B1} \naput[npos=1]{$X$}
\end{pspicture}\]
Closed monoidal categories don't really have a cap unless they are 
compact.  So, we drew a {\bf bubble} enclosing $f$ and the cap, 
to keep us from doing any illegal manipulations.  In the compact case,
both the bubble and the clasp are unnecessary, so we can draw 
$\tilde{f}$ like this:
\[\begin{pspicture}(0,0)(2,4)
  \psset{angleA=-90,angleB=90,ArrowInside=->,arrowscale=2}
  \pnode(0,0){A}
  \pnode(1,4){B}
  \rput(1,2){\ovalnode{C}{$f$}}
  \pnode(2,0){D}
  \nccurve[angleA=90,angleB=135]{A}{C} \naput[npos=.25]{$X$}
  \nccurve[angleB=45]{B}{C} \nbput[npos=.25]{$Y$}
  \nccurve{C}{D} \naput[npos=.75]{$Z$}
\end{pspicture}\]

An important special case of currying gives the {\bf name} of a morphism
$f \maps X\to Y$, 
\[          \name{f} \maps I \to X \lhom Y .\]
This is obtained by currying the morphism
\[ f r_x \maps I \tensor X \to Y . \]
In string diagrams, we draw $\name{f}$ as follows:
\[\begin{pspicture}(0,0)(1,3)
    \psset{angleA=-90,angleB=90,ArrowInside=->,arrowscale=2}
    \pnode(0,0){A}
    \rput(0.5,2){\ovalnode{C}{$f$}}
    \pnode(1,0){D}
    \nccurve[angleA=90, angleB=135]{A}{C} \naput[npos=.25]{$X$} \ncput[npos=.25]{\pnode{E}}
    \nccurve[angleA=-45]{C}{D} \naput[npos=.75]{$Y$} \ncput[npos=.75]{\cnode{4pt}{F}}
    \nccurve[linestyle=none,ArrowInside=-]{C}{C} \ncput{\pscircle{.8}}
    \nccurve[angleA=0,angleB=180,ArrowInside=-]{E}{F}
\end{pspicture}\]
In the category $\Set$, the unit object is the one-element set, $1$.
So, a morphism from this object to a set $Q$ picks out a point of $Q$.
In particular, the name $\name{f} \maps 1 \to X \lhom Y$ picks out the
element of $X \lhom Y$ corresponding to the function $f \maps X \to
Y$.  More generally, in any cartesian closed category the unit object
is the terminal object $1$, and a morphism from $1$ to an object $Q$
is called a {\bf point} of $Q$.  So, even in this case, we can say the
name of a morphism $f \maps X \to Y$ is a point of $X \lhom Y$.

Something similar works for $\Hilb$, though this example
is compact rather than cartesian.  In $\Hilb$, the unit object 
$I$ is just $\CC$.  So, a nonzero morphism from $I$ to any Hilbert 
space $Q$ picks out a nonzero vector in $Q$, which we can normalize
to obtain a {\bf state} in $Q$: that is, a unit vector.  In particular, 
the name of a nonzero morphism $f \maps X \to Y$ gives a state of
$X^* \tensor Y$.   This method of encoding operators as states is 
the basis of `gate teleportation' \cite{GC}.

Currying is a bijection, so we can also {\bf uncurry}:
\[ \begin{array}{rrcl}
 c_{X,Y,Z}^{-1} \maps & \hom(Y, X \lhom Z) & \isoto & \hom(X \tensor Y, Z)  \\
                      &         g            & \mapsto    & \utilde{g}   .
\end{array}
\]
Since we draw a morphism $g \maps Y \to X \lhom Z$ like this:
\[\begin{pspicture}(0,0)(3,4)
  \psset{angleA=-90,angleB=90,ArrowInside=->,arrowscale=2}
  \pnode(1,0){A}
  \pnode(2,4){B}
  \rput(2,2){\ovalnode{C}{$g$}}
  \pnode(3,0){D}
  \nccurve[angleA=90,angleB=-135]{A}{C} \naput[npos=.25]{$X$} \ncput[npos=.25]{\pnode{E}}
  \nccurve{B}{C} \naput[npos=.25]{$Y$}
  \nccurve[angleA=-45]{C}{D} \naput{$Z$} \ncput[npos=.75]{\cnode{4pt}{F}}
  \nccurve[angleA=0,angleB=180,ArrowInside=-]{E}{F}
\end{pspicture}\]
we draw its `uncurried' version $\utilde{g} \maps X \tensor Y\to Z $ 
by bending the output $X$ up to become an input:
\[\begin{pspicture}(0,0)(2,4)
  \psset{angleA=-90,angleB=90,ArrowInside=->,arrowscale=2}
  \pnode(0,4){A}
  \pnode(2,4){B}
  \rput(1,2){\ovalnode{C}{$g$}}
  \pnode(1,0){D}
  \nccurve[angleB=-135]{A}{C} \nbput{$X$} \ncput[npos=.9]{\pnode{E}}
  \nccurve{B}{C} \naput{$Y$} 
  \nccurve[angleA=-45,ArrowInsidePos=.75]{C}{D} \nbput[npos=.75]{$Z$} \ncput[npos=.2]{\cnode{4pt}{F}}
  \nccurve[linestyle=none,ArrowInside=-]{C}{C} \ncput{\pscircle{.8}}
  \nccurve[angleA=-90,angleB=180,ArrowInside=-]{E}{F}
\end{pspicture}\]
Again, we must put a bubble around the `cup' formed when we bend
down the wire labelled $Y$, unless we are in a compact monoidal category.

A good example of uncurrying is the {\bf evaluation} morphism:
\[ \ev_{X,Y} \maps X \tensor (X\lhom Y) \to Y . \]
This is obtained by uncurrying the identity 
\[          1_{X \lhom Y} \maps (X \lhom Y) \to (X \lhom Y). \]
In $\Set$, $\ev_{X,Y}$ takes any function from $X$ to $Y$ and evaluates it
at any element of $X$ to give an element of $Y$.  
In terms of string diagrams, the evaluation morphism looks like this:
\[\begin{pspicture}(0,0)(3,4)
  \psset{angleA=-90,angleB=90,ArrowInside=->,arrowscale=2}
  \pnode(0,4){A}
  \pnode(2,4){B}
  \pnode(3,4){C}
  \rput(1.5,2){\ovalnode{D}{$\ev$}}
  \pnode(1.5,0){E}
  \nccurve[angleB=144]{A}{D} \nbput{$X$}
  \nccurve[angleA=72,angleB=-90]{D}{B} \naput{$X$}\ncput[npos=0.75]{\pnode{F}}
  \nccurve[angleB=36]{C}{D} \naput{$Y$} \ncput[npos=0.25]{\cnode{4pt}{G}}
  \nccurve{D}{E} \nbput{$Y$}
  \nccurve[angleA=0,angleB=180,ArrowInside=-]{F}{G}
  \end{pspicture}
  \begin{pspicture}(0,0)(2,4)\rput(1,2){=}\end{pspicture}
  \begin{pspicture}(0,0)(3,4)
  \psset{angleA=-90,angleB=90,ArrowInside=->,arrowscale=2}
  \pnode(0,4){A}
  \pnode(2,4){B}
  \pnode(3,4){C}
  \rput(1.5,2){\cnode{24pt}{D}}
  \pnode(1.5,0){E}
  \nccurve[angleB=144]{A}{D} \nbput{$X$} \ncput[npos=1]{\pnode{H}}
  \nccurve[angleA=72,angleB=-90]{D}{B} \naput{$X$}\ncput[npos=0.75]{\pnode{F}} \ncput[npos=0]{\pnode{I}}
  \nccurve[angleB=36]{C}{D} \naput{$Y$} \ncput[npos=0.25]{\cnode{4pt}{G}} \ncput[npos=1]{\pnode{J}}
  \nccurve{D}{E} \nbput{$Y$} \ncput[npos=0]{\pnode{K}}
  \nccurve[angleA=0,angleB=180,ArrowInside=-]{F}{G}
  \psset{ArrowInside=-}
  \nccurve[angleA=-36, angleB=-72]{H}{I}
  \nccurve[angleA=-144]{J}{K}
\end{pspicture}\]

In any closed monoidal category, we can recover a morphism 
from its name using evaluation.  More precisely, this
diagram commutes:
\[
\begin{diagram}
\node{X \tensor I} \arrow{s,l}{1_X \tensor \name{f}}
\node{X} \arrow{w,t}{r^{-1}} \arrow{s,r}{f} \\
\node{X \tensor (X \lhom Y)} \arrow{e,b}{\ev_{X,Y}} \node{Y}
\end{diagram}
\]
Or, in terms of string diagrams:
\[\begin{pspicture}(0,0)(3,8)
  \psset{angleA=-90,angleB=90,ArrowInside=->,arrowscale=2}
  \pnode(0,8){id}
  \rput(2.5,6){\ovalnode{func}{$f$}}
  \rput(2.5,6){\pscircle{.8}}
  \pnode(0,4){A}
  \pnode(2,4){B}
  \pnode(3,4){C}
  \rput(1.5,2){\cnode{24pt}{D}}
  \pnode(1.5,0){E}
  \nccurve{id}{A}
  \nccurve[angleA=90, angleB=135]{B}{func}
  \nccurve[angleA=-45]{func}{C}
  \nccurve[angleB=144]{A}{D} \nbput{$X$} \ncput[npos=1]{\pnode{H}}
  \nccurve[angleA=72,angleB=-90]{D}{B} \naput{$X$}\ncput[npos=0.75]{\pnode{F}} \ncput[npos=0]{\pnode{I}}
  \nccurve[angleB=36]{C}{D} \naput{$Y$} \ncput[npos=0.25]{\cnode{4pt}{G}} \ncput[npos=1]{\pnode{J}}
  \nccurve{D}{E} \nbput{$Y$} \ncput[npos=0]{\pnode{K}}
  \nccurve[angleA=0,angleB=180,ArrowInside=-]{F}{G}
  \psset{ArrowInside=-}
  \nccurve[angleA=-36, angleB=-72]{H}{I}
  \nccurve[angleA=-144]{J}{K}
  \end{pspicture}
  \begin{pspicture}(0,0)(2,8)\rput(2,4){=}\end{pspicture}
  \begin{pspicture}(0,0)(2,8)
  \psset{angleA=-90,angleB=90,ArrowInside=->,arrowscale=2}
  \pnode(1,8){A}
  \rput(1,4){\ovalnode{B}{$f$}}
  \pnode(1,0){C}
  \nccurve{A}{B} \nbput{$X$}
  \nccurve{B}{C} \nbput{$Y$}
\end{pspicture}\]
We leave the proof of this as an exercise.  In general, one must
use the naturality of currying.  In the special case of a compact
monoidal category, there is a nice picture proof!  Simply pop the bubbles
and remove the clasp:
\[\begin{pspicture}(0,0)(3,7)
  \psset{angleA=-90,angleB=90,ArrowInside=->,arrowscale=2}
  \pnode(0,7){id}
  \rput(2.5,6){\ovalnode{func}{$f$}}
  \pnode(0,4){A}
  \pnode(2,4){B}
  \pnode(3,4){C}
  \rput(1.5,2){\cnode[linestyle=none]{24pt}{D}}
  \pnode(1.5,0){E}
  \nccurve{id}{A}
  \nccurve[angleA=90, angleB=135]{B}{func}
  \nccurve[angleA=-45]{func}{C}
  \nccurve[angleB=144]{A}{D} \nbput{$X$} \ncput[npos=1]{\pnode{H}}
  \nccurve[angleA=72,angleB=-90]{D}{B} \naput{$X$}\ncput[npos=0.75]{\pnode{F}} \ncput[npos=0]{\pnode{I}}
  \nccurve[angleB=36]{C}{D} \naput{$Y$} \ncput[npos=1]{\pnode{J}}
  \nccurve{D}{E} \nbput{$Y$} \ncput[npos=0]{\pnode{K}}
  \psset{ArrowInside=-}
  \nccurve[angleA=-36, angleB=-72]{H}{I}
  \nccurve[angleA=-144]{J}{K}
  \end{pspicture}
  \begin{pspicture}(0,0)(2,7)\rput(2,3){=}\end{pspicture}
  \begin{pspicture}(0,0)(2,7)
  \psset{angleA=-90,angleB=90,ArrowInside=->,arrowscale=2}
  \pnode(1,7){A}
  \rput(1,3){\ovalnode{B}{$f$}}
  \pnode(1,0){C}
  \nccurve{A}{B} \nbput{$X$}
  \nccurve{B}{C} \nbput{$Y$}
\end{pspicture}\]
The result then follows from one of the zig-zag identities.  

In our rapid introduction to string diagrams, we have not had
time to illustrate how these diagrams become a powerful tool for
solving concrete problems.  So, here are some starting points
for further study:

\begin{itemize}
\item 
Representations of Lie groups play a fundamental
role in quantum physics, especially gauge field theory.  
Every Lie group has a compact symmetric monoidal
category of finite-dimensional representations.  In his book
{\sl Group Theory}, Cvitanovic \cite{Cvitanovic} develops 
detailed string diagram descriptions 
of these representation categories for the classical Lie groups 
$\mathrm{SU}(n)$, $\mathrm{SO}(n)$, $\mathrm{SU}(n)$ and 
also the more exotic `exceptional' Lie groups.
His book also illustrates how this technology can be used to 
simplify difficult calculations in gauge field theory.
\item 
Quantum groups are a generalization of groups which show
up in 2d and 3d physics.  The big difference is that a quantum
group has compact {\em braided} monoidal category of finite-dimensional
representations.  Kauffman's {\sl Knots and Physics} \cite{Kauffman} 
is an excellent introduction to how quantum groups show up 
in knot theory and physics; it is packed with string diagrams.  
For more details on quantum groups and braided monoidal categories, 
see the book by Kassel \cite{Kassel}.
\item 
Kauffman and Lins \cite{KL} have written a beautiful string 
diagram treatment of the category of representations of the simplest 
quantum group, $SU_q(2)$.  They also use it to construct some famous 
3-manifold invariants associated to 3d and 4d topological quantum field 
theories: the Witten--Reshetikhin--Turaev, Turaev--Viro and Crane--Yetter 
invariants.  In this example, string diagrams are often called 
`$q$-deformed spin networks' \cite{Smolin}.  For generalizations to 
other quantum groups, see the more advanced texts by Turaev 
\cite{Turaev} and by Bakalov and Kirillov \cite{BK}.
The key ingredient is a special class of compact braided monoidal 
categories called `modular tensor categories'.
\item Kock \cite{Kock} has written a nice introduction to 2d 
topological quantum field theories which uses diagrammatic
methods to work with $2\Cob$.
\item 
Abramsky, Coecke and collaborators \cite{Abramsky,AC,AD,Coecke1,
CP1,CP2} have developed string diagrams as a tool for understanding
quantum computation.  The easiest introduction is Coecke's 
`Kindergarten quantum mechanics' \cite{Coecke2}.
\end{itemize}

\subsection{Dagger Categories}
\label{dagger}

Our discussion would be sadly incomplete without an important
admission: {\it nothing we have done so far with Hilbert spaces 
used the inner product!}   So, we have not yet touched on
the essence of quantum theory.

Everything we have said about $\Hilb$ applies equally well to $\Vect$:
the category of finite-dimensional {\it vector spaces} and linear
operators.  Both $\Hilb$ and $\Vect$ are compact symmetric monoidal
categories.  In fact, these compact symmetric monoidal categories are
`equivalent' in a certain precise sense \cite{MacLane2}.

So, what makes $\Hilb$ different?  In terms of category theory, the
special thing is that we can take the Hilbert space adjoint of any 
linear operator $f \maps X \to Y$ between finite-dimensional Hilbert 
spaces, getting an operator $f^\dagger \maps Y \to X$.  This ability 
to `reverse' morphisms makes $\Hilb$ into a `dagger category':

\begin{definition}
A {\bf dagger category} is a category $C$ such that for any morphism
$f \maps X \to Y$ in $C$ there is a specified morphism $f^\dagger \maps 
Y \to X$ such that
\[  (gf)^\dagger = f^\dagger g^\dagger  \]
for every pair of composable morphisms $f$ and $g$, and 
\[       (f^\dagger)^\dagger = f  \]
for every morphism $f$.
\end{definition}
Equivalently, a dagger category is one
equipped with a functor $\dagger \maps C \to C^\op$ 
that is the identity on objects and satisfies 
$(f^\dagger)^\dagger = f$ for every morphism.

In fact, all our favorite examples of categories can be made into
dagger categories, except for $\Set$:

\begin{itemize}
\item There is no way to make $\Set$ into a dagger category,
since there is a function from the empty set to the 1-element
set, but none the other way around.
\item The category $\Hilb$ becomes a dagger category as follows.
Given any morphism $f \maps X \to Y$ in $\Hilb$, there is a 
morphism $f^\dagger \maps Y \to X$, the {\bf Hilbert space adjoint} 
of $f$, defined by 
\[         \langle f^\dagger \psi  , \phi \rangle = 
           \langle \psi, f \phi \rangle  \]
for all $\phi \in X$, $\psi \in Y$.  
\item For any $k$, the category $\Tang_k$ becomes a dagger
category where we obtain $f^\dagger \maps Y \to X$ by reflecting
$f \maps X \to Y$ in the vertical direction, and then switching 
the direction of the little arrows denoting the orientations of
arcs and circles.
\item For any $n$, the category $n\Cob$ becomes a 
dagger category where we obtain $f^\dagger \maps Y \to X$
by switching the input and output of $f \maps X \to Y$, 
and then switching the orientation of each connected component
of $f$.  Again, a picture speaks a thousand words:
\[
\begin{pspicture}[.5](4,2.5)
  \rput(2,0){\comultc}
\end{pspicture}
\quad
 \xy
 {\ar_{f} (0,10)*+{X}; (0,-11)*+{Y}};
 \endxy
\qquad
\begin{pspicture}[.5](4,2.5)
  \rput(2,0){\multc}
\end{pspicture}
\quad
 \xy
 {\ar_{f^\dagger} (0,10)*+{Y}; (0,-11)*+{X}};
 \endxy
\]
In applications to physics, this dagger operation amounts to
`switching the future and the past'.
\end{itemize}

In all the dagger categories above, the dagger structure interacts in a 
nice way with the monoidal structure and also, when it exists, the 
braiding.  One can write a list of axioms characterizing how this
works \cite{Abramsky,AC,Selinger}.  So, it seems that the
ability to `reverse' morphisms is another way in which categories 
of a quantum flavor differ from the category of sets and functions.  
This has important implications for the foundations of quantum theory 
\cite{B4} and also for topological quantum field theory \cite{BD},
where dagger categories seem to be part of larger story involving
`$n$-categories with duals' \cite{HDA4}.  However, this story is still 
poorly understood --- there is much more work to be done.

\section{Logic}
\label{logic}
\EnableBpAbbreviations 

\subsection{Background}
\label{logic_overview}

Symmetric monoidal closed categories show up not only in physics and
topology, but also in logic.  We would like to explain how.
To set the stage, it seems worthwhile to sketch a few ideas from
20th-century logic.  

Modern logicians study many systems of reasoning beside ordinary
classical logic.  Of course, even classical logic comes in 
various degrees of strength.  First there is the `propositional 
calculus', which allows us to reason with abstract propositions 
$X, Y, Z, \dots$ and these logical connectives:
\[  
\begin{array}{cc}
\text{and}        & \wedge   \\
\text{or}         & \vee  \\
\text{implies}    & \Rightarrow \\
\text{not}        & \neg  \\
\text{true}       & \top     \\
\text{false}      & \bot
\end{array}
\]
Then there is the `predicate calculus', which also 
allows variables like $x,y,z, \dots$, predicates like 
$P(x)$ and $Q(x,y,z)$, and the symbols `for all' ($\forall$) 
and `there exists' ($\exists$), which allow us to quantify over variables.
There are also higher-order systems that allow us to quantify over
predicates, and so on.  To keep things simple, we mainly confine ourselves 
to the propositional calculus in what follows.  But even here, there are
many alternatives to the `classical' version!

The most-studied of these alternative systems are {\it weaker} 
than classical logic: they make it harder or even impossible to prove 
things we normally take for granted.  One reason is that some logicians
deny that certain familiar principles are actually valid.  But
there are also subtler reasons.   One is that
studying systems with rules of lesser strength allows for a fine-grained 
study of precisely which methods of reasoning are needed to prove which 
results.  Another reason --- the one that concerns us most here --- 
is that dropping familiar rules and then adding them back in one at 
at time sheds light on the connection between logic and category theory.

For example, around 1907 Brouwer \cite{Heyting} began advocating
`intuitionism'.  As part of this, he raised doubts about the law 
of excluded middle, which amounts to a rule saying that from $\neg \neg X$ 
we can deduce $X$.  One problem with this principle is that proofs 
using it are not `constructive'.   For example, we may 
prove by contradiction that some equation has a solution, but 
still have no clue how to construct the solution.   For Brouwer, this
meant the principle was invalid.

Anyone who feels the law of excluded middle is invalid is 
duty-bound to study intuitionistic logic.  But, there is 
another reason for studying this system.  Namely: we do not 
really {\it lose} anything by dropping the law
of excluded middle!  Instead, we {\em gain} a 
fine-grained distinction: the distinction between a direct proof 
of $X$ and a proof by contradiction, which yields merely $\neg \neg X$. 
If we do not care about this distinction we are free to ignore it,
but there is no harm in having it around.  

In the 1930's, this idea was made precise by G\"odel \cite{Goedel} and
Gentzen \cite{Gentzen}.  They showed that we can embed classical logic
in intuitionistic logic.  In fact, they found a map sending any
formula $X$ of the propositional calculus to a new formula $X^\circ$,
such that $X$ is provable classically if and only if $X^\circ$ is
provable intuitionistically.  (More impressively, this map also works
for the predicate calculus.)

Later, yet another reason for being interested in intuitionistic logic
became apparent: its connection to category theory.  In its 
very simplest form, this connection works as follows.  Suppose we 
have a set of propositions $X, Y, Z, \dots$ obeying the laws of
the intuitionistic propositional calculus.  We can create a category
$C$ where these propositions are objects and there is at most one morphism
from any object $X$ to any object $Y$: a single morphism when $X$ implies $Y$, 
and none otherwise!  

A category with at most one morphism from any object to any other 
is called a {\bf preorder}.  In the propositional calculus, we often
treat two propositions as equal when they both imply each other.
If we do this, we get a special sort of preorder: one where
isomorphic objects are automatically equal.  This special sort of
preorder is called a {\bf partially ordered set}, or {\bf poset}
for short.  Posets abound in logic, precisely because they offer
a simple framework for understanding implication.  

If we start from a set of propositions obeying the intuitionistic
propositional calculus, the resulting category $C$ is better
than a mere poset.  It is also cartesian, with $X \wedge Y$ as the product 
of $X$ and $Y$, and $\top$ as the terminal object!  To see this, note
that any proposition $Q$ has a unique morphism to $X \wedge Y$ whenever 
it has morphisms to $X$ and to $Y$.  This is simply a fancy way of
saying that $Q$ implies $X \wedge Y$ when it implies $X$ and implies $Y$.  
It is also easy to see that $\top$ is terminal: anything implies 
the truth.  

Even better, the category $C$ is cartesian closed, with 
$ X \Rightarrow Y$ as the internal hom.  The reason is that 
\[  X \wedge Y \; \text{implies}\; Z \quad \text{iff} \quad Y 
\; \text{implies} \; X \Rightarrow Z .\]
This automatically yields the basic property of the internal
hom:
\[ \hom(X\tensor Y, Z) \cong \hom(Y,X \lhom Z)  . \]
Indeed, if the reader is puzzled by the difference between 
`$X$ implies $Y$' and $X \Rightarrow Y$, we can now explain
this more clearly: the former involves the homset $\hom(X,Y)$ 
(which has one element when $X$ implies $Y$ and none otherwise), 
while the latter is the internal hom, an object in $C$.

So, $C$ is a cartesian closed poset.  But, it also has
one more nice property, thanks to the presence of $\vee$ and $\bot$
We have seen that $\wedge$ and $\top$ make the category $C$ cartesian; 
$\vee$ and $\bot$ satisfy exactly analogous rules, but with the 
implications turned around, so they make $C^\op$ cartesian.  

And that is all!  In particular, negation gives nothing more, since we
can define $\neg X$ to be $X \Rightarrow \bot$, and all its
intuitionistically valid properties then follow.  So, the kind of
category we get from the intuitionistic propositional calculus by
taking propositions as objects and implications as morphisms is
precisely a {\bf Heyting algebra}: a cartesian closed poset $C$ such
that $C^\op$ is also cartesian.

Heyting, a student of Brouwer, introduced Heyting algebras in
intuitionistic logic before categories were even invented.  
So, he used very different language to define them.   But, the
category-theoretic approach to Heyting algebras illustrates the
connection between cartesian closed categories and logic.
It also gives more evidence that dropping the law of 
excluded middle is an interesting thing to try.

Since we have explained the basics of cartesian closed categories,
but not said what happens when the {\it opposite} of such a category 
is {\it also} cartesian, in the sections to come we will take a
drastic step and limit our discussion of logic even further.   
We will neglect $\vee$ and $\bot$, and concentrate only on the 
fragment of the propositional calculus involving $\wedge$, $\top$ and 
$\Rightarrow$.  

Even here, it turns out, there are interesting things to say --- and
interesting ways to modify the usual rules.  This will be the
main subject of the sections to come.  But to set the stage, we 
need to say a bit about proof theory.

Proof theory is the branch of mathematical logic that treats proofs
as mathematical entities worthy of study in their own right.
It lets us dig deeper into the propositional calculus by studying 
not merely {\it whether or not} some assumption $X$ implies some 
conclusion $Y$, but the whole {\it set of proofs} leading from $X$ to 
$Y$.  This amounts to studying not just posets (or preorders), but
categories that allow many morphisms from one object to another.

In Hilbert's approach to proof, there were many axioms and just one rule 
to deduce new theorems: {\em modus ponens}, which says that from $X$ 
and `$X$ implies $Y$' we can deduce $Y$.  Most of modern proof theory 
focuses on another approach, the `sequent calculus', due to 
Gentzen \cite{Gentzen}.  In this approach there are few axioms 
but many inference rules.  

An excellent introduction to the sequent calculus is the book 
{\it Proofs and Types} by Girard, Lafont and Taylor, freely available
online \cite{GLT}.   Here we shall content ourselves with some sketchy
remarks.  A `sequent' is something like this:
\[          X_1, \dots, X_m \lHom Y_1, \dots , Y_n  \]
where $X_i$ and $Y_i$ are propositions.  We read this sequent as saying
that {\em all} the propositions $X_i$, taken together, can be used
to prove at least {\em one} of the propositions $Y_i$.  This 
strange-sounding convention gives the sequent calculus a nice symmetry, 
as we shall soon see.

In the sequent calculus, an `inference rule' is something that 
produces new sequents from old.  For example, here is the
{\bf left weakening} rule:
\begin{center}
\AXC{$X_1, \dots, X_m \lHom Y_1, \dots, Y_n$} 
\UIC{$X_1, \dots , X_m, A \lHom Y_1, \dots, Y_n$} \DP 
\end{center}
This says that from the sequent above the line we can get the sequent below 
the line: we can throw in the extra assumption $A$ without harm.  Thanks 
to the strange-sounding convention we mentioned, this rule has a 
mirror-image version called {\bf right weakening}:
\begin{center}
\AXC{$X_1, \dots, X_m \lHom Y_1, \dots, Y_n$} 
\UIC{$X_1, \dots , X_m  \lHom Y_1, \dots, Y_n, A$} \DP 
\end{center}
In fact, Gentzen's whole setup has this mirror symmetry!
For example, his rule called {\bf left contraction}:
\begin{center}
\AXC{$X_1, \dots, X_m, A, A \lHom Y_1, \dots, Y_n$} 
\UIC{$X_1, \dots , X_m, A \lHom Y_1, \dots, Y_n$} \DP 
\end{center}
has a mirror partner called {\bf right contraction}:
\begin{center}
\AXC{$X_1, \dots, X_m \lHom Y_1, \dots, Y_n, A, A$} 
\UIC{$X_1, \dots , X_m \lHom Y_1, \dots, Y_n, A$} \DP 
\end{center}
Similarly, this rule for `and'
\begin{center}
\AXC{$X_1, \dots, X_m, A \lHom Y_1, \dots, Y_n$} 
\UIC{$X_1, \dots , X_m, A \wedge B \lHom Y_1, \dots, Y_n$} \DP 
\end{center}
has a mirror partner for `or':
\begin{center}
\AXC{$X_1, \dots, X_m \lHom Y_1, \dots, Y_n, A$} 
\UIC{$X_1, \dots , X_m  \lHom Y_1, \dots, Y_n, A \vee B$} \DP 
\end{center}
Logicians now realize that this mirror symmetry can be understood
in terms of the duality between a category and its opposite.  

Gentzen used sequents to write inference rules for the classical 
propositional calculus, and also the classical predicate calculus.  
Now, in these forms of logic we have
\[          X_1, \dots, X_m \lHom Y_1, \dots, Y_n  \]
if and only if we have
\[ X_1 \wedge \cdots \wedge X_m \lHom Y_1 \vee \cdots \vee Y_n  .\]
So, why did Gentzen use sequents with a {\it list} of propositions
on each side of the $\lHom$ symbol, instead just a single proposition?
The reason is that this let him use only inference rules having the 
`subformula property'.  This says that every proposition 
in the sequent above the line appears as part of
some proposition in the sequent below the line.  So, a proof built
from such inference rules becomes a `tree' where all the propositions
further up the tree are subformulas of those below.

This idea has powerful consequences.  For example, in 1936 Gentzen
was able prove the consistency of Peano's axioms of arithmetic!
His proof essentially used induction on trees  (Readers familiar 
with G\"odel's second incompleteness theorem should be reassured 
that this sort of induction cannot itself be carried out in Peano arithmetic.)

The most famous rule {\em lacking} the subformula property is the `cut rule':
\begin{center}
\AXC{$X_1, \dots, X_m \lHom Y_1, \dots, Y_k, A$} 
\AXC{$X_{m+1}, \dots, X_n, A \lHom Y_{k+1}, \dots, Y_{\ell}$} 
\BIC{$X_1, \dots , X_n \lHom Y_1, \dots, Y_\ell$} \DP 
\end{center}
From the two sequents on top, the cut rule gives us the sequent
below.  Note that the intermediate step $A$ does not appear
in the sequent below.  It is `cut out'.  So, the cut rule lacks the 
subformula property.  But, one of Gentzen's great achievements was to show 
that any proof in the classical propositional (or even predicate) calculus
that can be done {\it with} the cut rule can also be done {\it without}
it.  This is called `cut elimination'.  

Gentzen also wrote down inference rules suitable for the intuitionistic
propositional and predicate calculi.  These rules lack the mirror symmetry 
of the classical case.  But in the 1980s, this symmetry was restored by
Girard's invention of `linear logic' \cite{Girard1}.  

Linear logic lets us keep track of how many times we use a given premise 
to reach a given conclusion.   To accomplish this, Girard introduced some 
new logical connectives!   For starters, he introduced `linear' connectives 
called $\tensor$ and $\lhom$, and a logical constant called $I$.  These 
act a bit like $\wedge$, $\Rightarrow$ and $\top$.  However, they satisfy 
rules corresponding to a symmetric monoidal category instead of a cartesian 
closed category.  In particular, from $X$ we can prove neither $X \tensor X$ 
nor $I$.  So, we cannot freely `duplicate' and `delete' propositions using 
these new connectives.   This is reflected in the fact that linear logic 
drops Gentzen's contraction and weakening rules. 

By itself, this might seem unbearably restrictive.  However, Girard 
{\it also} kept the connectives $\wedge$, $\Rightarrow$ and $\top$ in 
his system, still satisfying the usual rules.  And, he introduced an 
operation called the `exponential', $!$, which takes a proposition $X$ 
and turns it into an `arbitrary stock of copies of $X$'.  So, for 
example, from $!X$ we can prove $1$, and $X$, and $X \tensor X$, and 
$X \tensor X \tensor X$, and so on.  

Full-fledged linear logic has even more connectives than we have
described here.  It seems baroque and peculiar at first 
glance.   It also comes in both classical and intuitionistic 
versions!  But, {\it just as classical logic can
be embedded in intuitionistic logic, intuitionistic logic can be
embedded in intuitionistic linear logic} \cite{Girard1}.  So, we 
do not lose any 
deductive power.  Instead, we gain the ability to make even more
fine-grained distinctions.

In what follows, we discuss the fragment of intuitionistic linear
logic involving only $\tensor, \lhom$ and $I$.  This is called
`multiplicative intuititionistic linear logic' \cite{Hasegawa,Schalk}.
It turns out to be the system of logic suitable for closed symmetric
monoidal categories --- nothing more or less.

\subsection{Proofs as Morphisms}
\label{proof_theory}

In Section \ref{physics_topology} we described categories with various
amounts of extra structure, starting from categories pure and simple,
and working our way up to monoidal categories, braided monoidal
categories, symmetric monoidal categories, and so on.  Our treatment
only scratched the surface of an enormously rich taxonomy.  In fact,
each kind of category with extra structure corresponds to a system of
logic with its own inference rules!

To see this, we will think of {\em propositions} as {\em objects} in
some category, and {\em proofs} as giving {\em morphisms}.  Suppose
$X$ and $Y$ are propositions.  Then, we can think of a proof starting
from the assumption $X$ and leading to the conclusion $Y$ as giving a
morphism $f \maps X \to Y$.  (In Section \ref{theories} we shall see
that a morphism is actually an equivalence class of proofs --- but for
now let us gloss over this issue.)

Let us write $X \lHom Y$ when, starting from the assumption 
$X$, there is a proof leading to the conclusion $Y$.   An inference 
rule is a way to get new proofs from old.  For example, in almost every 
system of logic, if there is a proof leading from $X$ to $Y$, and a 
proof leading from $Y$ to $Z$, then there is a proof leading from $X$ 
to $Z$.  We write this inference rule as follows:

\begin{center}
\AXC{$X \lHom Y$} \AXC{$Y \lHom Z$} \BIC{$X \lHom Z$} \DP 
\end{center}

\noindent
We can call this {\bf cut rule}, since it lets us `cut out'
the intermediate step $Y$.   It is a special case of Gentzen's 
cut rule, mentioned in the previous section.  It should remind us of 
composition of morphisms in a category: if we have a morphism 
$f \maps X \to Y$ and a morphism $g \maps Y \to Z$, we get a morphism 
$gf \maps X \to Z$.

Also, in almost every system of logic there is a 
proof leading from $X$ to $X$.  We can write this as an inference
rule that starts with {\em nothing} and concludes the existence of a
proof of $X$ from $X$:

\begin{center}
\AXC{} \UIC{$X\lHom X$} \DP 
\end{center}

\noindent
This rule should remind us of how every object in category has
an identity morphism: for any object $X$, we automatically get a 
morphism $1_X \maps X \to X$.  Indeed, this rule is sometimes called
the {\bf identity rule}.

If we pursue this line of thought, we can take the definition of 
a closed symmetric monoidal category and extract a collection of inference 
rules.  Each rule is a way to get new morphisms from old in a 
closed symmetric monoidal category.  There are various superficially different 
but ultimately equivalent ways to list these rules.  Here is one:

\cent{\begin{tabular}{ccc}
     \AXC{} \UIC{$X\lHom X$} \DP {\scriptsize ($\id$)} &\hbox{\qquad}&
     \AXC{$X \lHom Y$} \AXC{$Y \lHom Z$} \BIC{$X \lHom Z$} \DP 
{\scriptsize ($\cut$)} \\ \\
     \AXC{$W \lHom X$} \AXC{$Y \lHom Z$} \BIC{$W \tensor Y \lHom X \tensor Z$} \DP {\scriptsize ($\tensor$)}  &&
     \AXC{$W \lHom (X \tensor Y) \tensor Z$} \doubleLine \UIC{$W \lHom X \tensor (Y \tensor Z)$} \DP 
{\scriptsize ({\rm a})} \\ \\
     \AXC{$X \lHom I \tensor Y$} \doubleLine \UIC{$X \lHom Y$} \DP 
{\scriptsize ({\rm l})} &&
     \AXC{$X \lHom Y \tensor I$} \doubleLine \UIC{$X \lHom Y$} \DP 
{\scriptsize ({\rm r})} \\  \\  
     \AXC{$W \lHom X \tensor Y$} \doubleLine \UIC{$W \lHom Y \tensor X$} \DP 
{\scriptsize ({\rm b})} &&
     \AXC{$X \tensor Y \lHom Z$} \doubleLine \UIC{$Y \lHom X \lhom Z$} \DP 
{\scriptsize ({\rm c})} 
\end{tabular}}

\noindent
Double lines mean that the inverse rule also holds.  We have
given each rule a name, written to the right in parentheses.
As already explained, rules ($\id$) and 
($\cut$) come from the presence of identity morphisms and 
composition in any category.  Rules ($\tensor$), 
(a), (l), and (r) come from tensoring, the associator, and the left and 
right unitors in a monoidal category.  Rule (b) comes from
the braiding in a braided monoidal category, and rule (c) 
comes from currying in a closed monoidal category.

Now for the big question: {\it what does all this mean in terms of logic?}
These rules describe a small fragment of the propositional calculus.
To see this, we should read the connective $\tensor$ as `and', the 
connective $\lhom$ as `implies', and the proposition $I$ as `true'.  

In this interpretation, rule (c) says we can turn a proof leading from 
the assumption `$Y$ and $X$' to the conclusion $Z$ into a proof leading 
from $X$ to `$Y$ implies $Z$'.  It also says we can do the reverse.  
This is true in classical, intuitionistic and linear logic, and so are 
all the other rules.  Rules (a) and (b) say that `and' is associative 
and commutative.  
Rule (l) says that any proof leading from the assumption $X$ to the 
conclusion `true and $Y$' can be converted to a proof leading from 
$X$ to $Y$, and vice versa.  Rule (r) is similar.

What do we do with these rules?  We use them to build `deductions'.  
Here is an easy example:
\begin{center}
\AXC{}
\RightLabel{\scriptsize ($\id$)} 
\UIC{$X \lhom Y \lHom X \lhom Y$} 
\RightLabel{\scriptsize (${\rm c}^{-1})$} 
\UIC{$X \tensor (X \lhom Y) \lHom Y$} 
\DP
\end{center}
First we use the identity rule, and then the inverse
of the currying rule.  At the end, we obtain
$$   X \tensor (X \lhom Y) \lHom Y . $$
This should remind us of
the evaluation morphisms we have in a closed monoidal category:
$$\ev_{X,Y} \maps X \tensor (X \lhom Y) \to Y. $$
In terms of logic, the point is that we can prove 
$Y$ from $X$ and `$X$ implies $Y$'.  This fact comes in handy so 
often that we may wish to abbreviate the above deduction as an extra 
inference rule --- a rule derived from our basic list:
\begin{center}
\AXC{}
\RightLabel{\scriptsize (ev)} 
\UIC{$X \tensor (X \lhom Y) \lHom Y$} 
\DP
\end{center}
This rule is called {\bf modus ponens}.

In general, a deduction is a tree built from inference rules.
Branches arise when we use the ($\circ$) or ($\tensor$) rules.  
Here is an example:
\begin{center}
\AXC{}
\RightLabel{\scriptsize (i)}
\UIC{$(A \tensor B) \tensor C \lHom
(A \tensor B) \tensor C$}
\RightLabel{\scriptsize (a)} 
\UIC{$(A \tensor B) \tensor C \lHom
A \tensor (B \tensor C)$}
\AXC{$A \tensor (B \tensor C) \lHom D$} 
\RightLabel{\scriptsize ($\circ$)}
\BIC{$(A \tensor B) \tensor C \lHom D$}
\DP
\end{center}
Again we can abbreviate this deduction as a derived rule.
In fact, this rule is reversible:
\begin{center}
\AXC{$A \tensor (B \tensor C) \lHom D$} 
\RightLabel{\scriptsize ($\alpha$)} 
\doubleLine
\UIC{$(A \tensor B) \tensor C \lHom D$}
\DP
\end{center}

For a more substantial example, suppose we want to show
$$  (X \lhom Y) \tensor (Y \lhom Z) \lHom X \lhom Z  . $$
The deduction leading to this will not even fit on the page unless
we use our abbreviations: 
\begin{center}
\AXC{}
\RightLabel{\scriptsize (ev)}
\UIC{$X \tensor (X \lhom Y) \lHom Y$}
\AXC{}
\RightLabel{\scriptsize ($\id$)}
\UIC{$Y \lhom Z \lHom Y \lhom Z$}
\RightLabel{\scriptsize ($\tensor$)}
\BIC{$(X \tensor (X \lhom Y)) \tensor (Y \lhom Z) \lHom Y \tensor (Y \lhom Z)$}
\AXC{}
\RightLabel{\scriptsize (ev)}
\UIC{$Y \tensor (Y \lhom Z) \lHom Z$}
\BIC{$(X \tensor (X \lhom Y)) \tensor (Y \lhom Z) \lHom Z$}
\RightLabel{\scriptsize ($\alpha^{-1}$)}
\UIC{$X \tensor ((X \lhom Y) \tensor (Y \lhom Z)) \lHom Z$}
\RightLabel{\scriptsize (c)}
\UIC{$(X \lhom Y) \tensor (Y \lhom Z) \lHom X \lhom Z$}
\DP
\end{center}
Since each of the rules used in this deduction came from a way to get
new morphisms from old in a closed monoidal category (we never used
the braiding), it follows that in every such category we have 
{\bf internal composition} morphisms:
\[ \bullet_{X,Y,Z}\, \maps (X \lhom Y) \tensor (Y \lhom Z) \to X \lhom Z . \]
These play the same role for the internal hom that ordinary composition
\[  \circ \maps \hom(X,Y) \times \hom(Y,Z) \to \hom(X,Z)  \]
plays for the ordinary hom.

We can go ahead making further deductions in this system of  
logic, but the really interesting thing is what it omits.   
For starters, it omits the connective `or' and 
the proposition `false'.  It also omits two inference rules we normally
take for granted --- namely, {\bf contraction}: 

\begin{center}
\AXC{$X \lHom Y$} \RightLabel{\scriptsize $(\Delta)$} 
\UIC{$X \lHom Y \tensor Y$}\DP 
\end{center}

\noindent
and {\bf weakening}:

\begin{center}
\AXC{$X \lHom Y$} \RightLabel{\scriptsize $(!)$} \UIC{$X \lHom I$} \DP
\end{center}

\noindent
which are closely related to duplication and deletion in a cartesian
category.  Omitting these rules is a distinctive feature of 
linear logic \cite{Girard1}.  The word `linear' should
remind us of the category $\Hilb$.  As noted in Section
\ref{monoidal}, this category with its usual tensor product
is noncartesian, so it does not permit duplication and deletion.  
But, what does omitting these rules mean {\it in terms of logic?}

Ordinary logic deals with propositions, so we have been thinking
of the above system of logic in the same way.  Linear logic
deals not just with propositions, but also other resources --- for 
example, physical things!  Unlike propositions in ordinary logic, 
we typically can't duplicate or delete these other resources.  In 
classical logic, if we know that a proposition $X$ is true, we can 
use $X$ as many or as few times as we like when trying to prove some 
proposition $Y$.   But if we have a cup of milk, we can't use it to make 
cake and then use it again to make butter.  Nor can we make it disappear 
without a trace: even if we pour it down the drain, it must go somewhere.

In fact, these ideas are familiar in chemistry.  Consider
the following resources:
\[
\begin{array}{ccl}
          \H_2  &=& \text{one molecule of hydrogen} \\
          \O_2  &=& \text{one molecule of oxygen} \\
          \H_2\O &=& \text{one molecule of water} \\
\end{array}
\]
We can burn hydrogen, combining one molecule of oxygen with two
of hydrogen to obtain two molecules of water.
A category theorist might describe this reaction as a morphism:
\[        f \maps \O_2 \tensor (\H_2 \tensor \H_2) \to 
                  \H_2\O \tensor \H_2\O .  \]
A linear logician might write:
\[        \O_2 \tensor (\H_2 \tensor \H_2) \lHom \H_2\O \tensor \H_2\O    \]
to indicate the existence of such a morphism.  But, we cannot
duplicate or delete molecules, so for example
\[         \H_2 \not{\lHom} \H_2 \tensor \H_2 \]
and
\[         \H_2 \not{\lHom} I \]
where $I$ is the unit for the tensor product: not iodine, but
`no molecules at all'.

In short, ordinary chemical reactions are morphisms
in a symmetric monoidal category where objects are collections 
of molecules.  As chemists normally conceive of it, this category 
is not closed.  So, it obeys an even more limited system of logic than 
the one we have been discussing, a system lacking the connective $\lhom$.  
To get a closed category --- in fact 
a compact one --- we need to remember one of the great discoveries 
of 20th-century physics: {\em antimatter}.  This lets us define 
$Y \lhom Z$ to be `anti-$Y$ and $Z$':
\[         Y \lhom Z = Y^* \tensor Z . \]
Then the currying rule holds:

\begin{center}
 \AXC{$Y \tensor X \lHom Z$} \doubleLine \UIC{$X \lHom Y^\ast \tensor Z$} \DP 
\end{center}
 
\noindent
Most chemists don't think about antimatter very often
--- but particle physicists do.  They don't use the notation 
of linear logic or category theory, but they know perfectly well that 
since a neutrino and a neutron can collide and turn into
a proton and an electron:
\[   \nu \tensor n \lHom p \tensor e , \]
then a neutron can turn into a antineutrino together
with a proton and an electron:
\[   n \lHom \nu^* \tensor (p \tensor e) . \]
This is an instance of the currying rule, rule (c).

\subsection{Logical Theories from Categories}
\label{theories}

We have sketched how different systems of logic naturally arise
from different types of categories.  To illustrate this idea,
we introduced a system of logic with inference rules coming from 
ways to get new morphisms from old in a {\it closed 
symmetric monoidal category}.  One could substitute many other types of 
categories here, and get other systems of logic.  

To tighten the connection between proof theory and category theory,
we shall now describe a recipe to get a logical theory from any
closed symmetric monoidal category.  For this, we shall now use 
$X\lHom Y$ to denote the {\em set} of proofs --- or actually, 
equivalence classes of proofs --- leading from the assumption $X$ 
to the conclusion $Y$.  This is a change of viewpoint.  Previously
we would write $X \lHom Y$ when this set of proofs was nonempty;
otherwise we would write $X \not{\lHom} Y$.  The advantage of 
treating $X \lHom Y$ as a set is that this set is precisely 
what a category theorist would call $\hom(X,Y)$: a homset in a category.

If we let $X \lHom Y$ stand for a homset, an inference rule 
becomes a function from a product of homsets to a single homset.  
For example, the cut rule

\begin{center}
\AXC{$X \lHom Y$} \AXC{$Y \lHom Z$} 
\RightLabel{\scriptsize ($\circ$)} 
\BIC{$X \lHom Z$} \DP 
\end{center}

\noindent
becomes another way of talking about the composition function
$$  \circ_{X,Y,Z} \maps \hom(X,Y) \times \hom(Y,Z) \to \hom(X,Z), $$
while the identity rule

\begin{center}
\AXC{} 
\RightLabel{\scriptsize ($\id$)}
\UIC{$X\lHom X$} \DP 
\end{center}

\noindent
becomes another way of talking about the function
$$  \id_X \maps 1 \to \hom(X,X) $$
that sends the single element of the set 1 to the identity morphism 
of $X$.  (Note: the set 1 is a {\em zero-fold} product of homsets.)

Next, if we let inference rules be certain functions 
from products of homsets to homsets, deductions become
more complicated functions of the same sort built from these
basic ones.  For example, this deduction:
\begin{center}
\AXC{} 
\RightLabel{\scriptsize ($\id$)} 
\UIC{$X\tensor I \lHom X \tensor I$} 
\RightLabel{\scriptsize (r)}   
\UIC{$X\tensor I \lHom X$}
\AXC{}
\RightLabel{\scriptsize ($\id$)} 
\UIC{$Y \lHom Y$}
\RightLabel{\scriptsize ($\tensor$)} 
\BIC{$(X \tensor I) \tensor Y \lHom X \tensor Y$}
\DP
\end{center}
specifies a function from $1$ to
$\hom((X \tensor I) \tensor Y, X \tensor Y)$, built from 
the basic functions indicated by the labels at each step.
This deduction: 
\begin{center}
                                   \AXC{} 
    \RightLabel{\scriptsize ($\id$)} \UIC{$(X \tensor I) \tensor Y \lHom (X \tensor I) \tensor Y$} 
    \RightLabel{\scriptsize (a)}   \UIC{$(X \tensor I) \tensor Y \lHom X \tensor (I \tensor Y)$}
                                   \AXC{} 
    \RightLabel{\scriptsize ($\id$)} \UIC{$I \tensor Y \lHom I \tensor Y$} 
    \RightLabel{\scriptsize (r)}   \UIC{$I \tensor Y \lHom Y$}
                                   \AXC{} 
    \RightLabel{\scriptsize ($\id$)} \UIC{$X \lHom X$} 
    \RightLabel{\scriptsize ($\tensor$)} \BIC{$X \tensor (I \tensor Y) \lHom X \tensor Y$}
    \RightLabel{\scriptsize ($\circ$)} \BIC{$(X \tensor I) \tensor Y \lHom X \tensor Y$}
    \DP
\end{center}
gives another function from 1 to 
$\hom((X \tensor I) \tensor Y, X \tensor Y)$.

If we think of deductions as giving functions this way, the question
arises when two such functions are equal.  In the example just mentioned,
the triangle equation in the definition of monoidal category
(Definition \ref{mon.cat}): 
\di{
\node{(X\tensor I)\tensor Y}\arrow[2]{e,t}{a_{X,I,Y}}\arrow{se,b}{r_X\tensor 1_Y}
\node[2]{X\tensor(I\tensor Y)}\arrow{sw,b}{1_X\tensor l_Y}\\
\node[2]{X\tensor Y}
}
says these two functions {\em are} equal.  Indeed, the triangle equation 
is precisely the statement that these two functions agree!  (We leave 
this as an exercise for the reader.)

So: even though two deductions may look quite different, they may give
the same function from a product of homsets to a homset if we demand
that these are homsets in a closed symmetric monoidal category.  This
is why we think of $X \lhom Y$ as a set of {\em equivalence classes}
of proofs, rather than proofs: it is forced on us by our desire to use
category theory.  We could get around this by using a 2-category with
proofs as morphisms and `equivalences between proofs' as 2-morphisms
\cite{Seely}.  This would lead us further to the right in the Periodic
Table (Table \ref{periodic_table}).  But let us restrain ourselves and
make some definitions formalizing what we have done so far.

From now on we shall call the objects $X,Y, \dots$ 
`propositions', even though we have seen they may represent more
general resources.  Also, purely for the sake of brevity, we use
the term `proof' to mean `equivalence class of proofs'.
The equivalence relation must be coarse enough to make the
equations in the following definitions hold:

\begin{definition}
A {\bf closed monoidal theory} consists of the following:
\begin{itemize}
    \item A collection of {\bf propositions}.  The collection must
contain a proposition $I$, and if $X$ and $Y$ are propositions, then
so are $X \tensor Y$ and $X \lhom Y$. 
    \item For every pair of propositions $X,Y,$ a set $X \lHom Y$ of 
{\bf proofs} leading from $X$ to $Y$.  If $f \in X \lHom Y,$ then we 
write $f \maps X\to Y$.
     \item Certain functions, written as {\bf inference rules}:
{\rm 
\cent{\begin{tabular}{ccc}
     \AXC{} \UIC{$X\lHom X$} \DP {\scriptsize ($\id$)} &\hbox{\qquad}&
     \AXC{$X \lHom Y$} \AXC{$Y \lHom Z$} \BIC{$X \lHom Z$} \DP 
{\scriptsize ($\cut$)} \\ \\
     \AXC{$W \lHom X$} \AXC{$Y \lHom Z$} \BIC{$W \tensor Y \lHom X \tensor Z$} 
\DP {\scriptsize ($\tensor$)}  &&
     \AXC{$W \lHom (X \tensor Y) \tensor Z$} \doubleLine \UIC{$W \lHom X \tensor (Y \tensor Z)$} \DP 
{\scriptsize ({\rm a})} \\ \\
     \AXC{$X \lHom I \tensor Y$} \doubleLine \UIC{$X \lHom Y$} \DP 
{\scriptsize ({\rm l})} &&
     \AXC{$X \lHom Y \tensor I$} \doubleLine \UIC{$X \lHom Y$} \DP 
{\scriptsize ({\rm r})} \\  \\  
     \AXC{$X \tensor Y \lHom Z$} \doubleLine \UIC{$Y \lHom X \lhom Z$} \DP 
{\scriptsize ({\rm c})} &&
\end{tabular}}
}
A double line means that the function is invertible.  So, for
example, for each triple $X,Y,Z$ we have a function 
\[     \circ_{X,Y,Z} \maps (X \lHom Y) \times (Y \lHom Z) \; \to \; 
(X \lHom Z) \]
and a bijection
\[     c_{X,Y,Z} \maps (X \tensor Y \lHom Z) \; \to \; (Y \lHom X \lhom Z) .\]
    \item Certain equations that must be obeyed by the inference rules.
The inference rules $(\circ)$ and $(\id)$ must obey
equations describing associativity and the left and right unit laws.
Rule $(\tensor)$ must obey an equation saying it is a functor.
Rules {\rm (a)}, {\rm (l)}, {\rm (r)}, and {\rm (c)} must
obey equations saying they are natural transformations.
Rules {\rm (a)}, {\rm (l)}, {\rm (r)} and $(\tensor)$ must also
obey the triangle and pentagon equations.  
\end{itemize}
\end{definition}

\begin{definition}
A {\bf closed braided monoidal theory} is a closed monoidal theory
with this additional inference rule:
{\rm 
\begin{center}
     \AXC{$W \lHom X \tensor Y$} \doubleLine \UIC{$W \lHom Y \tensor X$} \DP 
{\scriptsize (\rm b)} 
\end{center}
}
\noindent We demand that this rule give a natural
transformation satisfying the hexagon equations.  
\end{definition}

\begin{definition}
A {\bf closed symmetric monoidal theory} is a closed braided
monoidal theory where the rule {\rm (b)} is its own inverse.
\end{definition}

These are just the usual definitions of various kinds of closed
category --- monoidal, braided monoidal and symmetric monoidal ---
written in a new style.  This new style lets us \emph{build such
categories from logical systems}.  To do this, we take the objects to
be propositions and the morphisms to be equivalence classes of proofs,
where the equivalence relation is generated by the equations listed in
the definitions above.

However, the full advantages of this style only appear when we dig deeper
into proof theory, and generalize the expressions we have been considering: 
\[   X \lHom Y \]
to `sequents' like this: 
\[   X_1, \dots, X_n \lHom Y . \]
Loosely, we can think of such a sequent as meaning
\[ X_1 \tensor \cdots \tensor X_n \lHom Y .\]
The advantage of sequents is that they let us use inference rules that
--- except for the cut rule and the identity rule --- have the
`subformula property' mentioned near the end of Section
\ref{logic_overview}.

Formulated in terms of these inference rules, the logic of closed
symmetric monoidal categories goes by the name of `multiplicative
intuitionistic linear logic', or MILL for short \cite{Hasegawa,
Schalk}.  There is a `cut elimination' theorem for MILL, which says
that with a suitable choice of other inference rules, the cut rule
becomes redundant: any proof that can be done with it can be done
without it.  This is remarkable, since the cut rule corresponds to
{\it composition of morphisms} in a category.  One consequence is that
in the free symmetric monoidal closed category on any set of objects,
the set of morphisms between any two objects is {\it finite}.  There
is also a decision procedure to tell when two morphisms are equal.
For details, see Trimble's thesis \cite{Trimble} and the papers by Jay
\cite{Jay1990} and Soloviev \cite{Soloviev}.  Also see Kelly and Mac
Lane's coherence theorem for closed symmetric monoidal categories
\cite{KM}, and the related theorem for compact symmetric monoidal
categories \cite{KellyLaplaza}.

MILL is just one of many closely related systems of logic.  Most 
include extra features, but some {\it subtract} features.  
Here are just a few examples:

\begin{itemize}
\item Algebraic theories.   In his famous thesis, Lawvere \cite{Lawvere}
defined an {\bf algebraic theory} to be a cartesian category where 
every object is an $n$-fold cartesian power $X \times \cdots \times X$ 
($n \ge 0$) of a specific object $X$.  He showed how such categories 
regarded as logical theories of a simple sort --- the sort that had 
previously been studied in `universal algebra' \cite{BurrisS}.  This 
work initiated the categorical approach to logic which we have been
sketching here.  Crole's book \cite{Crole} gives a gentle introduction 
to algebraic theories as well as some richer logical systems.
More generally, we can think of any cartesian category as a
generalized algebraic theory. 
\item Intuitionistic linear logic (ILL).  ILL
supplements MILL with the operations familiar from intuitionistic
logic, as well as an operation $!$ turning any proposition (or 
resource) $X$ into an `indefinite stock of copies of $X$'.  Again 
there is a nice category-theoretic interpretation.  Bierman's 
thesis \cite{BiermanThesis} gives a good overview, including a proof 
of cut elimination for ILL and a proof of the result, originally due 
to Girard, that intuitionistic logic can be be embedded in ILL.
\item Linear logic (LL).  For full-fledged linear logic, 
the online review article by Di Cosmo and Miller \cite{dCM} is
a good place to start.  For more, try the original
paper by Girard \cite{Girard1} and the book by Troelstra \cite{Troelstra}.
Blute and Scott's review article \cite{BS} serves as a Rosetta Stone 
for linear logic and category theory, and so do the lectures notes
by Schalk \cite{Schalk}.  
\item Intuitionistic Logic (IL).   Lambek and 
Scott's classic book \cite{LS} is still an excellent introduction to 
intuitionistic logic and cartesian closed categories.  The online review 
article by Moschovakis \cite{Moschovakis} contains many suggestions for 
further reading.
\end{itemize}

To conclude, let us say precisely what an `inference rule' amounts to
in the setup we have described.  We have said it gives a function from
a product of homsets to a homset.  While true, this is not the last
word on the subject.  After all, instead of treating the propositions
appearing in an inference rule as {\it fixed}, we can treat them as
{\it variable}.  Then an inference rule is really a `schema' for
getting new proofs from old.  How do we formalize this idea?

First we must realize that $X \lHom Y$ is not just a set: it is a set 
{\em depending in a functorial way} on $X$ and $Y$.   As noted in
Definition \ref{hom.functor}, there is a functor, the `hom functor'
$$    \hom \maps C^\op \times C \to \Set, $$ 
sending $(X,Y)$ to the homset $\hom(X,Y) = X \lHom Y$.  To look like
logicians, let us write this functor as $\lHom$.

Viewed in this light, most of our inference rules are {\it natural 
transformations}.  For example, rule (a) is a natural transformation 
between two functors from $C^\op \times C^3$ to $\Set$, namely
the functors 
\[        (W,X,Y,Z) \mapsto W \lHom (X \tensor Y) \tensor Z \]
and 
\[        (W,X,Y,Z) \mapsto W \lHom X \tensor (Y \tensor Z) .\]
This natural transformation turns any proof
\[        f \maps W \to (X \tensor Y) \tensor Z \]
into the proof
\[  a_{X,Y,Z} f \maps W \to X \tensor (Y \tensor Z)  .\]
The fact that this transformation is {\it natural} means that it
changes in a systematic way as we vary $W,X,Y$ and $Z$.  The
commuting square in the definition of natural transformation,
Definition \ref{naturality}, makes this precise.

Rules (l), (r), (b) and (c) give natural transformations in a 
very similar way.   The $(\tensor)$ rule gives a natural transformation 
between two functors from $C^\op \times C \times C^\op \times C$ 
to $\Set$, namely
\[       (W,X,Y,Z) \mapsto (W \lHom X) \; \times \; (Y \lHom Z) \]
and 
\[       (W,X,Y,Z) \mapsto W \tensor Y \lHom X \tensor Z .\]
This natural transformation sends any element $(f,g) \in 
\hom(W,X) \times \hom(Y,Z)$ to $f \tensor g$. 

The identity and cut rules are different: they {\it do not} give 
natural transformations, because the top line of these rules has a 
different number of variables than the bottom line!
Rule (i) says that for each $X \in C$ there is a function 
\[ \id_X \maps 1 \; \to \; X \lHom X \] 
picking out the identity morphism $1_X.$  
What would it mean for this to be natural in $X$?
Rule $(\circ)$ says that for each triple $X,Y,Z \in C$ there is
a function
\[     \circ \maps (X \lHom Y) \; \times \; (Y \lHom Z) \; \to \; X \lHom Z .\] 
What would it mean for this to be natural in $X,Y$ and $Z$? 
The answer to both questions involves a generalization of natural
transformations called `dinatural' transformations \cite{MacLane}.

As noted in Definition \ref{naturality},
a natural transformation $\alpha \maps F\Rightarrow G$ between two functors 
$F, G \maps C \to D$ makes certain squares in $D$ commute.  
If in fact $C=C_1^{\op} \times C_2,$ then we actually obtain commuting 
cubes in $D.$  Namely, the natural transformation $\alpha$ assigns to each 
object $(X_1, X_2)$ a morphism $\alpha_{X_1,X_2}$ such that for any 
morphism $(f_1 \maps Y_1\to X_1, f_2 \maps X_2\to Y_2)$ in $C$, 
this cube commutes:

\begin{center}
\dgARROWLENGTH=3.75em
\di{
\node[2]{G(Y_1,X_2)}\arrow[2]{e,t,1}{G(1_{Y_1},f_2)}
\node[2]{G(Y_1,Y_2)}\arrow[2]{s,r,1}{G(f_1,1_{Y_2})}
\\
\node{F(Y_1,X_2)}\arrow[2]{s,r,1}{F(f_1,1_{X_2})}\arrow[2]{e,t,1}{F(1_{Y_1},f_2)}\arrow{ne,l}{\alpha_{Y_1,X_2}}
\node{}\arrow{n,r,-}{G(f_1,1_{X_2})}\arrow{s}
\node{F(Y_1,Y_2)}\arrow[2]{s,r,1}{F(f_1,1_{Y_2})}\arrow{ne,l}{\alpha_{Y_1,Y_2}}
\\
\node[2]{G(X_1,X_2)}\arrow{e,t,-}{G(1_{X_1},f_2)}
\node{}\arrow{e}
\node{G(X_1,Y_2)}
\\
\node{F(X_1,X_2)}\arrow[2]{e,t}{F(1_{X_1},f_2)}\arrow{ne,r,3}{\alpha_{X_1,X_2}}
\node[2]{F(X_1,Y_2)}\arrow{ne,r,3}{\alpha_{X_1,Y_2}}
}
\end{center}

\begin{figure}
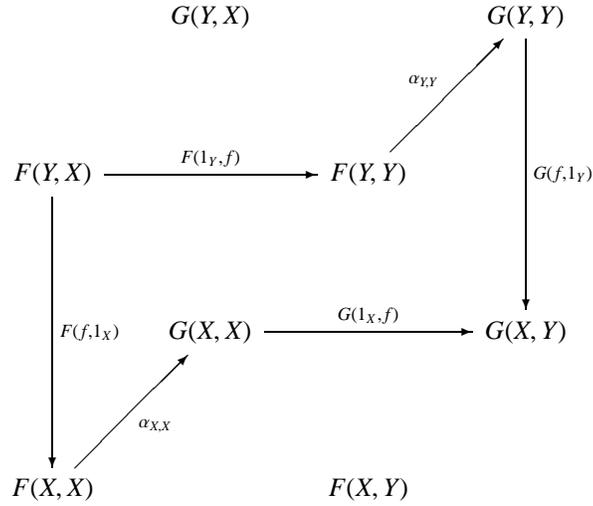

\begin{center}
\dgARROWLENGTH=3.75em
\di{
\node[2]{G(Y,X)}
\node[2]{G(Y,Y)}\arrow[2]{s,r}{G(f,1_{Y})}
\\
\node{F(Y,X)}\arrow[2]{s,r}{F(f,1_{X})}\arrow[2]{e,t}{F(1_{Y},f)}
\node{}\arrow{n,!}{}\arrow{s,!}
\node{F(Y,Y)}\arrow{ne,l}{\alpha_{Y,Y}}
\\
\node[2]{G(X,X)}\arrow[2]{e,t}{G(1_{X},f)}
\node{}
\node{G(X,Y)}
\\
\node{F(X,X)}\arrow{ne,r}{\alpha_{X,X}}
\node[2]{F(X,Y)}
}
\caption{A natural transformation between functors 
${F,G\maps C^{\op}\times C \to D}$ gives a commuting cube in 
$D$ for any morphism $f \maps X \to Y$, and there are two paths 
around the cube that only involve $\alpha$ for repeated arguments.}
\label{dinaturalcube}
\end{center}
\end{figure}

If $C_1 = C_2,$ we can choose a single object $X$ and a single 
morphism $f \maps X\to Y$ and use it in both slots.  As shown
in Figure \ref{dinaturalcube}, there are then two paths from one 
corner of the cube to the antipodal corner that only involve $\alpha$ 
for repeated arguments: that is, $\alpha_{X,X}$ and $\alpha_{Y,Y}$, 
but not $\alpha_{X,Y}$ or $\alpha_{Y,X}$.   These paths give a commuting 
hexagon.  

This motivates the following:

\begin{definition}
    A {\bf dinatural transformation} $\alpha \maps F\Rightarrow G$ between
    functors $F,G \maps C^{\op}\times C \to D$ assigns to every object $X$
    in $C$ a morphism $\alpha_X \maps F(X,X) \to G(X,X)$ in $D$ such that
    for every morphism $f \maps X\to Y$ in $C$, the hexagon in Figure
    \ref{dinaturalcube} commutes.
\end{definition}

\begin{figure}
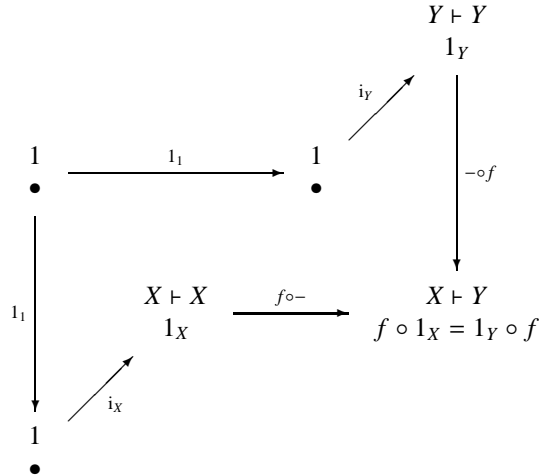

\dgARROWLENGTH=3.75em
\di{
\node[4]{\begin{array}{c}Y\lHom Y\\1_Y\end{array}}
\arrow[2]{s,r}{- \circ f}
\\
\node{\begin{array}{c} 1 \\ \bullet \end{array}}\arrow[2]{s,l}{1_{1}}\arrow[2]{e,t}{1_{1}}
\node{}\arrow{n,!}
\node{\begin{array}{c} 1 \\ \bullet \end{array}}\arrow{ne,l}{\id_Y}
\\
\node[2]{\begin{array}{c}X\lHom X\\1_X\end{array}}\arrow[2]{e,t}{f \circ -}
\node{}
\node{\begin{array}{c}X \lHom Y\\f \circ 1_X = 1_Y \circ f\end{array}}
\\
\node{\begin{array}{c} 1 \\ \bullet \end{array}}\arrow{ne,r}{\id_X}
}
\caption{Dinaturality of the (i) rule, where $f\maps X\to Y$.  Here
$\bullet \in 1$ denotes the one element of the one-element set.}
\label{hypcube}
\end{figure}

In the case of the identity rule, this commuting hexagon follows
from the fact that the identity morphism is a left and right unit 
for composition: see Figure \ref{hypcube}.  For the cut rule,
this commuting hexagon says that composition is associative:
see Figure \ref{cutcube}.

{
So, in general, the sort of logical theory we are discussing involves:
\begin{itemize}
\item
A \emph{category} $C$ of propositions and proofs.
\item 
A \emph{functor} $\lHom \maps C^\op \times C \to \Set$
sending any pair of propositions to the set of 
proofs leading from one to the other.
\item
A set of \emph{dinatural transformations} describing inference
rules.
\end{itemize}
}

\begin{figure}
\dgARROWLENGTH=3.75em
\di{
\node[4]{\begin{array}{c}X\lHom Z\\h \circ (f \circ g)\end{array}}\arrow[2]{s,r}{1_{X\lHom Z}}
\\
\node{\begin{array}{c}(X\lHom W) \; \times \; (Y\lHom Z)\\(g, h)\end{array}}\arrow[2]{s,l}{(1_{X\lHom W}, - \circ f)}\arrow[2]{e,t}{(f \circ -, 1_{Y\lHom Z})}
\node{}\arrow{n,!}
\node{\begin{array}{c}(X\lHom Y) \; \times \; (Y\lHom Z)\\(f \circ g, h)\end{array}}\arrow{ne,l}{\circ}
\\
\node[2]{\begin{array}{c}X\lHom Z\\(h \circ f) \circ g\end{array}}\arrow[2]{e,t}{1_{X\lHom Z}}
\node{}
\node{\begin{array}{c}X\lHom Z\\(h \circ f) \circ g = h \circ (f \circ g)\end{array}}
\\
\node{\begin{array}{c}(X\lHom W) \; \times \; (W\lHom Z)\\(g, h \circ f)\end{array}}\arrow{ne,r}{\circ}
}
\caption{Dinaturality of the cut rule, where 
${f\maps W\to Y,}$ ${g\maps X\to W,}$ ${h\maps Y\to Z.}$
}
\label{cutcube}
\end{figure}

\section{Computation}
\label{computation}

\subsection{Background}
\label{computation_overview}

In the 1930s, while Turing was developing what are now called `Turing
machines' as a model for computation, Church and his student Kleene
were developing a different model, called the `lambda calculus'
\cite{Church, Kleene}.  While a Turing machine can be seen as an
idealized, simplified model of computer {\it hardware}, the lambda
calculus is more like a simple model of {\it software}.

By now the are many careful treatments of the lambda calculus in 
the literature, from Barendregt's magisterial tome \cite{Barendregt}
to the classic category-theoretic treatment of Lambek and Scott
\cite{LS}, to Hindley and Seldin's user-friendly introduction 
\cite{HS} and Selinger's elegant free online notes \cite{Selinger2}.  So, 
we shall content ourselves with a quick sketch.

Poetically speaking, the lambda calculus describes a universe where
everything is a program and everything is data: {\it programs are
data}.  More prosaically, everything is a `$\lambda$-term', or `term'
for short.  These are defined inductively:
\begin{itemize}
\item {\bf Variables:}
there is a countable set of `variables' $x, y, z, \dots$
which are all terms. 
\item {\bf Application:}
if $f$ and $t$ are terms, we can `apply' $f$
to $t$ and obtain a term $f(t)$.
\item {\bf Lambda-abstraction:} if 
$x$ is a variable and $t$ is a term, there is a term 
$(\lambda x . t)$.
\end{itemize}

Let us explain the meaning of application and lambda-abstraction.
Application is simple.  Since `programs are data', we can think of any
term either as a program or a piece of data.  Since we can apply
programs to data and get new data, we can apply any term $f$ to any
other term $t$ and get a new term $f(t)$.

Lambda-abstraction is more interesting.  We think of $(\lambda x.t)$
as the program that, given $x$ as input, returns $t$ as output.  
For example, consider
\[       (\lambda x. x(x)) . \]
This program takes any program $x$ as input and returns $x(x)$ as 
output.  In other words, it applies any program to itself. 
So, we have
\[       (\lambda x . x(x))(s) = s(s) \]
for any term $s$.  

More generally, if we apply $(\lambda x . t)$ to any term $s$, 
we should get back $t$, but with $s$ substituted for each free
occurrence of the variable $x$.  This fact is codified in a rule 
called {\bf beta reduction}:
\[       (\lambda x.t)(s) = t[s/x] \]
where $t[s/x]$ is the term we get by taking $t$ and substituting $s$
for each free occurrence of $x$.  But beware: this rule is not an
equation in the usual mathematical sense.  Instead, it is a `rewrite
rule': given the term on the left, we are allowed to rewrite it and
get the term on the right.  Starting with a term and repeatedly 
applying rewrite rules is how we take a program and let it run!

There are two other rewrite rules in the lambda calculus.  If $x$ is a
variable and $t$ is a term, the term
\[    (\lambda x . t(x)) \]
stands for the program that, given $x$ as input, returns $t(x)$ as
output.  But this is just a fancy way of talking about the program
$t$.  So, the lambda calculus has a rewrite rule called {\bf eta
reduction}, saying
\[        (\lambda x . t(x)) = t. \]

The third rewrite rule is {\bf alpha conversion}.  This allows
us to replace a bound variable in a term by another variable.  
For example:
\[        (\lambda x . x(x)) = (\lambda y . y(y))  \]
since $x$ is `bound' in the left-hand expression by its appearance
in `$\lambda x$'.   In other words, $x$ is just a dummy variable;
its name is irrelevant, so we can replace it with $y$.  On the other 
hand,
\[        (\lambda x . y(x)) \ne (\lambda x . z(x)).  \]
We cannot replace the variable $y$ by the variable $z$ here, since 
this variable is `free', not bound.  Some care must be taken to make
the notions of free and bound variables precise, but we shall gloss over 
this issue, referring the reader to the references above for details.

The lambda calculus is a very simple formalism.  Amazingly, starting
from just this, Church and Kleene were able to build up Boolean logic,
the natural numbers, the usual operations of arithmetic, and so on.
For example, they defined `Church numerals' as follows:
\[  \begin{array}{ccl}
       \overline{0} &=& (\lambda f . (\lambda x . x))          \\
       \overline{1} &=& (\lambda f . (\lambda x . f(x)))   \\
       \overline{2} &=& (\lambda f . (\lambda x . f(f(x))))   \\
       \overline{3} &=& (\lambda f . (\lambda x . f(f(f(x)))))
\end{array}   \]
and so on.  Note that $f$ is a variable above.  Thus, the Church
numeral $\overline{n}$ is the program that `takes any program to the
$n$th power': if you give it any program $f$ as input, it returns the
program that applies $f$ $n$ times to whatever input $x$ it receives.

To get a feeling for how we can define arithmetic operations on 
Church numerals, consider
\[          \lambda g . \overline{3}(\overline{2}(g))  . \]
This program takes any program $g$, squares it, and then cubes the result.
So, it raises $g$ to the sixth power.  This suggests that
\[       \lambda g . \overline{3}(\overline{2}(g)) = \overline{6} . \]
Indeed this is true.  If we treat the definitions of Church numerals
as reversible rewrite rules, then we can start with the left side of
the above equation and grind away using rewrite rules until we reach
the right side:
\[
\begin{array}{ccll}
 (\lambda g . \overline{3} (\overline{2} (g))
 &=& (\lambda g . \overline{3} ((\lambda f . (\lambda x . f(f(x))))) (g)) 
  & \textrm{def.\ of } \overline{2} \\
 &=& (\lambda g . \overline{3} (\lambda x . g(g(x))))  
  & \textrm{beta} \\
 &=& (\lambda g . (\lambda f . (\lambda x . f(f(f(x))))) 
     (\lambda x . g(g(x)))) 
  & \textrm{def.\ of } \overline{3} \\
 &=& (\lambda g . (\lambda x . (\lambda x . g(g(x)))
     ((\lambda x . g(g(x))) ((\lambda x . g(g(x))) (x))))) 
  & \textrm{beta} \\
 &=& (\lambda g . (\lambda x . (\lambda x. g(g(x)))
     ((\lambda g . g(g(x))) (g(g(x))))))
  & \textrm{beta} \\
 &=& (\lambda g . (\lambda x . (\lambda x . g(g(x))) (g(g(g(g(x))))))) 
  & \textrm{beta} \\
 &=& (\lambda g . (\lambda x . g(g(g(g(g(g(x))))))))  
  & \textrm{beta} \\
 &=& \overline{6} 
   & \textrm{def.\ of } \overline{6} 
\end{array}
\]
If this calculation seems mind-numbing, that is precisely the point:
it resembles the inner workings of a computer.  We see here how the
lambda calculus can serve as a programming language, with each step of
computation corresponding to a rewrite rule.

Of course, we got the answer $\overline{6}$ because $3 \times 2 = 6$. 
Generalizing from this example, we can define a program called `times' that
multiplies Church numerals:
\[    \Times = (\lambda a . (\lambda b . (\lambda x . a(b(x))))) .\]
For example, 
\[
\Times(\overline{3})(\overline{2}) = \overline{6} .
\]
The enterprising reader can dream up similar programs for the other
basic operations of arithmetic.  With more cleverness, Church and
Kleene were able to write terms corresponding to more complicated
functions.  They eventually came to believe that {\it all} computable
functions $f \maps \NN \to \NN$ can be defined in the lambda calculus.

Meanwhile, G\"odel was developing another approach to computability,
the theory of `recursive functions'.  Around 1936, Kleene proved that
the functions definable in the lambda calculus were the same as
G\"odel's recursive functions.  In 1937 Turing described his `Turing
machines', and used these to give yet another definition of computable
functions.  This definition was later shown to agree with the other
two.  Thanks to this and other evidence, it is now widely accepted
that the lambda calculus can define {\it any} function that can be
computed by {\it any} systematic method.  We say it is `Turing
complete'.

After this burst of theoretical work, it took a few decades for
programmable computers to actually be built.  It took even longer for
computer scientists to profit from Church and Kleene's insights. 
This began around 1958, when McCarthy invented the programming
language Lisp, based on the lambda calculus \cite{McCarthy}.  
In 1965, an influential paper by Landin \cite{Landin} pointed 
out a powerful analogy between the lambda calculus and the language 
ALGOL.    These developments led to a renewed interest in the lambda 
calculus which continues to this day.  By now, a number of computer 
languages are explicitly based 
on ideas from the lambda calculus.  The most famous of these include Lisp, 
ML and Haskell.  These languages, called `functional programming languages',
are beloved by theoretical computer scientists for their conceptual
clarity.  In fact, for many years, everyone majoring in computer
science at MIT has been required to take an introductory course that
involves programming in Scheme, a dialect of Lisp.  The cover of the
textbook for this course \cite{Sussman} even has a big $\lambda$ on
the cover!

We should admit that languages of a different sort --- `imperative
programming languages' --- are more popular among working programmers.
Examples include FORTRAN, BASIC, and C.  In imperative programming, a
program is a series of instructions that tell the computer what to do.
By constrast, in functional programming, a program simply describes a
function.  To run the program, we apply it to an input.  So, as in the
lambda calculus, `application' is a fundamental operation in functional 
programming.  If we combine application with lambda abstraction, we
obtain a language powerful enough to compute any computable function.  

However, most functional programming languages are more regimented
than the original lambda calculus.  As we have seen, in the lambda
calculus as originally developed by Church and Kleene, any term can be
applied to any other.  In real life, programming involves many kinds of 
data.  For example, suppose we are writing a program that involves days 
of the week.  It would not make sense to write 
\[       \Times(\overline{3})(\Tuesday)  \]
because Tuesday is not a number.  We might choose to represent Tuesday
by a number in some program, but doubling that number doesn't have a
good interpretation: is the first day of the week Sunday or Monday?
Is the week indexed from zero or one?  These are arbitrary choices
that affect the result.  We could let the programmer make the choices,
but the resulting unstructured framework easily leads to mistakes.  

It is better to treat data as coming in various `types', such as
integers, floating-point numbers, alphanumeric strings, and so on.
Thus, whenever we introduce a variable in a program, we should make
a `type declaration' saying what type it is.  For example, we might
write:
\[ \begin{array}{l}

  \Tuesday:\Day
\end{array}
\]
This notation is used in Ada, Pascal and some other languages.  Other
notations are also in widespread use.  Then, our system should have a
`type checker' (usually part of the compiler) that complains if we try
to apply a program to a piece of data of the wrong type.

Mathematically, this idea is formalized by a more sophisticated
version of the lambda calculus: the `typed' lambda calculus, 
where every term has a type.  This idea is also fundamental to
category theory, where every morphism is like a black box with 
input and output wires of specified types:
\[\begin{pspicture}(0,0)(2,4)
\psset{angleA=-90,angleB=90,ArrowInside=->,arrowscale=2}
\pnode(1,4){A}
\rput(1,2){\ovalnode{B}{$f$}}
\pnode(1,0){C}
\nccurve{A}{B} \nbput{$X$}
\nccurve{B}{C} \nbput{$Y$}
\end{pspicture}\]
and it makes no sense to hook two black boxes together unless
the output of the first has the same type as the input of the next:
\begin{center}
    \begin{pspicture}(0,0)(2,6)
    \psset{angleA=-90,angleB=90,ArrowInside=->,arrowscale=2}
    \pnode(1,6){A}
    \rput(1,4){\ovalnode{B2}{$f$}}
    \rput(1,2){\ovalnode{B1}{$g$}}
    \pnode(1,0){C}
    \nccurve{A}{B2} \nbput{$X$}
    \nccurve{B2}{B1} \nbput{$Y$}
    \nccurve{B1}{C} \nbput{$Z$}
    \end{pspicture}
\end{center}

Indeed, there is a deep relation between the typed lambda calculus and
cartesian closed categories.  This was discovered by Lambek in 1980
\cite{Lambek}.  Quite roughly speaking, a `typed lambda-theory' is a
very simple functional programming language with a specified
collection of basic data types from which other more complicated types
can be built, and a specified collection of basic terms from which
more complicated terms can be built.  The data types of this language
are {\it objects} in a cartesian closed category, while the programs ---
that is, terms --- give {\it morphisms!}

Here we are being a bit sloppy.  Recall from Section \ref{theories}
that in logic we can build closed monoidal categories where the
morphisms are equivalence classes of proofs.  We need to take
equivalence classes for the axioms of a closed monoidal category to
hold.  Similarly, to get closed monoidal categories from computer
science, we need the morphisms to be equivalence classes of terms.
Two terms count as equivalent if they differ by rewrite rules such as
beta reduction, eta reduction and alpha conversion.  As we have seen,
these rewrites represent the steps whereby a program carries out its
computation.  For example, in the original `untyped' lambda calculus,
the terms $\Times(\overline{3})(\overline{2})$ and $\overline{6}$
differ by rewrite rules, but they give the same morphism.  So, when we
construct a cartesian closed category from a typed lambda-theory, we
{\it neglect the actual process of computation}.  To remedy this we
should work with a cartesian closed 2-category which has:
\begin{itemize}
\item types as objects,
\item terms as morphisms,
\item equivalence classes of rewrites as 2-morphisms.
\end{itemize}
For details, see the work of Seely \cite{Seely}, Hilken \cite{Hilken},
and Melli\'es \cite{Mellies}.  Someday this work will be part of the
larger $n$-categorical Rosetta Stone mentioned at the end of Section
\ref{symmetric}.

In any event, Lambek showed that every typed lambda-theory gives a
cartesian closed category --- and conversely, every cartesian closed
category gives a typed lambda-theory.  This discovery led to a rich
line of research blending category theory and computer science.  There
is no way we can summarize the resulting enormous body of work, though
it constitutes a crucial aspect of the Rosetta Stone.  Two good
starting points for further reading are the textbook by Crole
\cite{Crole} and the online review article by Scott \cite{Scott}.

In what follows, our goal is more limited.  First, in Section
\ref{lambda}, we explain how every `typed lambda-theory' gives a
cartesian closed category, and conversely.  We follow the treatment of
Lambek and Scott \cite{LS}, in a somewhat simplified form.  Then, in
Section \ref{linear_type_theories}, we describe how every `linear type
theory' gives a closed symmetric monoidal category, and conversely.

The idea here is roughly that a `linear type theory' is a programming 
language suitable for {\it both classical and quantum computation}.  This
language differs from the typed lambda calculus in that it forbids
duplication and deletion of data except when expressly permitted.  The
reason is that while every object in a cartesian category comes
equipped with `duplication' and `deletion' morphisms:
\[   \Delta_X \maps X \to X \tensor X, \qquad !_X \maps X \to 1 , \]
a symmetric monoidal category typically lacks these.  As we saw in
Section \ref{monoidal}, a great example is the category $\Hilb$ with
its usual tensor product.  So, a programming language suitable for
quantum computation should not assume we can duplicate all types
of data \cite{ChuangNielsen,WZ}.

Various versions of `quantum' or `linear' lambda calculus have already
been studied, for example by Benton, Bierman de Paiva and Hyland
\cite{BBPH}, Dorca and van Tonder \cite{vT}, and Selinger and Valiron
\cite{SV}.  Abramsky and Tzevelekos sketch a version in their paper in
this volume \cite{AT}.  We instead explain the `linear type theories'
developed by Simon Ambler in his 1991 thesis \cite{Ambler}.

\subsection{The Typed Lambda Calculus}
\label{lambda}

Like the original `untyped' lambda calculus explained above, the typed
lambda calculus uses terms to represent both programs and data.
However, now every term has a specific type.  A program that inputs
data of type $X$ and outputs data of type $Y$ is said to be of type $X
\lhom Y$.  So, we can only apply a term $s$ to a term $t$ of type $X$
if $s$ is of type $X \lhom Y$ for some $Y$.  In this case $s(t)$ is a
well-defined term of type $Y$.  We call $X \lhom Y$ a {\bf function
type}.

Whenever we introduce a variable, we must declare its type.  We write
$t\!:\!X$ to mean that $t$ is a term of type $X$.  So, in lambda
abstraction, we no longer simply write expressions like $(\lambda
x\, . \,t)$.  Instead, if $x$ is a variable of type $X$, we write
\[      (\lambda x\!:\!X \; .\; t) . \] 
For example, here is a simple program that
takes a program of type $X \lhom X$ and `squares' it:
\[       (\lambda f\!:\!X \lhom X \;.\;(\lambda \!x:\!X \; . \; f(f(x)))) .\]

In the original lambda calculus, all programs take a single piece of
data as input.  In other words, they compute unary functions.  This is
no real limitation, since we can handle functions that take more than
one argument using a trick called `currying', discussed in Section
\ref{closed}  This turns a function of several arguments into a
function that takes the first argument and returns a function of the
remaining arguments.  We saw an example in the last section: the
program `$\Times$'.  For example, $\Times(\overline{3})$ is a program
that multiplies by 3, so $\Times(\overline{3})(\overline{2}) =
\overline{6}$.

While making all programs compute unary functions is economical, it is
not very kind to the programmer.  So, in the typed lambda calculus we
also introduce products: given types $X$ and $Y$, there is a type $X
\times Y$ called a {\bf product type}.  We can think of a datum of
type $X \times Y$ as a pair consisting of a datum of type $X$ and a
datum of type $Y$.  To make this intuition explicit, we insist that
given terms $s:X$ and $t:Y$ there is a term $(s,t):X \times Y$.  We
also insist that given a term $u:X \times Y$ there are terms $p(u):X$
and $p'(u):Y$, which we think of as the first and second components of
the pair $t$.  We also include rewrite rules saying:
\[      
\begin{array}{ccll}
(p(u),p'(u)) &=& u & \textrm{for all } u:X \times Y, \\
p(s,t) &=& s  & \textrm{for all } s\!:\!X \textrm{ and } t\!:\!Y, \\
p'(s,t) &=& t & \textrm{for all } s\!:\!X \textrm{ and } t\!:\!Y .
\end{array}
\]

Product types allow us to write programs that take more than one
input.  Even more importantly, they let us deal with programs that
produce more than one output.  For example, we might have a type
called `integer'.  Then we might want a program that takes an integer
and duplicates it:
\[ 
  \duplicate: \integer \lhom (\integer \times \integer)
\]
Such a program is easy to write:
\[  \duplicate = (\lambda x\!:\!\integer \;.\; (x,x))  .\]
Of course this a program we should {\it not} be allowed to write
when duplicating information is forbidden, but in this section
our considerations are all `classical', i.e., suited to cartesian
closed categories.

The typed lambda calculus also has a special type called the `unit
type', which we denote as $1$.  There is a single term of this type,
which we denote as $()$.  From the viewpoint of category theory, the
need for this type is clear: a category with finite products must have
not only binary products but also a terminal object (see Definition
\ref{finite_products}).  For example, in the category $\Set$, the
terminal object can be taken as any one-element set, and $()$ is the
unique element of this set.  It may be less clear why this type is
useful in programming.  One reason is that it lets us think of a
constant of type $X$ as a function of type $1 \lhom X$ --- that is,
a `nullary' function, one that takes no arguments.  There are some
other reasons, but they go beyond the scope of this discussion.  
Suffice it to say that Haskell, Lisp and even widely used imperative 
languages such as C, C++ and Java include the unit type.

Having introduced the main ingredients of the typed lambda calculus,
let us give a more formal treatment.  As we shall see, a `typed
lambda-theory' consists of types, terms and rewrite rules.  From a
typed lambda-theory we can get a cartesian closed category.  The types
will give objects, the terms will give morphisms, and the rewrite
rules will give equations between morphisms.

First, the {\bf types} are given inductively as follows:
\begin{itemize}
\item {\bf Basic types:}
There is an arbitarily chosen set of types called {\bf basic types}.
\item {\bf Product types:}
Given types $X$ and $Y$, there is a type $X \times Y$.
\item {\bf Function types:}
Given types $X$ and $Y$, there is a type $X \lhom Y$.
\item {\bf Unit type:} There is a type $1$.
\end{itemize}
There may be unexpected equations between types: for example we may 
have a type $X$ satisfying $X \times X = X$.  However,
we demand that:
\begin{itemize}
\item
If $X = X'$ and $Y = Y'$ then $X \times Y = X' \times Y'$.
\item
If $X = X'$ and $Y = Y'$ then $X \lhom Y = X' \lhom Y'$.
\end{itemize}

Next we define {\bf terms}.  Each term has a specific type, and if $t$
is a term of type $X$ we write $t\!:\!X$.  The rules for building terms
are as follows:
\begin{itemize}
  \item {\bf Basic terms:} For each type $X$ there is a set of
{\bf basic terms} of type $X$.
  \item {\bf Variables:} For each type $X$ there is a countably 
infinite collection of terms of type $X$ called {\bf variables}
of type $X$.
  \item {\bf Application:} 
If $f:X \lhom Y$ and $t:X$ then there is a term $f(t)$ of type
$Y$.
  \item {\bf Lambda abstraction:} 
If $x$ is a variable of type $X$ and $t:Y$ 
then there is a term $(\lambda x\! : \!X \, .\, t)$ of type $X \lhom Y$.
  \item {\bf Pairing:} If $s:X$ and $t:Y$ then there is a term $(s,t)$ of
type $X \times Y$.
  \item {\bf Projection:}
If $t:X \times X'$ then there is a term $p(t)$ of $X$ and a term
$p'(t)$ of type $X'$.
  \item {\bf Unit term:} There is a term $()$ of type $1$.
\end{itemize}

Finally there are {\bf rewrite rules} going between terms of the same
type.  Given any fixed set of variables ${\cal S}$, there will be
rewrite rules between terms of the same type, all of whose free
variables lie in the set ${\cal S}$.  For our present purposes, we
only need these rewrite rules to decide when two terms determine the
same morphism in the cartesian closed category we shall build.  So,
what matters is not really the rewrite rules themselves, but the
equivalence relation they generate.  We write this equivalence
relation as $s {\; {\sim}_{\cal S} \; } t$.

The relation $\sim_{\cal S}$ can be any equivalence relation
satisfying the following list of rules.  In what follows, $t[s/x]$
denotes the result of taking a term $t$ and replacing every free
occurence of the variable $x$ by the term $s$.  Also, when when we say
`term' without further qualification, we mean `term all of whose free
variables lie in the set ${\cal S}$'.
\begin{itemize}
\item {\bf Type preservation:} If $t \sim_{\cal S} t'$ then 
$t$ and $t'$ must be terms of the same type, all of whose free variables
lie in the set $\cal S$.
\item {\bf Beta reduction:} Suppose $x$ is a variable of
type $X$, $s$ is a term of type $X$, and $t$ is any term.
If no free occurrence of a variable 
in $s$ becomes bound in $t[s/x]$, then:
\[    (\lambda x\!:\!X\,.\, t)(s) { \;\; {\sim}_{\cal S} \;\; } t[s/x] .\]
\item {\bf Eta reduction:} Suppose the variable $x$ does not appear in the 
term $f$.  Then:
\[        (\lambda x\! : \! X \, .\, f(x)) {\; {\sim}_{\cal S} \; } f . \]
\item {\bf Alpha conversion:}  Suppose $x$ and $y$ are variables of
type $X$, and no free occurrence of any variable in $t$ becomes bound 
in $t[x/y]$.  Then:
\[        (\lambda x\! : \! X \; . \; t) \;\; {\sim}_{\cal S} \;\;
(\lambda y\!:\!X \; . \; t[x/y]). \]
\item {\bf Application:} 
Suppose $t$ and $t'$ are terms of type $X$ with $t \sim_{\cal S} t'$, 
and suppose that $f\! :\! X \lhom Y$.  Then:
\[       f(t) \;\; \sim_{\cal S} \;\;  f(t').  \]
\item {\bf Lambda abstraction:}
Suppose $t$ and $t'$ are terms of type $Y$, all of whose 
free variables lie in the set ${\cal S} \cup \{x\}$.  Suppose 
that $t \sim_{{\cal S} \cup \{ x \}} t'$.  Then:
\[        (\lambda x\! : \! X \; . \; t) \;\; {\sim}_{\cal S} \;\;
(\lambda x\!:\!X \; . \; t') \]
\item {\bf Pairing:} If $u$ is a term of type $X \times Y$ then:
\[  (p(u),p'(u)) \;\; {\sim}_{\cal S} \;\; u .\]
\item {\bf Projection:} if  $s$ is a term of type $X$ and $t$
is a term of type $Y$ then:
\[
\begin{array}{c}
p(s,t) \;\; {\sim}_{\cal S} \;\;  s  \\
p'(s,t) \;\; {\sim}_{\cal S} \;\;  t.
\end{array}
\]
\item {\bf Unit term:} If $t$ is a term of type $1$ then:
\[          t \;\; {\sim}_{\cal S} \;\;  ().  \]
\end{itemize}

Now we can describe Lambek's classic result relating typed
lambda-theories to cartesian closed categories.  From a typed
lambda-theory we get a cartesian closed category $C$ for which:
\begin{itemize}
\item
The objects of $C$ are the types.
\item 
The morphisms $f \maps X \to Y$ of $C$
are equivalence classes of pairs
$(x,t)$ consisting of a variable $x\!:\!X$ and a term $t\!:\!Y$ with
no free variables except perhaps $x$.  Here $(x,t)$ is equivalent
to $(x,t')$ if and only if:
\[     t \; \sim_{\{x\}} \; t'[x/x'] . \]
\item Given a morphism $f \maps X \to Y$
coming from a pair $(x,t)$ and a morphism $g \maps Y \to Z$
coming from a pair $(y,u)$ as above, the composite 
$gf \maps X \to Y$ comes from the pair $(x,u[t/y])$.
\end{itemize}
We can also reverse this process and get a typed lambda-theory from a
cartesian closed category.  In fact, Lambek and Scott nicely explain
how to construct a category of category of cartesian closed categories
and a category of typed-lambda theories.  They construct functors
going back and forth between these categories and show these functors
are inverses up to natural isomorphism.  We thus say these categories
are `equivalent' \cite{LS}.

\subsection{Linear Type Theories}
\label{linear_type_theories}

In his thesis \cite{Ambler}, Ambler described how to generalize
Lambek's classic result from cartesian closed categories to closed
symmetric monoidal categories.  To do this, he replaced typed
lambda-theories with `linear type theories'.  A linear type theory can
be seen as a programming language suitable for both classical and
quantum computation.  As we have seen, in a noncartesian category like
$\Hilb$, we cannot freely duplicate or delete information.  So linear
type theories must prevent duplication or deletion of data {\it except
when it is expressly allowed}.

To achieve this, linear type theories must not allow us to write a
program like this:
\[    (\lambda x\!:\!X \; . \; (x,x)) .      \]
Even a program that `squares' another program, like this:
\[   
 (\lambda f\!:\!X \lhom X \;.\;(\lambda \!x:\!X \; . \; f(f(x)))), \]
is not allowed, since it `reuses' the variable $f$.  On the other
hand, a program that composes two programs is allowed!

To impose these restrictions, linear type theories treat variables
very differently than the typed lambda calculus.  In fact, in a linear
type theory, any term will contain a given variable at most {\it
once}.  But linear type theories depart even more dramatically from
the typed lambda calculus in another way.  They make no use of lambda
abstraction!  Instead, they use `combinators'.

The idea of a combinator is very old: in fact, it predates the lambda
calculus.  Combinatory logic was born in a 1924 paper by Sch\"onfinkel
\cite{Schoenfinkel}, and was rediscovered and extensively developed
by Curry \cite{Curry} starting in 1927.  In retrospect, we can see
their work as a stripped-down version of the untyped lambda calculus
that completely avoids the use of variables.  Starting from a basic
stock of terms called `combinators', the only way to build new ones is
application: we can apply any term $f$ to any term $t$ and get a term
$f(t)$.

To build a Turing-complete programming language in such an impoverished
setup, we need a sufficient stock of combinators.  Remarkably, it
suffices to use three.  In fact it is possible to use just {\it one}
cleverly chosen combinator --- but this tour de force is not
particularly enlightening, so we shall describe a commonly used set of
three.  The first, called $I$, acts like the identity, since it comes
with the rewrite rule:
\[     I(a) = a  \]
for every term $a$.  The second, called $K$, gives a constant function
$K(a)$ for each term $a$.  In other words, it comes with a rewrite
rule saying
\[   K(a)(b) = a  \]
for every term $b$.  The third, called $S$, is the tricky one.  It
takes three terms, applies the first to the third, and applies the
result to the second applied to the third:
\[   S(a)(b)(c) = a(c)(b(c)).  \]

Later it was seen that the combinator calculus can be embedded in
the untyped lambda calculus as follows:
\[
\begin{array}{ccl}
  I &=& (\lambda x . x)  \\
  K &=& (\lambda x . (\lambda y . x)) \\
  S &=& (\lambda x . (\lambda y . (\lambda z . x(z)(y(z)) ))) .
\end{array}
\]
The rewrite rules for these combinators then follow from rewrite rules
in the lambda calculus.  More surprisingly, any function computable
using the lambda calculus can also be computed using just $I, K$ and
$S$!  While we do not need this fact to understand linear type
theories, we cannot resist sketching the proof, since it is a classic
example of using combinators to avoid explicit use of lambda
abstraction.

Note that all the variables in the lambda calculus formulas for $I,
K,$ and $S$ are bound variables.  More generally, in the lambda
calculus we define a {\bf combinator} to be a term in which all
variables are bound variables.  Two combinators $c$ and $d$ are {\bf
extensionally equivalent} if they give the same result on any input:
that is, for any term $t$, we can apply lambda calculus rewrite rules
to $c(t)$ and $d(t)$ in a way that leads to the same term.  There is
a process called `abstraction elimination' that takes any combinator
in the lambda calculus and produces an extensionally equivalent one
built from $I, K,$ and $S$.

Abstraction elimination works by taking a term $t = (\lambda x.u)$
with a single lambda abstraction and rewriting it into the form
$(\lambda x.f(x))$, where $f$ has no instances of lambda abstraction.
Then we can apply eta reduction, which says $(\lambda x. f(x)) = f$.
This lets us rewrite $t$ as a term $f$ that does not involve lambda
abstraction.  We shall use the notation $[[u]]_x$ to mean `any term
$f$ satisfing $f(x) = u$'.

There are three cases to consider; each case justifies the definition 
of one combinator:
\begin{enumerate}
  \item $t = (\lambda x.x)$. 
    We can rewrite this as $t = (\lambda x.I(x)),$ so $t = [[x]]_x = I$.  
  \item $t = (\lambda x. u)$, where $u$ does not depend on $x$.
    We can rewrite this as $t = (\lambda x. K(u)(x)),$ so 
   $t = [[u]]_x = K(u).$
  \item $t = (\lambda x. u(v)),$ where $u$ and $v$ may depend
    on $x$.  We can rewrite this as 
    $t = (\lambda x.(([[u]]_x x) ([[v]]_x x))$ or
    $t = (\lambda x. S([[u]]_x)([[v]]_x)(x))$, so $t = S([[u]]_x)([[v]]_x).$
\end{enumerate}

We can eliminate all use of lambda abstraction from any term by
repeatedly using these three rules `from the inside out'.  To see
how this works, consider the lambda term $t = (\lambda
x. (\lambda y. y)),$ which takes two inputs and returns the second.
Using the rules above we have:
\[
\begin{array}{ccl}
  (\lambda x. (\lambda y.y)) 
    &=& (\lambda x. (\lambda y. [[y]]_y(y))) \\
    &=& (\lambda x. (\lambda y. I(y))) \\
    &=& (\lambda x.I) \\
    &=& (\lambda x. [[I]]_x(x)) \\
    &=& (\lambda x. K(I)(x) \\
    &=& K(I) .
\end{array}
\]
We can check that it works as desired: $K(I)(x)(y) = I(y) = y.$  

Now let us return to our main theme: linear type theories.  Of the
three combinators described above, only $I$ is suitable for use in an
arbitrary closed symmetric monoidal category.  The reason is that $K$
deletes data, while $S$ duplicates it.  We can see this directly from
the rewrite rules they satisfy:
\[
\begin{array}{ccl}
 K(a)(b) &=& a   \\
 S(a)(b)(c) &=& a(c)(b(c)).
\end{array}
\]

Every linear type theory has a set of `basic combinators', which
neither duplicate nor delete data.  Since linear type theories
generalize {\it typed} lambda-theories, these basic combinators are
typed.  Ambler writes them using notation resembling the
notation for morphisms in category theory.   

For example, given two types $X$ and $Y$ in a
linear type theory, there is a {\bf tensor product type} $X \otimes
Y$.  This is analogous to a product type in the typed lambda calculus.
In particular, given a term $s$ of type $X$ and a term $t$ of type
$Y$, we can combine them to form a term of type $X \tensor Y$, which
we now denote as $(s \tensor t)$.  We reparenthesize iterated
tensor products using the following basic combinator:
\[   \assoc_{X,Y,Z}
      \maps (X\tensor Y)\tensor Z \to X \tensor (Y \tensor Z). \]
This combinator comes with the following rewrite rule:
\[   \assoc_{X,Y,Z}((s \tensor t) \tensor u) = (s \tensor (t \tensor u)) \]
for all terms $s:X$, $t:Y$ and $u:Z$.  

Of course, the basic combinator $\assoc_{X,Y,Z}$ is just a mildly
disguised version of the associator, familiar from category theory.
Indeed, all the basic combinators come from natural or dinatural
transformations implicit in the definition of `closed symmetric
monoidal category'.  In addition to these, any given
linear type theory also has combinators called `function symbols'.
These come from the morphisms particular to a given category.
For example, suppose in some category the tensor product $X \otimes X$
is actually the cartesian product.  Then the corresponding linear type
theory should have a function symbol
\[        \Delta_X \maps X \to X \tensor X \]
which lets us duplicate data of type $X$, together with function symbols
\[         p \maps X \tensor X \to X, \qquad p' \maps X \tensor X \to X \]
that project onto the first and second factors.  To make sure
these work as desired, we can include rewrite rules:
\[ 
\begin{array}{ccc}
      \Delta(s) &=& (s \otimes s)  \\
      p(s \otimes t) &=& s \\
      p'(s \otimes t) &=& t  .
\end{array}
\]
So, while duplication and deletion of data is not a `built-in feature'
of linear type theories, we can include it when desired.  

Using combinators, we could try to design a programming language
suitable for closed symmetric monoidal categories that completely
avoid the use of variables.  Ambler follows a different path.  He
retains variables in his formalism, but they play a very different ---
and much {\it simpler} ---- role than they do in the lambda calculus.
Their only role is to help decide which terms should count as
equivalent.  Furthermore, lambda abstraction plays no role in linear
type theories, so the whole issue of free versus bound variables does
not arise!  In a sense, all variables are free.  Moreover, every term
contains any given variable at most once.

After these words of warning, we hope the reader is ready for a more
formal treatment of linear type theories.  A {\bf linear type theory}
has types, combinators, terms, and rewrite rules.  The types will
correspond to objects in a closed symmetric monoidal category, while
equivalence classes of combinators will correspond to morphisms.  
Terms and rewrite rules are only used to define the equivalence relation.

First, the set of {\bf types} is defined inductively as follows:
\begin{itemize}
\item {\bf Basic types:}
There is an arbitarily chosen set of types called {\bf basic types}.
\item {\bf Product types:}
Given types $X$ and $Y$, there is a type $(X \tensor Y)$.
\item {\bf Function types:}
Given types $X$ and $Y$, there is a type $(X \lhom Y)$.
\item {\bf Trivial type:} There is a type $I$.
\end{itemize}
There may be equations between types, but we require that:
\begin{itemize}
\item
If $X = X'$ and $Y = Y'$ then $X \tensor Y = X' \tensor Y'$.
\item
If $X = X'$ and $Y = Y'$ then $X \lhom Y = X' \lhom Y'$.
\end{itemize}

Second, a linear type theory has for each pair of types $X$ and $Y$
a set of \textbf{combinators} of the form $f\maps X \to Y$.
These are defined by the following inductive rules:
\begin{itemize}
  \item Given types $X$ and $Y$ there is an arbitrarily chosen set
        of combinators $f\maps X\to Y$ called \textbf{function symbols}.
  \item Given types $X,Y,$ and $Z$ we have the following combinators,
        called \textbf{basic combinators}:
    \begin{itemize}
      \item $\Id_X \maps X \to X$
      \item $\assoc_{X,Y,Z}
      \maps (X\tensor Y)\tensor Z \to X \tensor (Y \tensor Z)$
      \item $\unassoc_{X,Y,Z}
      \maps X \tensor (Y \tensor Z) \to (X \tensor Y)\tensor Z$
      \item $\braid_{X,Y}\maps X \tensor Y \to Y \tensor X$ 
      \item $\Left_X \maps I \tensor X \to X$
      \item $\unleft_X \maps X \to I \tensor X$
      \item $\Right_X \maps I \tensor X \to X$
      \item $\unright_X \maps X \to I \tensor X$
      \item $\eval_{X,Y} \maps X \tensor (X \lhom Y) \to Y$      
    \end{itemize}
  \item If ${f\maps X \to Y}$ and ${g\maps Y \to Z}$ are combinators, then 
    $(g \circ f) \maps X \to Z$ is a combinator.
  \item If ${f\maps X \to Y}$ and ${g\maps X' \to Y'}$ are combinators, then
    ${(f \tensor g) \maps X \tensor X' \to Y \tensor Y'}$ is a combinator.
  \item If ${f\maps X \tensor Y \to Z}$ is a combinator, then we can
   {\bf curry} $f$ to obtain a combinator
    $\tilde{f} \maps Y \to (X \lhom Z)$.
\end{itemize}
It will generally cause no confusion if we leave out the subscripts on
the basic combinators.  For example, we may write simply `$\assoc$'
instead of $\assoc_{X,Y,Z}$.

Third, a linear type theory has a set of {\bf terms} of any given type.
As usual, we write $t:X$ to say that $t$ is a term of type $X$.  Terms
are defined inductively as follows:
\begin{itemize}
  \item  For each type $X$ there is a countably infinite collection of 
\textbf{variables} of type $X$.  If $x$ is a variable of type $X$ then 
$x:X$.
  \item There is a term $1$ with $1:I$.
  \item If $s:X$ and $t:Y,$ then there is a term $(s \tensor t)$ with
$(s \tensor t): X \tensor Y$, as long as no variable appears in both $s$
and $t$.
  \item If $f \maps X \to Y$ is a combinator and $t:X$ then there 
is a term $f(t)$ with $f(t):X$.
\end{itemize}
Note that any given variable may appear at most once in a term.

Fourth and finally, a linear type theory has {\bf rewrite rules} going
between terms of the same type.  As in our treatment of the typed
lambda calculus, we only care here about the equivalence relation
$\sim$ generated by these rewrite rules.  This equivalence relation
must have all the properties listed below.  In what follows, we say a
term is {\bf basic} if it contains no combinators.  Such a term is
just an iterated tensor product of distinct variables, such as
\[              (z \tensor ((x \tensor y) \tensor w)) .\]

These are the properties that the equivalence relation $\sim$ must have:
\begin{itemize}
\item
If $t \sim t'$ then $t$ and $t'$ must be terms of the same type,
containing the same variables.
\item The equivalence relation is {\bf substitutive}: 
\begin{itemize} 
\item 
Given terms $s \sim s'$, a variable $x$ of type $X$,
and terms $t \sim t'$ of type $X$ whose variables appear in 
neither $s$ nor $s'$, then $s[t/x] \sim s'[t'/x]$.  
\item 
Given a basic term $t$ with the same type as a variable
$x$, if none of the variables of $t$ appear in the terms
$s$ or $s'$, and $s[t/x] \sim s'[t/x]$, then $s \sim s'$.
\end{itemize}
\item The equivalence relation is {\bf extensional}: if 
$f : X \lhom Y$, $g: X \lhom Y$ and 
$\eval(t \tensor f) = \eval(t \tensor g)$ for all basic terms 
$t:X$, then $f = g$.  
\item We have:
    \begin{itemize}
      \item $\Id(s) \sim s$
      \item $(g \circ f)(s) \sim g(f(s))$
      \item $(f \tensor g)(s \tensor t) \sim (f(s)\tensor g(t))$
      \item $\assoc((s \tensor t) \tensor u) 
             \sim (s \tensor (t \tensor u))$
      \item $\unassoc(s \tensor (t \tensor u))
             \sim ((s \tensor t) \tensor u)$
      \item $\braid(s \tensor t) \sim (t \tensor s)$
      \item $\Left(1 \tensor s) \sim s$
      \item $\unleft(s) \sim (1 \tensor s)$
      \item $\Right(1 \tensor s) \sim s$
      \item $\unright(s) \sim (1 \tensor s)$
      \item $\eval(s \tensor \tilde{f}(t)) \sim f(s \tensor t)$
    \end{itemize}
\end{itemize}

Note that terms can have variables appearing anywhere within them.
For example, if $x,y,z$ are variables of types $X,Y$ and $Z$, and
$f\maps Y \tensor Z \to W$ is a function symbol, then
\[ \braid(x \tensor f(y \tensor z)) \]
is a term of type $W \tensor X$.  However, every term $t$ is
equivalent to a term of the form $\cp(t)(\vp(t))$, where $\cp(t)$
is the {\bf combinator part} of $t$ and $\vp(t)$ is a basic term
called the {\bf variable part} of $t$.  For example, the above term is
equivalent to
\[ \braid \circ (\Id \tensor (f \circ (\Id \tensor \Id)))
(x \tensor (y \tensor z)). \]
The combinator and variable parts can be computed inductively as follows:
\begin{itemize}
  \item If $x$ is a variable of type $X$, $\cp(x) = \Id\maps X \to X$.
  \item $\cp(1) = \Id\maps I \to I$.
  \item For any terms $s$ and $t$, $\cp(s \tensor t) = \cp(s) \tensor \cp(t)$.
  \item For any term $s:X$ and any combinator $f : X \to Y$, 
$\cp(f(s)) = f \circ \cp(s)$. \\
  \item If $x$ is a variable of type $X$, $\vp(x) = x$.
  \item $\vp(1) = 1$.
  \item For any terms $s$ and $t$, $\vp(s \tensor t) = \vp(s) \tensor \vp(t)$.
  \item For any term $s:X$ and any combinator $f \maps X \to Y$, 
$\vp(f(s)) = \vp(s)$.
\end{itemize}

Now, suppose that we have a linear type theory.  Ambler's first main
result is this: there is a symmetric monoidal category where objects
are types and morphisms are equivalence classes of
combinators.  The equivalence relation on combinators is defined as
follows: two combinators $f, g \maps X \to Y$ are equivalent if and
only if
\[                    f(t) \sim g(t)  \]
for some basic term $t$ of type $X$.  In fact, Ambler shows that
$f(t) \sim g(t)$ for {\it some} basic term $t:X$ if and only if 
$f(t) \sim g(t)$ for {\it all} such basic terms.

Ambler's second main result describes how we can build a linear type
theory from any closed symmetric monoidal category, say $C$.  Suppose $C$ 
has composition $\square$, tensor product $\bullet$, internal hom
$\multimapdot$, and unit object $\iota$.  We let the basic types of
our linear type theory be the objects of $C$.  We take as equations
between types those generated by:
\begin{itemize}
  \item $\iota = I$
  \item $A \bullet B = A \tensor B$
  \item $A \multimapdot B = A \lhom B$
\end{itemize}
We let the function symbols be all the morphisms of $C$.  We 
take as our equivalence relation on terms the smallest allowed
equivalence relation such that:
\begin{itemize}
  \item $1_A(x) \sim A$
  \item $(g \,\square\, f)(x) \sim g (f(x))$
  \item $(f \bullet g)(x \tensor y) \sim (f(x) \tensor g(y))$
  \item $a_{A,B,C}((x \tensor y) \tensor z) \sim (x \tensor (y \tensor z))$
  \item $b_{A, B}(x \tensor y) \sim (y \tensor x)$
  \item $l_A(1 \tensor x) \sim x$
  \item $r_A(x \tensor 1) \sim x$
  \item $\ev_{A, B}(x \tensor \tilde{f}(y)) \sim f(x \tensor y)$
\end{itemize}
Then we define
\begin{itemize}
  \item $\Id = 1$
  \item $\assoc = a$
  \item $\unassoc = a^{-1}$
  \item $\braid = b$
  \item $\Left = l$
  \item $\unleft = l^{-1}$
  \item $\Right = r$
  \item $\unleft = r^{-1}$
  \item $\eval = \ev$
  \item $g \circ f = g \,\square\, f$
\end{itemize}
and we're done!

Ambler also shows that this procedure is the `inverse' of his
procedure for turning linear type theories into closed symmetric
monoidal categories.  More precisely, he describes a category of
closed symmetric monoidal categories (which is well-known), and also a
category of linear type theories.  He constructs functors going back
and forth between these, based on the procedures we have sketched, and
shows that these functors are inverses up to natural isomorphism.
So, these categories are `equivalent'.

In this section we have focused on closed symmetric monoidal
categories.  What about closed categories that are just braided
monoidal, or merely monoidal?  While we have not checked the details,
we suspect that programming languages suited to these kinds of
categories can be obtained from Ambler's formalism by removing various
features.  To get the braided monoidal case, the obvious guess is to
remove Ambler's rewrite rule for the `$\braid$' combinator
and add two rewrite rules corresponding to the hexagon equations (see
Section \ref{braided} for these).  To get the monoidal case, the
obvious guess is to completely remove the combinator `$\braid$' and
all rewrite rules involving it.  In fact, Jay \cite{Jay1989} gave 
a language suitable for closed monoidal categories in 1989; Ambler's
work is based on this.  

\section{Conclusions}
\label{conclusions}

In this paper we sketched how category theory can serve to clarify the
analogies between physics, topology, logic and computation.  Each
field has its own concept of `thing' (object) and `process' (morphism)
--- and these things and processes are organized into categories that
share many common features.  To keep our task manageable, we focused
on those features that are present in every closed symmetric monoidal
category.  Table \ref{analogy_detailed}, an expanded version of the
Rosetta Stone, shows some of the analogies we found.

\begin{table}[h]
\begin{center}
\begin{tabular}{|c|c|c|c|c|}
\hline
Category Theory & Physics & Topology  & Logic & Computation 
\\
\hline
object $X$       &
Hilbert space $X$  &
manifold $X$ &
proposition $X$ &
data type $X$
\\
\hline
morphism &
operator &
cobordism &
proof  &
program  \\
$f\maps X \to Y$ &
$f\maps X \to Y$ &
$f \maps X \to Y$ &
$f \maps X \to Y$ &
$f \maps X \to Y$
\\
\hline
tensor product  &
Hilbert space   &
disjoint union &
conjunction &
product 
\\
of objects: &
of joint system: &
of manifolds: &
of propositions: &
of data types: 
\\
$X \tensor Y$ &
$X \tensor Y$ &
$X \tensor Y$ &
$X \tensor Y$ &
$X \tensor Y$
\\
\hline
tensor product of &
parallel  &
disjoint union of &
proofs carried out & 
programs executing
\\
morphisms: $f \tensor g$ & 
processes: $f \tensor g$ & 
cobordisms: $f \tensor g$ & 
in parallel: $f \tensor g$ &
in parallel: $f \tensor g$
\\
\hline
internal hom:   &
Hilbert space of &
disjoint union of  &
conditional &
function type:
\\
$X \lhom Y$    &
`anti-$X$ and $Y$':  &
orientation-reversed &
proposition: &
$X \lhom Y$
\\
    &
$X^* \tensor Y$    &
$X$ and $Y$: $X^* \tensor Y$ &
$X \lhom Y$ &
\\
\hline
\end{tabular}
\\
\caption{The Rosetta Stone (larger version)}
\label{analogy_detailed}
\end{center}
\end{table}

However, we only scratched the surface!   There is much
more to say about categories equipped with extra structure, and how
we can use them to strengthen the ties between physics, topology,
logic and computation ---
not to mention what happens when we go from categories to $n$-categories.
But the real fun starts when we exploit these analogies to come up
with new ideas and surprising connections.  Here is an example.

In the late 1980s, Witten \cite{Witten}
realized that string theory was deeply connected to a 3d topological 
quantum field theory and thus the theory of knots and tangles 
\cite{Kohno}.   This led to a huge explosion of work, which was 
ultimately distilled into a beautiful body of results focused on a 
certain class of compact braided monoidal categories called `modular 
tensor categories' \cite{BK,Turaev}.  

All this might seem of purely theoretical interest, were it not
for the fact that superconducting thin films in magnetic fields
seem to display an effect --- the `fractional quantum Hall effect' --- 
that can be nicely modelled with the help of such categories
\cite{Stern,Stone}.  In a nutshell, the idea is that excitations 
of these films can act like particles, called `anyons'.  When two 
anyons trade places, the result depends on how they go about
it:
\[\begin{pspicture}(0,0)(1,2)
\psset{angleA=-90,angleB=90,ArrowInside=->,arrowscale=2}
\pnode(0,2){A1}
\pnode(1,2){A2}
\pnode(0,0){B1}
\pnode(1,0){B2}
\nccurve[ArrowInsidePos=.25]{A1}{B2} \nbput[npos=0]{$ $} \ncput{\cnode[linecolor=white,fillstyle=solid,fillcolor=white]{4pt}{C}}
\nccurve[ArrowInsidePos=.25]{A2}{B1} \naput[npos=0]{$ $}
\end{pspicture}
\begin{pspicture}(0,0)(2,2)
\rput(1,1){$\ne$}
\end{pspicture}
\begin{pspicture}(0,0)(1,2)
\psset{angleA=-90,angleB=90,ArrowInside=->,arrowscale=2}\pnode(0,2){A1}
\pnode(1,2){A2}
\pnode(0,0){B1}
\pnode(1,0){B2}
\nccurve[ArrowInsidePos=.25]{A2}{B1} \naput[npos=0]{$ $} \ncput{\cnode[linecolor=white,fillstyle=solid,fillcolor=white]{4pt}{C}}
\nccurve[ArrowInsidePos=.25]{A1}{B2} \nbput[npos=0]{$ $} 
\end{pspicture}\]

So, collections of anyons are described by objects in a braided 
monoidal category!  The details depend on things like the strength 
of the magnetic field; the range of possibilities can be worked out 
with the help of modular tensor categories \cite{MR,RSW}.

So far this is all about physics and topology.
Computation entered the game around 2000, when Freedman, Kitaev,
Larsen and Wang \cite{FKW} showed that certain systems of anyons 
could function as `universal quantum computers'.  This means that,
in principle, arbitrary computations can be carried out by moving 
anyons around.  Doing this {\it in practice} will be far from easy.
However, Microsoft has set up a research unit called Project Q 
attempting to do just this.  After all, a working quantum 
computer could have huge practical consequences.  

But regardless of whether topological quantum computation ever becomes
practical, the implications are marvelous.  A simple diagram 
like this:
\[
  \begin{pspicture}(0,0)(2,4)
  \psset{angleA=-90,angleB=90,ArrowInside=->,arrowscale=2}
  \pnode(0,4){A1}
  \pnode(1,4){B1}
  \pnode(2,4){C1}
  \pnode(0,2){A2}
  \pnode(1,2){B2}
  \pnode(2,2){C2}
  \pnode(0,0){A3}
  \pnode(1,0){B3}
  \pnode(2,0){C3}
  \nccurve[ArrowInsidePos=1]{A1}{B2} \nbput[npos=0]{$ $} \ncput{\cnode[linecolor=white,fillstyle=solid,fillcolor=white]{4pt}{C}}
  \nccurve[ArrowInsidePos=1]{B1}{A2} \naput[npos=0]{$ $} 
  \nccurve[ArrowInsidePos=1]{C1}{C2} \naput[npos=0]{$ $} 
  \nccurve[ArrowInside=]{A2}{A3} \nbput[npos=1]{$ $} 
  \nccurve[ArrowInside=]{B2}{C3} \naput[npos=1]{$ $} \ncput{\cnode[linecolor=white,fillstyle=solid,fillcolor=white]{4pt}{C}}
  \nccurve[ArrowInside=]{C2}{B3} \nbput[npos=1]{$ $}
\end{pspicture}\]
can now be seen as a {\it quantum process}, a {\it tangle}, a
{\it computation} --- or an abstract morphism in any braided 
monoidal category!  This is just the sort of thing one would hope for 
in a general science of systems and processes.

\subsection*{Acknowledgements}

We owe a lot to participants of the seminar at UCR where some of this
material was first presented: especially David Ellerman, Larry Harper,
Tom Payne --- and Derek Wise, who took notes \cite{Seminar}.  This
paper was also vastly improved by comments by Andrej Bauer, Tim
Chevalier, Derek Elkins, Greg Friedman, Matt Hellige, Robin Houston,
Theo Johnson--Freyd, J\"urgen Koslowski, Todd Trimble, Dave Tweed, and
other regulars at the $n$-Category Caf\'e.  MS would like to thank
Google for letting him devote 20\% of his time to this research, and
Ken Shirriff for helpful corrections.  This work was supported by the
National Science Foundation under Grant No.\ 0653646.

\end{document}